\pdfoutput=1 
\documentclass[11pt,a4paper]{article}

\usepackage{fullpage}
\usepackage{xcolor}

\usepackage{mathtools,amssymb,amsfonts,graphicx,setspace}

\usepackage[nosort]{cite}
\usepackage{youngtab} 

\DeclareMathAlphabet{\mathsfit}{T1}{\sfdefault}{\mddefault}{\sldefault}
\SetMathAlphabet{\mathsfit}{bold}{T1}{\sfdefault}{\bfdefault}{\sldefault}

\usepackage{booktabs} 
\usepackage{array}    
\newcolumntype{L}{>{$}l<{$}} 
\newcolumntype{R}{>{$}r<{$}} 
\newcolumntype{C}{>{$}c<{$}} 

\usepackage[unicode,pdfborder={0 0 0},colorlinks=false,linktocpage=true]{hyperref} 
\AtBeginDocument{\hypersetup{unicode,pdfborder={0 0 0},colorlinks=false,linktocpage=true}}

\numberwithin{equation}{section}

\makeatletter
\g@addto@macro\bfseries{\boldmath}
\makeatother

\DeclareFontFamily{U}{bbold}{}
\DeclareFontShape{U}{bbold}{m}{n}
 {  <-5.5> s*[1.04] bbold5
    <5.5-6.5> s*[1.04] bbold6
    <6.5-7.5> s*[1.04] bbold7
    <7.5-8.5> s*[1.04] bbold8
    <8.5-9.5> s*[1.04] bbold9
    <9.5-11.5> s*[1.04] bbold10
    <11.5-16> s*[1.04] bbold12
    <16-> s*[1.04] bbold17
 }{}

\usepackage[bbgreekl]{mathbbol}
\DeclareSymbolFontAlphabet{\mathbbm}{bbold}
\DeclareSymbolFontAlphabet{\mathbb}{AMSb}%

\usepackage{dsfont}

\newcommand{\nn}{\nonumber}

\newcommand{\cA}{{\mathcal{A}}\xspace}

\newcommand{\cE}{{\mathcal{E}}\xspace}

\newcommand{\cG}{{\mathcal{G}}\xspace}

\newcommand{\cL}{{\mathcal{L}}\xspace}
\newcommand{\cM}{{\mathcal{M}}\xspace}
\newcommand{\cN}{{\mathcal{N}}\xspace}

\newcommand{\cP}{{\mathcal{P}}\xspace}

\newcommand{\cS}{{\mathcal{S}}\xspace}

\newcommand{\cU}{{\mathcal{U}}\xspace}
\newcommand{\cV}{{\mathcal{V}}\xspace}

\newcommand{\fG}{{\mathfrak{G}}\xspace}

\newcommand{\bbP}{{\mathbb{P}}\xspace}

\newcommand{\bbT}{{\mathbb{T}}\xspace}

\newcommand{\bbmE}{{\mathbbm{E}}\xspace}

\newcommand{\bbmV}{{\mathbbm{V}}\xspace}

\newcommand{\dsR}{{\mathds{R}}\xspace}

\newcommand{\dsT}{{\mathds{T}}\xspace}

\newcommand{\dsZ}{{\mathds{Z}}\xspace}

\newcommand{\sfC}{{\mathsf{C}}\xspace}

\newcommand{\sfF}{{\mathsf{F}}\xspace}

\newcommand{\bbLambda}{\mathbbm{\Lambda}}

\newcommand{\mf}[1]{{\mathfrak{#1}}}
\newcommand{\ul}[1]{{\underline{#1}}}
\newcommand{\dd}{{\mathrm{d}}}
\newcommand{\Tr}{{\mathrm{Tr}}}

\usepackage{xspace} 

\newcommand{\twL}[1]{\cL^{\raisebox{.5ex}{$\scriptscriptstyle[#1]$}}}

\makeatletter

\DeclareRobustCommand{\SO}{\@ifnextchar({\@SO}{\@@SO}}
\def\@SO(#1){\ensuremath{\mathrm{SO}(#1)}\xspace}
\def\@@SO#1(#2){\ensuremath{\mathrm{SO}^{#1}\kern-1pt(#2)}\xspace}

\DeclareRobustCommand{\so}{\@ifnextchar({\@so}{\@@so}}
\def\@so(#1){\ensuremath{\mathfrak{so}(#1)}\xspace}
\def\@@so#1(#2){\ensuremath{\mathfrak{so}^{#1}\kern-1pt(#2)}\xspace}

\DeclareRobustCommand{\Spin}{\@ifnextchar({\@Spin}{\@@Spin}}
\def\@Spin(#1){\ensuremath{\mathrm{Spin}(#1)}\xspace}
\def\@@Spin#1(#2){\ensuremath{\mathrm{Spin}^{#1}\kern-1pt(#2)}\xspace}

\DeclareRobustCommand{\spin}{\@ifnextchar({\@spin}{\@@spin}}
\def\@spin(#1){\ensuremath{\mathfrak{spin}(#1)}\xspace}
\def\@@spin#1(#2){\ensuremath{\mathfrak{spin}^{#1}\kern-1pt(#2)}\xspace}

\DeclareRobustCommand{\CSO}{\@ifnextchar({\@CSO}{\@@CSO}}
\def\@CSO(#1){\ensuremath{\mathrm{CSO}(#1)}\xspace}
\def\@@CSO#1(#2){\ensuremath{\mathrm{CSO}^{#1}\kern-1pt(#2)}\xspace}

\DeclareRobustCommand{\O}{\@ifnextchar({\@O}{\@@O}}
\def\@O(#1){\ensuremath{\mathrm{O}(#1)}\xspace}
\def\@@O#1(#2){\ensuremath{\mathrm{O}^{#1}\kern-1pt(#2)}\xspace}

\DeclareRobustCommand{\cso}{\@ifnextchar({\@cso}{\@@cso}}
\def\@cso(#1){\ensuremath{\mathfrak{cso}(#1)}\xspace}
\def\@@cso#1(#2){\ensuremath{\mathfrak{cso}^{#1}\kern-1pt(#2)}\xspace}

\DeclareRobustCommand{\SU}{\@ifnextchar({\@SU}{\@@SU}}
\def\@SU(#1){\ensuremath{\mathrm{SU}(#1)}\xspace}
\def\@@SU#1(#2){\ensuremath{\mathrm{SU}^{#1}\kern-1pt(#2)}\xspace}

\DeclareRobustCommand{\su}{\@ifnextchar({\@su}{\@@su}}
\def\@su(#1){\ensuremath{\mathfrak{su}(#1)}\xspace}
\def\@@su#1(#2){\ensuremath{\mathfrak{su}^{#1}\kern-1pt(#2)}\xspace}

\DeclareRobustCommand{\SL}{\@ifnextchar({\@SL}{\@@SL}}
\def\@SL(#1){\ensuremath{\mathrm{SL}(#1)}\xspace}
\def\@@SL#1(#2){\ensuremath{\mathrm{SL}^{#1}\kern-1pt(#2)}\xspace}

\let\sl\relax
\DeclareRobustCommand{\sl}{\@ifnextchar({\@sl}{\@@sl}}
\def\@sl(#1){\ensuremath{\mathfrak{sl}(#1)}\xspace}
\def\@@sl#1(#2){\ensuremath{\mathfrak{sl}^{#1}\kern-1pt(#2)}\xspace}

\DeclareRobustCommand{\GL}{\@ifnextchar({\@GL}{\@@GL}}
\def\@GL(#1){\ensuremath{\mathrm{GL}(#1)}\xspace}
\def\@@GL#1(#2){\ensuremath{\mathrm{GL}^{#1}\kern-1pt(#2)}\xspace}

\DeclareRobustCommand{\gl}{\@ifnextchar({\@gl}{\@@gl}}
\def\@gl(#1){\ensuremath{\mathfrak{gl}(#1)}\xspace}
\def\@@gl#1(#2){\ensuremath{\mathfrak{gl}^{#1}\kern-1pt(#2)}\xspace}

\makeatother

\def\Sp(#1){\ensuremath{\mathrm{Sp}(#1)}\xspace}
\def\USp(#1){\ensuremath{\mathrm{USp}(#1)}\xspace}
\def\U(#1){\ensuremath{\mathrm{U}(#1)}\xspace}
\def\ISO(#1){\ensuremath{\mathrm{ISO}(#1)}\xspace}

\def\E#1{\ensuremath{\mathrm{E}_{#1(#1)}}\xspace}
\def\e#1{\ensuremath{\mathfrak{e}_{#1(#1)}}\xspace}

\def\En{\ensuremath{\mathrm{E}_{n(n)}}\xspace}
\def\en{\ensuremath{\mathfrak{e}_{n(n)}}\xspace}

\def\KEn{\ensuremath{K(\mathrm{E}_{n(n)})}\xspace}

\makeatother

\makeatletter
\usepackage{tensind}
\tensordelimiter{;}
\tensorformat{c}
\begingroup\lccode`~=``
\lowercase{\endgroup\def~}{\@ifnextchar_{\@tenssub}{\@tensother}}
\def\@tenssub_#1`{;{}_#1;}
\def\@tensother{\@ifnextchar^{\@tenssup}{\@tens}}
\def\@tenssup#1`{;{}^#1;}
\def\@tens#1`{;#1;}
\mathcode``="8000
\makeatother

\newcommand{\str}{\scriptscriptstyle\rm\xspace}
\newcommand{\de}{\partial}

\newcommand{\Rv}{\mathbf{R}_{\rm v}}
\newcommand{\RTheta}{\mathbf{R}_{\Theta}}

\newcommand{\weitz}[1]{W^{\raisebox{.5ex}{$\scriptscriptstyle[#1]$}}}
\newcommand{\tors}[1]{T^{\raisebox{.5ex}{$\scriptscriptstyle[#1]$}}}

\makeatletter
\DeclareRobustCommand{\dorf}{\@ifnextchar[{\@dorf}{\circ}}
\def\@dorf[#1]{\overset{\smash{#1}}{\circ}}
\makeatother

\makeatletter
\let\oldvartheta\vartheta
\undef\vartheta
\DeclareRobustCommand{\vartheta}{\@ifnextchar[{\@vartheta}{\oldvartheta}}
\def\@vartheta[#1]{\oldvartheta^{\raisebox{.5ex}{$\scriptscriptstyle[#1]$}}}
\makeatother

\makeatletter
\let\oldTheta\Theta
\undef\Theta
\DeclareRobustCommand{\Theta}{\@ifnextchar[{\@Theta}{\oldTheta}}
\def\@Theta[#1]{\oldTheta^{\raisebox{.5ex}{$\scriptscriptstyle[#1]$}}}
\makeatother

\newcommand{\hGgauge}{\ensuremath{\widehat{\rm{G}}_{\str gauge}}\xspace}%
\newcommand{\Ggauge}{\ensuremath{{\rm{G}}_{\str gauge}}\xspace}%
\newcommand{\hHgauge}{\ensuremath{\widehat{\rm{H}}_{\str gauge}}\xspace}%
\newcommand{\Hgauge}{\ensuremath{{\rm{H}}_{\str gauge}}\xspace}%
\newcommand{\hhgauge}{\ensuremath{\widehat{\mf{h}}_{\str gauge}}\xspace}%

\newcommand{\hggauge}{\ensuremath{\widehat{\mf{g}}_{\str gauge}}\xspace}
\newcommand{\ggauge}{\ensuremath{{\mf{g}}_{\str gauge}}\xspace}

\begin{document}

\titlepage

\vspace*{1.5cm}

\begin{center}
{\bf \Large How to uplift $D=3$ maximal supergravities}

\vspace*{2cm} \textsc{%
Gianluca Inverso$^{a}$ and
Davide Rovere$^{b,a}$
} \\

\vspace*{0.5cm} $^a$ 
INFN, Sezione di Padova\\ Via Marzolo 8, 35131 Padova, Italy\\

\vspace*{0.5cm} $^b$
Dipartimento di Fisica e Astronomia `G. Galilei', Universit\`a di Padova,\\ Via Marzolo 8, 35131 Padova, Italy\\

\vspace*{1cm}
\texttt{gianluca.inverso@pd.infn.it},\ \ \texttt{davide.rovere@studenti.unipd.it}

\end{center}

\vspace*{2.5cm}

\begin{abstract}

We prove necessary and sufficient algebraic conditions to determine whether a $D=3$ gauged maximal supergravity can be obtained from consistent Kaluza--Klein truncation of ten- or eleven-dimensional supergravity.
We describe the procedure to identify the internal geometry and explicitly construct the frame encoding the reduction ansatz.
As byproducts, we derive several results on twistings, deformations and global aspects of \E8 exceptional geometry and define \E8 generalised diffeomorphism for massive IIA supergravity.
We devise simple algebraic conditions for imposing compactness of the internal space and derive no-go results for the uplift of compact gaugings and of a large class of gauged maximal supergravities with $\cN=(8,0)$ AdS$_3$ solutions.

\end{abstract}

\vspace*{0.5cm}

\newpage
\tableofcontents
\newpage

\section{Introduction}
\label{sec: intro}

In the last decade there has been a lot of progress in the study of consistent Kaluza--Klein truncations of ten- and eleven-dimensional supergravities.
In such truncations one puts the original supergravity theory on some $d$-dimensional internal manifold and identifies a slice of the space of field configurations whose restricted dynamics reproduce a $(11{-}d)$- or $(10{-}d)$-dimensional gauged supergravity theory.
The truncation is consistent if solving the gauged supergravity equations of motion also solves the ones of the higher-dimensional parent theory.
This makes gauged supergravities obtained through consistent trucation a inestimable tool for generating new solutions and studying their physical properties, especially in the context of black-hole physics and holography.

It is of special interest to study consistent truncations preserving maximal supersymmetry.
The resulting gauged maximal supergravities present a plethora of different physical properties depeding on their gauge groups and embeddings, and often are the starting point for carrying out further truncations to smaller models with fewer degrees of freedom. 
Supersymmetry is naturally preserved if one considers so-called Scherk--Schwarz trucations on tori and group manifolds \cite{Scherk:1978ta,Scherk:1979zr}, but it is extremely interesting to look beyond this class of manifolds.
Some relevant examples are the reduction of 11d supergravity on $S^7$ which can be truncated to \SO(8) gauged maximal supergravity in 4 dimensions~\cite{deWit:1986mz,deWit:1986oxb,deWit:1982bul,Nicolai:2011cy}, and similar examples for consistent truncations of 11d supergravity on $S^4$~\cite{Nastase:1999cb} and IIB supergravity on $S^5$~\cite{Lee:2014mla,Hohm:2014qga}.
It is indicative of the complexity of these reduction ansaetze that the proof of the latter's consistent truncation to $D=5$ \SO(6) gauged maximal supergravity was completed only relatively recently.
Many more examples of consistent truncations have been developed since~\cite{Guarino:2015jca,Guarino:2015vca,Inverso:2016eet,Malek:2017cle,DallAgata:2019klf,Samtleben:2019zrh,Bossard:2022wvi,Bossard:2023jid,Eloy:2023zzh}.
Some examples with immediate holographic applications have been the truncation of massive IIA supergravity on $S^6$~\cite{Guarino:2015jca,Guarino:2015vca}, a class of S-folds of IIB supergravity on $S^5\times S^1$, each giving a certain $D=4$ `dyonic' CSO gauged supergravity \cite{Inverso:2016eet}, new truncations of type IIB supergravity on $S^3\times S^3\times S^1$~\cite{Eloy:2023zzh}, as well as the truncation of 11d supergravity on $S^8\times S^1$ to \SO(9) gauged maximal supergravity in two dimensions~\cite{Bossard:2022wvi,Bossard:2023jid}.

This surge of new results is fueled by the development of a general framework that captures the structure of consistent truncations to gauged supergravities.
Exceptional field theories~\cite{Hohm:2013pua,Hohm:2013vpa,Hohm:2013uia,Hohm:2014fxa,Abzalov:2015ega,Musaev:2015ces,Berman:2015rcc,Bossard:2018utw,Bossard:2019ksx,Bossard:2021jix,Bossard:2021ebg} and exceptional generalised geometries~\cite{Coimbra:2011ky,Coimbra:2012af}, organise the full field content, gauge symmetries and dynamics of 11d and Type II supergravities in terms of \En covariant objects.
The group \En is the global symmetry group of $(11{-}n)$-dimensional maximal \emph{un}gauged supergravity, as obtained from a standard Kaluza--Klein reduction on a torus.
In these frameworks, consistent Kaluza--Klein truncations are obtained by factorising the dependence on the internal manifold of all fields and gauge parameters in terms of a coordinate-dependent element of $\En\times\dsR^+$, often called the `twist matrix' or `frame'.
Provided some differential conditions are solved, the factorisation leads to a gauged maximal supergravity in $D=(11{-}n)$ dimensions, such that the gauge couplings are encoded into a so-called \emph{generalised torsion} constructed from the frame.
This procedure is called a \emph{generalised} Scherk--Schwarz reduction~\cite{Grana:2008yw,Aldazabal:2011nj,Geissbuhler:2011mx,Grana:2012rr,Berman:2012uy,Aldazabal:2013mya,Berman:2013cli,Aldazabal:2013via,Lee:2014mla,Hohm:2014qga}.
The exceptional field theory framework also allows to efficently compute the complete Kaluza--Klein mass spectrum for any vacuum solution found through such truncations~\cite{Malek:2019eaz,Malek:2020yue}.

\smallskip
Maximal supergravity theories admit many different gaugings, best encoded in terms of the embedding tensor formalism~\cite{Nicolai:2000sc,deWit:2004nw,deWit:2007kvg} (see \cite{Samtleben:2008pe} and \cite{Trigiante:2016mnt} for reviews, and references therein).
Not only there exists no full classification of inequivalent gaugings for $D<8$, but it is well-known that only a subset of gaugings can be obtained from consistent Kaluza--Klein truncations.
Perhaps the most prominent example is the no-go result of~\cite{Lee:2015xga}  which rules out an higher-dimensional supergravity origin for the one-parameter family of inequivalent \SO(8) gauged maximal supergravities discovered in~\cite{DallAgata:2012mfj}.
It is therefore highly desirable to determine conditions on the gauge couplings (i.e., on the embedding tensor) of a maximal supergravity that can differentiate between those models admitting an embedding in eleven- or ten-dimensional supergravity, and those which do not.
One may as well hope that classifying gauged maximal supergravities with such uplifts should be easier than classifying all gaugings altogether.

In \cite{Inverso:2017lrz}, it was proved that the requirement for existence of an uplift can be cast in a duality invariant way, by constructing a coset space $\hGgauge/\hHgauge$ from the gauge group $\hGgauge$ and imposing certain algebraic constraints on the associated embedding tensor.
In the same paper, the procedure is described to {explicitly} construct the $\En\times\dsR^+$ frame (see also~\cite{duBosque:2017dfc} for earlier progress in $D=7$).
The theorems in~\cite{Inverso:2017lrz} apply to $D\ge4$.
These results have also been reframed in terms of certain generalisations of the notion of an algebroid in \cite{Bugden:2021wxg,Bugden:2021nwl,Hulik:2022oyc,Hulik:2023aks}, see also \cite{Hassler:2022egz}.
The results in \cite{Inverso:2017lrz} also allowed to derive no-go results for the existence of a generalised Scherk--Schwarz uplift of some classes of gaugings of $D=4$ maximal supergravity, and to identify alternative uplifts for certain other models, such as electric and `dyonic' CSO gaugings, giving examples beyond group manifolds of what has been later labelled `Poisson--Lie U-duality'~\cite{Sakatani:2019zrs,Malek:2019xrf}.

The case of $D=3$ gauged maximal supergravities was excluded from~\cite{Inverso:2017lrz} because the gauge structure of \E8 exceptional field theory deviates from its lower--rank siblings.
The internal gauge symmetries of an exceptional field theory act through a generalised Lie derivative, and an important step in~\cite{Inverso:2017lrz} was to classify the most general way to locally `twist' or deform the generalised Lie derivative while preserving closure of its action on fields and other gauge parameters.
Such deformations were found earlier to be necessary to encode the Romans mass of Type IIA supergravity in exceptional field theory and exceptional generalised geometry \cite{Ciceri:2016dmd,Cassani:2016ncu}. 
In \E8 exceptional field theory, closure of the internal gauge symmetries of \E8 exceptional field theory requires the introduction of `ancillary' gauge parameters, which are absent for \En with $\le7$.
This affects not only the classification of twists and deformations of the generalised Lie derivative, but also the very definition of generalised torsion on which generalised Scherk--Schwarz reductions are based.
Last but not least, it complicates the way several objects patch together globally along the internal manifold.
In particular the `\emph{untwisting}' procedure usually employed in exceptional generalised geometry to encode the global properties of generalised vectors must be amended for \E8.

\smallskip

In this paper we address these complications. We then work out conditions analogous to~\cite{Inverso:2017lrz} for a $D=3$ gauged maximal supergravity to admit a geometric uplift to a higher-dimensional theory.
We also study a few covenient approaches that can be used to reframe and exploit these results and that can be applied to $D>3$ as well.
We show that one can rephrase the uplift conditions found in \cite{Inverso:2017lrz} and here in terms of constraints linear in the embedding tensor, at the price of explicitly breaking \En covariance.
These conditions are sufficient, and also necessary \emph{up to duality orbit}, meaning that it is enough to solve the above relations after $`X_AB^C`$ has been rotated by some \En element.
Such conditions are mapped to the conditions described in \cite{Bugden:2021wxg,Bugden:2021nwl,Hulik:2022oyc,Hulik:2023aks} in a rather different language (see also~\cite{Hassler:2022egz}, where some partial results for \E8 are also presented).
Here we extend them to \E8 and tabulate their solutions.%
\footnote{A set of necessary (up to duality orbit) linear uplift conditions for $D=2$ gauged supergravities has also been computed in \cite{Bossard:2023jid}. Determining sufficiency would require to repeat the analysis in this paper for E$_9$ exceptional field theory.}
This allows to greatly reduce the amount of independent entries of an embedding tensor and therefore drastically simplifies the classification of inequivalent models admitting an uplift.
Furthermore, we shall observe that there is a simple approach to requiring compactness of the internal manifold associated to a generalised Scherk--Schwarz reduction.
This can be phrased again in terms of linear algebraic constraints on the embedding tensor.

\smallskip

Let us summarise here the uplift conditions we derive in this paper.
One must consider coset spaces $\hGgauge/\hHgauge$ constructed out of the gauge group \hGgauge and takes the projection of the embedding tensor on the coset generators.
We denote the latter $\widehat\Theta_A{}^{\ul m}$, with $A$ running over \e8 and $\ul m$ over the coset generators.
This projection must satisfy the \emph{section constraint}
\begin{equation}\label{SC Theta INTRO}
  `Y^AB_CD`\,\widehat\Theta_A{}^{\ul m}\,\widehat\Theta_B{}^{\ul n}\ =\ 0\,,
\end{equation}
where $`Y^AB_CD`$ is a constant $\E8\times\dsR^+$ invariant tensor determined by exceptional field theory (see~\eqref{Ytens} and~\eqref{Ytens e8} below).
The geometric interpretation of this constraint is that one must be able to embed the vectors generating the transitive action of $\hGgauge$ on the internal manifold into the $\E8\times\dsR^+$ valued matrix that encodes the generalised Scherk--Schwarz ansatz.
A second constraint must also be imposed to guarantee consistency of the fluxes threading the internal space.
It reads
\begin{equation}\label{extracstr INTRO}
  `Y^AB_CD` \, \Big( `\vartheta_A` - \, \widehat\Theta_G{}^{\ul m} `t_{\ul m\,A}^G` \Big) \, \widehat\Theta_B{}^{\ul n}\ =\ 0\,,
\end{equation}
where $`\vartheta_A`$, if non-vanishing, is the component of the embedding tensor that encodes a gauging of the trombone $\dsR^+$ symmetry of $D=3$ maximal supergravity. We denoted $`t_{\ul m\,A}^B`$ the coset generators in the adjoint representation of \e8.
Whenever such uplift exists, we provide the explicit construction of the generalised frame that encodes the generalised Scherk--Schwarz reduction, which is obtained as the product of the $\hGgauge/\hHgauge$ coset representative, the \hHgauge frame constructed from its Maurer--Cartan form  and a matrix $`S_M^N`$ obtained from local integration of the background fluxes threading the internal geometry.
The latter are constructed explicitly from the embedding tensor and coset representative and we prove they satisfy the appropriate Bianchi identities to admit local integration.

Let us now summarise how these conditions can be rephrased as constraints linear in the embedding tensor.
We shall do so for gauged maximal supergravities in generic dimension~$D$.
The embedding tensor is a constant object $`X_AB^C`$ subject to certain representation and quadratic constraints (see \eqref{LCX} and \eqref{QCX} below).
The objects $\vartheta$ and $\widehat\Theta_A{}^{\ul m}$ above are extracted from $`X_AB^C`$.
We select a projector $\Pi`_A^B`$ satisfying the section constraint
\begin{equation}\label{SCPi INTRO}
  `Y^AB_CD`\,\Pi`_A^E`\,\Pi`_B^F`\ =\ 0\,,
\end{equation}
and its complement $\overline\Pi`_A^B` = \delta_A^B-\Pi`_A^B`$.
In exceptional field theory, solutions to the section constraint encode to which higher-dimensional theory we want to uplift a $D$-dimensional gauged maximal supergravity. This is usually a choice between 11d, type IIA or type IIB supergravity.
We shall find the following uplift constraints linear in the embedding tensor:
\begin{align}\label{linear uplift INTRO}\nonumber
  `X_(AB)^C`\Pi`_C^D`\  &=\  0\,,\\[1ex]
  \overline\Pi`_A^E`\,\overline\Pi`_B^F`\,`X_EF^G`\,\Pi`_G^C`\  &=\  0\,,\\[1ex]\nonumber
  `Y^AB_CD`\, \big( \vartheta_A+\omega\,`X_AF^G``\Pi_G^F` \big) \,\Pi`_B^E` \ &=\ 0\,,
\end{align}
where $\omega$ is the weight of a generalised vector in \En exceptional field theory.
These conditions are sufficient, and also necessary \emph{up to duality orbit}, meaning that it is enough to solve the above relations after $`X_AB^C`$ has been rotated by some \En element.

The linear conditions above are of better use when one wants to carry out a classification of models admitting an uplift.
If one is interested in finding an uplift for a \emph{specific} gauging, the duality invariant conditions \eqref{SC Theta INTRO}, \eqref{extracstr INTRO} are more convenient, but still require to look for different choices of the coset space $\hGgauge/\hHgauge$.
It is then desirable to look for duality-invariant necessary conditions that may allow to quickly rule out an uplift for certain classes of gaugings.
The embedding tensor $`X_AB^C`$ for Lagrangian gaugings of $D=3$  maximal supergravities is required to sit in the representations $\mathbf{1}+\mathbf{3875}$.
We shall find that the singlet component must vanish for a Lagrangian gauging to admit an uplift, and that traces of powers of the $\bf3875$ component must vanish as well.
The latter condition extends a constraint found in~\cite{Eloy:2023zzh}.\footnote{Necessary, duality invariant conditions for $D=2$ gauged maximal supergravities were found in~\cite{Bossard:2023wgg}.}
We will then prove no-go results for compact gaugings and for the class of gaugings admitting supersymmetric AdS$_3$ vacua described in~\cite{Deger:2019tem}.

\bigskip

The structure of this paper is as follows.
In section~\ref{sec: Exft and gSS} we review the relevant structures of \En exceptional field theories of rank lower than 8 as well as generalised Scherk--Schwarz reductions and the general results of~\cite{Inverso:2017lrz}.
The main computations and results of the paper are contained in sections~\ref{sec: 3d uplifts} and~\ref{ssec:gss uplift cond e8}, where we study the construction of generalised torsions, the notion of twisting/untwisting of generalised diffeomorphisms and the consistency conditions to introduce deformations to the generalised Lie derivative. 
A byproduct of this analysis is the costruction of the \E8 generalised Dorfman product encoding the gauge structure of massive type IIA supergravity.
With these results at hand, in section~\ref{ssec:gss uplift cond e8} we generalise to \E8 the construction of \cite{Inverso:2017lrz} and also rephrase the uplift conditions found so far as linear constraints on the embedding tensor. 
We then tabulate the available `geometric' gauge couplings.
In section~\ref{sec:examples} we show how to impose compactness of the internal space, discuss several examples and prove no-go theorems for some large classes of gaugings.
We make some final comments in section~\ref{sec:outro}.

\section{Exceptional field theory and consistent truncations}
\label{sec: Exft and gSS}

\begin{table}\centering\Yboxdim{.67em}
\begin{tabular}{CLCC}
\toprule
D & \text{Group} & \Rv& \RTheta \\
\midrule
9 & \GL+(2) & \mathbf2_3 + \mathbf 1_{-4} & \mathbf2_{-3} + \mathbf 3_{4}  \\
8 & \SL(3)\times\SL(2) & (\mathbf2,\,\mathbf3') 
                       & (\mathbf2,\,\mathbf3+\mathbf6')\\
7 & \SL(5) & \bf10' & \mathbf{15}+\mathbf{40}' \\
6 & \SO(5,5) & \mathbf{16}_c &\mathbf{144}_c \\
5 & \E6 & \bf27 & \bf351' \\
4 & \E7 & \bf56 &\bf912 \\ 
3 & \E8 & \bf248 &{\bf1}+{\bf3875} \\ 
\bottomrule
\end{tabular}
\label{tab:groups and irreps}
\caption{Summary of groups and representations for relevant instances of (super)gravity theories and associated double/exceptional/extended geometries. Trombone charges are not displayed but can be normalised to $+1$ for vector representations and $-1$ for the embedding tensor. The trombone component of the embedding tensor always sits in the conjugate of $\Rv$.}
\end{table}

We begin by summarising some basic facts about exceptional field theories (ExFTs) and generalised Scherk--Schwarz (gSS) reductions. We refer in particular to the papers \cite{Berman:2012vc,Hohm:2013vpa,Hohm:2013uia,Lee:2014mla,Hohm:2014qga} for the structure of ExFTs and \cite{Lee:2014mla,Hohm:2014qga,Inverso:2017lrz} (and references therein) for gSS reductions.

\subsection{Gauge structure of exceptional field theories}
\label{ssec: exft review}

ExFTs repackage the field content, gauge symmetries and complete dynamics of 11d and type II supergravities in a formally \En covariant fashion. 
The relation is obtained by a partial gauge fixing of the ten- or eleven-dimensional Lorentz symmetry and a dimensional split of spacetime coordinates into an `external' spacetime of dimension $D=11-n$ (with coordinates denoted $x^\mu$ thorughout this paper) and a $d$-dimensional internal space (with coordinates $y^m$).\footnote{In ExFT one usually introduces a set of `exceptional' internal coordinates $Y^M$ of which physical ones are a subset. We avoid this notation here.} 
We have $d=n$ for eleven-dimenional supergravity and $d=n-1$ for Type II supergravities.
Fields are rearranged in a way similar to Kaluza--Klein reductions and end up being encoded into the types of objects appearing in $D$-dimensional maximal supergravities.
In particular, the bosonic field content of \En ExFTs (with $n\le7$) consists of a $D$-dimensional metric $g_{\mu\nu}(x,y)$, scalar fields parametrising the symmetric space $\En/\KEn$,\footnote{We denote $K(\mathrm{G})$ the maximal compact subgroup of a Lie group $G$.} described by a (unimodular, symmetric) generalised metric $\cM_{MN}(x,y)$, vectors $A^M_\mu(x,y)$, and a hierarchy of higher $p$-form fields in diverse \En representations.
The indices $M,\,N,\ldots$ denote the specific \En representation in which vector fields $A^M_\mu$ transform. We denote it $\Rv$ and list it in table~\ref{tab:groups and irreps} for the relevant groups and dimensions, together with other relevant representations discussed below.

The structure of \En ExFTs is mainly dictated by their internal gauge symmetries, called \emph{generalised diffeomorphisms}.
Up to $n\le7$, they are parametrised by \emph{generalised vectors} $\Lambda^M(x,y)$ and act on the field content of the theory through a generalised Lie derivative.
For our purposes this is most conveniently defined by acting on another generalised vector $V^M(x,y)$:
\begin{equation}\label{genlie with P, upto7}
\cL_{\Lambda} V^M = \Lambda^N \partial_N V^M +\omega \, \partial_N\Lambda^N \, V^M 
- \alpha \, `\bbP^P_Q^M_N`\partial_P\Lambda^Q\,V^N\,,
\end{equation}
where $\omega$ is a characteristic density weight and $`\bbP^P_Q^M_N`$ projects the product $\Rv\otimes\overline{\Rv}$ on the \en Lie algebra, thus guaranteeing that $\cL_\Lambda$ preserves the \En representations in which the ExFT fields reside.
The coefficients $\omega$ and $\alpha$ depend on the specific \En ExFT under consideration \cite{Berman:2012vc}.
Definition \eqref{genlie with P, upto7} generalises in the obvious way to tensorial objects.
An equivalent expression reads
\begin{equation}\label{genlie with Y, upto7}
\cL_{\Lambda}V^M = \Lambda^N \partial_N V^M - V^N \partial_N \Lambda^M + `Y^MP_QN` \partial_P\Lambda^Q\,V^N\,,
\end{equation}
where we introduced the invariant tensor
\begin{equation}\label{Ytens}
`Y^MP_QN` = \delta^M_Q \delta^P_N + \omega\,\delta^M_N \delta^P_Q -\alpha\,`\bbP^P_Q^M_N`\,.
\end{equation}

The partial derivative operators $\partial_M$ sit in the conjugate representation $\overline\Rv$. 
They are not all independent, rather they are subject to an \En invariant section constraint which reads
\begin{equation}\label{SCpartial}
`Y^MN_PQ` \, \partial_M\otimes\partial_N = 0\,.
\end{equation}
In this expression it is understood that $\partial_M$ and $\partial_N$ may act on any fields, gauge parameters or products and derivatives thereof.
It is effectively an algebraic constraint that tells us that the differential operators $\partial_M$ encode the derivatives with respect to the internal coordinates $y^m$, with $m=1,\ldots,d$, through a constant, $\mathrm{dim}\Rv\otimes d$ dimensional `section matrix' of rank $d$ which we denote $\cE_M{}^m$:
\begin{equation}\label{SCE}
\partial_M = \cE_M{}^m \frac{\partial}{\partial y^m}\,,\qquad `Y^MN_PQ`\, \cE_M{}^m\otimes\cE_N{}^n = 0\,.
\end{equation}
This presentation in terms of a section matrix will prove convenient in the next sections.
Notice that the range $d$ of the internal indices $m,n,\ldots$, and hence the dimensions and rank of $\cE_M{}^m$, may differ for inequivalent solutions of the section constraint.

The section constraint is necessary and sufficient to guarantee that generalised diffeomorphisms close onto themselves.
In fact, having imposed the section constraint the generalised Lie derivative satisfies an even stronger condition, namely the Leibniz identity
\begin{equation}
\label{Leib id upto7}
\cL_{\Lambda_1}\cL_{\Lambda_2}\Phi - \cL_{\Lambda_2}\cL_{\Lambda_1}\Phi- \cL_{\cL_{\Lambda_1}\Lambda_2}\Phi\ =\ 0\,,
\end{equation}
where $\Phi$ denotes any field or gauge parameter.

Solutions to \eqref{SCE} are classified by their \En orbit. 
One finds two inequivalent solutions of maximal rank \cite{Blair:2013gqa}, corresponding to 11d (with $d=n$) and type IIB supergravity (with $d=n-1$), respectively.
Generalised diffeomorphisms capture the infinitesimal diffeomorphisms and $p-$form potential gauge transformations of such theories along the `internal' $d-$dimensional space.
The massive type IIA supergravity theory can also be described by this formalism, but requires a deformation of the generalised Lie derivative (and of the ExFT action as well) \cite{Ciceri:2016dmd,Cassani:2016ncu}. 
Solutions to the section constraint of non-maximal rank allow to capture maximal supergravities in $D+d<10$ dimensions, and their gauged deformations are captured in terms of a deformation of the generalised Lie derivative, in the same fashion as the type IIA Romans mass.
Such deformations are described below.

\subsubsection{Section solutions and relevant \texorpdfstring{$\En\times\dsR^+$}{Eₙ₍ₙ₎×ℝ⁺} subgroups}
\label{ssec: En subgroups}

As long as we do not commit to a specific solution to the section constraint \eqref{SCE}, all relevant expressions in ExFT can be cast in a (formally) covariant form under rigid \En transformations, provided we let \En also act on $\de_M$.
A non-trivial property of ExFTs is that, once we select a specific section, the $\En\times\dsR^+$ rigid group of transformations is broken to a subgroup $\GL(d) \ltimes \cS$, where $\cS$ is the stabiliser of the section matrix while \GL(d) reproduces linear transformations $y^m \to y^n (g^{-1})_n{}^m$ of the internal coordinates.
Their action on the section matrix is then \mbox{given by}
\begin{align}\label{GLd and S}
G_M{}^N \cE_N{}^m &= \cE_M{}^n g_n{}^m \,,\quad&& G_M{}^N\in\GL(d)\subset\En\times\dsR^+\,,\ g_m{}^n \in \GL(d)\\\nonumber
S_M{}^N \cE_N{}^m &= \cE_M{}^m \,,&& S\in\cS\,.
\end{align}
\def\Gup{\ensuremath{\cG_{\scriptscriptstyle\text{uplift}}}\xspace}%
\def\Rup{\ensuremath{\dsR^+_{\scriptscriptstyle\text{uplift}}}\xspace}%
Under a change of (internal) coordinates, the Jacobian of the transformation acts on tensorial objects as an element of the \GL(d) group so defined.
Indeed, the field content and gauge parameters of ExFT can be decomposed into \GL(d) representations to resurface the components of the original $D+d$ dimensional supergravity fields.
If for instance the $D+d$ dimensional theory contains a $p$-form potential, the representation $\Rv$ then decomposes as follows with respect to \GL(d), reflecting the fact that generalised vectors encode infinitesimal gauge transformations along the internal space:
\Yvcentermath1 
\Yboxdim.5em 
\begin{equation}\label{Rv deco youngtab}
\Rv\ \  \to\ \  \overline{{\yng(1)}}_{\scriptscriptstyle-1} \  +\  
{\scriptscriptstyle (p-1)}\!\!\left.\vphantom{\yng(1,1,1,1,1,1)}\right\{\!\!
\raisebox{-.8ex}{$\begin{array}{l}\smash{\yng(1,1)} \\\smash{\hspace{.25ex}\vdots}\\\smash{\raisebox{1.3ex}{$\yng(1)_{\scriptscriptstyle\,(p-1)}$}} \end{array}$}\hspace{-1ex} 
 + \ \ldots
 \vspace*{-2ex}
\end{equation}
Here and henceforth we denote $\yng(1)$ the basic representation of \SL(d), corresponding to an object $v_m$ with $m=1,\ldots,d$.
The conjugate irrep corresponds to objects $w^m$ and above we used the shortcut expression $\overline{\yng(1)}$ to denote it. 
Hence, $\overline{\yng(1)}_{-1}$ is identified with a vector and $\yng(1)_1$ with a one-form.

The stabiliser group $\cS$ is further decomposed into the semidirect product of two pieces. One corresponds to the global symmetries of the $(D+d)$-dimensional supergravity theory, henceforth denoted $\Gup\times\Rup$ with $\Rup$ the $(D+d)$-dimensional trombone.
The second piece, normalised by the former, is a solvable group $\cP$ of unipotent transformations which reproduce the transformation of the internal $p$-forms of the $(D+d)$-dimensional supergravity under their gauge transformations, with gauge parameters linear in $y^m$ (so that the $\cP$ element is constant).\footnote{For instance, in 11d supergravity $\Gup$ is trivial, in IIB supergravity $\Gup=\SL(2,\dsR)$ and in IIA supergravity $\Gup=\dsR^+$. 
In 11d supergravity, $\cP$ includes the transformation $C_{mnp} \to C_{mnp} + \Lambda_{mnp}$ of the internal components of the three-form potential, with $\Lambda_{mnp}=\Lambda_{[mnp]}$ and $\de_\mu\Lambda_{mnp}=0=\de_q\Lambda_{mnp}$. Notice that $\Lambda_{mnp}$ also appears in the transformation of the dual six-form. 
The group $\cP$ is embedded in $\En$ because such constant shifts of the $p$-forms appear as the $\En$ subgroup of manifest symmetries arising in a Kaluza--Klein reduction on the torus~$\dsT^d$.}
This is familiar from the standard Kaluza--Klein reductions of supergravity theories.
For instance, if we consider the solution of the section constraint corresponding to 11d supergravity, then the Lie algebra $\mf{p}$ generating $\cP$ decomposes as follows under \GL(d):
\Yboxdim.5em
\begin{equation}\label{p yountab deco 11d}
\en\supset\mf{p}\ =\ {\yng(1,1,1)}_{\scriptscriptstyle\,+3} \ +\ \ {\yng(1,1,1,1,1,1)}_{\scriptscriptstyle\,+6}  \ +\ \ \cdots\end{equation}
where the dots are only relevant for $n\geq8$.
These components are identified with shifts of the three- and dual six-form.

The following expression summarises our notation and how the choice of section induces a breaking of the exceptional group to a set of actual symmetries of the $(D+d)$-dimensional supergravity theory:\footnote{The conversion from the notation in \cite{Inverso:2017lrz} is $\cG_0\to\Gup$, $\dsR^+_0\to\Rup$, $\cP_0\to\cP$.}
\begin{equation}\label{GLxS schematic deco}
\En\times\dsR^+ \quad \overset{\text{\tiny choice of section}}{\vphantom{\sum}\longrightarrow}\quad
\big( \overset{\text{\tiny str. group}}{\ \GL(d)\ \vphantom{\sum}} \times \underbrace{
\overset{\text{\tiny global syms.}}{\  \Gup\times\Rup \ \ \vphantom{\sum}}\big)\  \ltimes
 \overset{\text{\!\!\!\!\!\tiny $p$-form shifts}\!\!\!\!\!}{\!\!\!\cP\vphantom{\sum}\!\!\!}}_{\text{\normalsize$\cS$}}\quad .
\end{equation}
This decomposition \eqref{GLxS schematic deco} reflects the parametrisation of the \En ExFT fields in terms of the field content of the $(D+d)$-dimensional supergravity theory and works exactly as in standard Kaluza--Klein reductions \cite{Cremmer:1997ct,Cremmer:1998px}.
In particular, those degrees of freedom that appear as scalars with respect to the $d$ dimensional external space parametrise a generalised metric $\fG_{MN}$, which in turn is better described by an $\En/\KEn\times\dsR^+$ `vielbein' $`\cV_M^{\ul M}`$, with underlined indices transforming under the local \KEn group, such that\footnote{It is understood that we parametrise \En so that $\delta_{\ul{MN}}$ is a \KEn invariant.}
\begin{equation}\label{cV param}
\fG_{MN} = `\cV_M^{\ul M}` `\cV_N^{\ul M}`\,,\qquad 
`[s]\cV_M^{\ul M}` = `[s]\sfC{-1\,}_M^N` \, `[s]\ell_N^P` \, `[s]e_P^{\ul M}`\,,
\end{equation}%
where $`e_P^{\ul M}`$ denotes the embedding of the internal vielbein $e_m{}^{\ul m}$ (where $\ul m$ denotes Lorentz flat indices) into $\GL(d)\subset\En\times\dsR^+$.
Scalar fields in the $(D+d)$-dimensional supergravity theory parametrise $\Gup/K(\Gup)$ and its coset representative is here denoted $\ell$, also embedded into \mbox{$\En\times\dsR^+$}.
For instance, $\ell$ encodes the dilaton in IIA supergravity and the axiodilaton in type IIB supergravity. 
Finally, the internal $p$-form content of the $(D+d)$-dimensional supergravity theory is encoded in an element $`\sfC_M^N` \in \cP$. 
Again using 11d supergravity as an example, denoting $t^{mnp}=t^{[mnp]}$ and $t^{mnpqrs}=t^{[mnpqrs]}$ the \en generators associated to \eqref{p yountab deco 11d}, one may write
\begin{equation}\label{11d Ctwist expansion upto7}
\sfC = \exp\Big( C_{mnp} t^{mnp} \Big) \exp\Big( C_{mnpqrs} t^{mnpqrs} \Big)
\end{equation}
in terms of the internal components $C_{mnp}$, $C_{mnpqrs}$ of the three-form  and six-form potentials. 
For the sake of brevity we have absorbed any proportionality coefficient into the definition of the generators.

We shall always encode the scalar fields into a unimodular version of the generalised metric, parametrising $\En/\KEn$:
\begin{equation}
  \cM_{MN} = \det(\fG)^{-\mathrm{dim}\Rv} \fG_{MN}\,.
\end{equation}
This amounts to rescaling $\cV`_M^{\ul M}`$ by a power of the determinant of the vielbein $`e_m^{\ul m}`$.

\subsection{Generalised Scherk--Schwarz reductions}
\label{sssec:gss upto7}

The dynamics of \En ExFTs can be consistently reduced to those of a gauged maximal supergravity by introducing a factorisation ansatz for the dependence on the internal coordinates $y^m$  of all fields and gauge parameters\cite{Grana:2008yw,Aldazabal:2011nj,Geissbuhler:2011mx,Grana:2012rr,Berman:2012uy,Musaev:2013rq,Aldazabal:2013mya,Berman:2013cli,Aldazabal:2013via,Lee:2014mla,Hohm:2014qga}.
The ansatz is encoded in a $y^m$-dependent element of $\En\times\dsR^+$ and the covariant ExFT fields decompose according to their $\En\times\dsR^+$ transformation properties. 
In order to take into account the different weights of the ExFT fields, we denote such gSS data as follows:\footnote{This is usually called a `twist matrix' but here we avoid this term to avoid confusion with the `twisting/untwisting' appearing later.}
\begin{equation}\label{gSS data upto7}
\rho(y)>0\,,\qquad\text{and}\qquad \cU(y)`_M^A` \in\En\,,
\end{equation}
where $\rho(y)$ denotes the $\dsR^+$ factor. 
We denote by $A,B,\ldots$ the $\Rv$ indices of the gauged maximal supergravity obtained after reduction. 
Under the action of the generalised Lie derivative \eqref{genlie with P, upto7}, such indices are treated as spectators.

The factorisation ansatz reads
\begin{align}\label{gSS ansatz}
\nonumber
\cA_\mu^M(x,y)     &= A_\mu^C(x)\, \rho(y)^{-1} \, `\cU_C^M`(y) = A_\mu^C(x) \hat E`[s]{}_C^M`(y) \\[1ex]
\cM_{MN}(x,y)  &= \cU`_M^A`(y)\,\cU`_N^B`(y)  \,M_{AB}(x)\,,\\[1ex]\nonumber
{g}`_\mu\nu`(x,y) &= \rho(y)^{-2} g`_\mu\nu`(x)\,.
\end{align}
Where $`\cU_M^A`\cU`_A^N` = \delta_M^N$ and $A_\mu^A(x) $, $M_{AB}(x)$ and $g_{\mu\nu}(x)$ are respectively the vectors, scalars and metric of a $D$-dimensional maximal supergravity.
The ansatz extends to higher $p$-form potentials but we do not need to display them here.
In the first line of \eqref{gss cond no F0} we have singled out a particular combination of the gSS data:
\begin{equation}\label{E vs U}
\hat E`_A^M`(y)  =   \rho(y)^{-1}\, `\cU_A^M`(y) \,.
\end{equation}
We refer to such combination as the (generalised) \emph{frame} defining the gSS reduction.
It can also be regarded as a collection of generalised vectors spanned by the spectator index $A$.

For the gSS reduction ansatz to consistently yield a consistent truncation to a $D$-dimensional gauged maximal supergravity, the frame must satisfy the condition
\begin{equation}\label{gss cond no F0}
\cL_{\hat E_A} \hat E`_B^M` = -`X_AB^C` \hat E`_C^M`\,,\qquad 
`X_AB^C`= \text{constant.}
\end{equation}
The constants $`X_AB^C`$ equal the \emph{embedding tensor} of the resulting gauged supergravity and entirely specify the $D$-dimensional theory.
By construction, they can be decomposed in terms of the duality algebra $\en+\dsR$:
\begin{equation}\label{XtoTheta upto7}
`X_AB^C` = \Big(`\Theta_A^\beta` 
- \frac{\alpha}{1+\omega}\,\vartheta_D \, t^{\beta}{}_{A}{}^D   \Big) `t_\beta\,B^C`
+ \vartheta_A \delta_B^C 
\end{equation}
where $\{t_{\alpha}\}$ denotes a basis of $\en$, written above in the $\Rv$ representation, and the Kronecker delta reproduces the action of the trombone generator in this representation.
The ungauged flavour of $D$-dimensional maximal supergravity (obtained by setting $r=\cU=\text{constant}$ above) exhibits a genuine $\En\times\dsR^+$ rigid symmetry.
The embedding tensor selects a subalgebra of $\en+\dsR$ to be gauged, using the available vector fields $A_\mu^A$ to build the gauge connection.
We refer the reader to the articles \cite{Nicolai:2000sc,deWit:2004nw}, reviews \cite{Samtleben:2008pe,Trigiante:2016mnt} and references therein for details on the construction.
Here it suffices to state that the embedding tensor is restricted to live in a subset of the irreps stemming from the tensor product of $\overline{\Rv}$ and the coadjoint of $\en+\dsR$.
\def\Rtheta{\mathbf{R}_{\Theta}}
The embedding tensor $\Theta_M{}^\alpha$ must sit in a representation $\Rtheta$, presented in table~\ref{tab:groups and irreps} for each dimension, so that
\begin{equation}\label{LCX}
`X_AB^C`\ \in\ \Rtheta+\overline{\Rv}\,,
\end{equation}
where $\vartheta_A$ sits in $\overline{\Rv}$.
Closure of the gauge algebra requires that the embedding tensor satisfies the quadratic constraint
\begin{equation}\label{QCX}
`X_AC^F``X_BF^D` - `X_BC^F``X_AF^D` + `X_AB^F``X_FC^D` = 0\,.
\end{equation}
Notice in particular that this constraint implies
\begin{equation}\label{QCXsym}
`X_(AB)^C` X`_CE^F` = 0\,,
\end{equation}
where the symmetrisation $`X_(AB)^C`$ is in general non-vanishing.
Equation \eqref{QCX} is the defining equation of the structure constants of a Leibniz algebra, generalising the Jacobi identity of Lie algebras.

The representation and quadratic constraints are automatically satisfied by $`X_AB^C`$ in \eqref{gss cond no F0}. 
The quadratic constraint \eqref{QCX} descends immediately from the combination of the Leibniz identity \eqref{Leib id upto7} with the gSS condition \eqref{gss cond no F0}.
The requirement that $\Theta`_M^\alpha`$ belongs to $\Rtheta$ descends from direct computation of the embedding tensor in terms of the frame.
For convenience, we first define the Weitzenb\"ock connection (with `spectator' indices)
\begin{equation}
\weitz{\hat E}_{AB}{}^C = \hat{E}_A{}^M \hat{E}_B{}^N \de_M \hat{E}_N{}^C\,,
\end{equation}
where $\hat{E}_M{}^A \hat{E}_A{}^N = \delta_M^N$.
Then, \eqref{gss cond no F0} gives
\begin{equation}\label{tors def upto7}
`X_AB^C` \ =\ \weitz{\hat E}_{AB}{}^C - \weitz{\hat E}_{BA}{}^C 
+`Y^CF_EB` \weitz{\hat E}_{FA}{}^E
\ =\ \bbT\big[ \weitz{\hat E}_{AB}{}^C\big]\,.
\end{equation}
This expression gives a linear combination of the projectors of $\overline\Rv\otimes({\en+\dsR})^*$ onto $\RTheta+\overline\Rv$.
On the right of \eqref{tors def upto7} we have introduced the notation $\bbT[\,\,]$ for such projection, which will prove convenient later on.
We refer to such projection as generalised torsion.

At this point it is important to stress that the only non-trivial condition imposed in \eqref{gss cond no F0} is that the generalised torsion associated to the frame $\hat{E}_A{}^M$, must be constant.
Given a generic (local) frame $`E_A^M`$, not necessarily satisfying the gSS condition \eqref{gss cond no F0}, we shall also use the shorter notation 
\begin{equation}\label{tors of any E}
\tors{E}_{AB}{}^C = \bbT\big[ \weitz{E}_{AB}{}^C\big]\,.
\end{equation}

A non-constant torsion $\tors{E}_{AB}{}^C$ still satisfies the linear constaint \eqref{LCX}.
However, the quadratic constraint is modified to a differential one
\begin{align}\label{torsion BI upto7}
&\tors{E}_{\,AC}{}^F\tors{E}_{\,BF}{}^D
-\tors{E}_{\,BC}{}^F\tors{E}_{\,AF}{}^D
+\tors{E}_{\,AB}{}^F\tors{E}_{\,FC}{}^D\\&\nonumber
+`[s]E_A^M`\de_M\tors{E}_{\,BC}{}^D
-`[s]E_B^M`\de_M\tors{E}_{\,AC}{}^D
+`[s]E_C^M`\de_M\tors{E}_{\,AB}{}^D
-`Y^DF_GC`\, `E_F^M`\de_M \tors{E}_{\,AB}{}^G
\ =\ 0\,,
\end{align}
where for later use we notice that the last three terms correspond to (minus) the projection defined in equation \eqref{tors def upto7}, acting on the indices $B,\,C$ and $D$.

\subsection{Massive IIA supergravity and other deformations}
\label{sssec:F0 defo upto7}

Exceptional field theories and generalised geometries as described so far do not capture the Romans mass of type IIA supergravity, nor the gaugings of other maximal supergravities.\footnote{For instance, one may consider a solution to the section constraint corresponding to a $D+d<10$ dimensional theory and then try to capture its gaugings within the ExFT formalism.}
These can be taken into account by a deformation of the generalised Lie derivative~\cite{Ciceri:2016dmd,Cassani:2016ncu}.

We begin by considering the following deformation of the local expression for the generalised Lie derivative:
\begin{equation}\label{genlie local F}
\cL_{\Lambda}V^M \to \twL{F}_{\Lambda} V^M = \cL_{\Lambda}V^M -\Lambda^P V^Q `F_PQ^M`\,,
\end{equation}
where we take $`F_PQ^M`$ to belong to the $\RTheta+\overline\Rv$ representations, just like an embedding tensor or generalised torsion.

The allowed deformations of the generalised Lie derivative introduced in \eqref{genlie local F} are severely restricted by the requirement of closure of generalised diffeomorphisms \cite{Ciceri:2016dmd}.
Given a solution of the section constraint \eqref{SCpartial}, $`F_MN^P`$ is only allowed to contain certain \GL(d) components within the $\RTheta+\overline\Rv$ representation.
Beyond any massive/gauged deformations (corresponding to some \GL(d) singlets in the branching), one can also introduce: background values for the $p$-form field strengths, which may be integrated out and absorbed into $\sfC_M{}^N$ of equation \eqref{cV param}; background values for the $\Gup/K(\Gup)$ coset space currents, which may be reabsorbed into a coordinate-dependent \Gup element acting on all fields; and a `trombone flux' also removable by a coordinate-dependent $\Rup$ field redefinition.%
\footnote{Just as for $p$-form fluxes, there may be obstructions to globally removing  current and trombone background deformations.
In such cases however the global patching of the locally redefined fields will involve elements in \mbox{$\Gup\times\Rup$}.}
The allowed components are determined by some linear algebraic constraints and depend on the solution of the section constraint because they are linear in $\de_M$ (or equivalently, in $\cE_M{}^m$).
The latter were computed in \cite{Ciceri:2016dmd,Inverso:2017lrz} for any extended field theory, including ones based duality groups other than \En, as long as they do not require ancillary parameters.\footnote{Examples are the half-maximal $D=4$ extended field theory \cite{Ciceri:2016hup} and extended geometries for the duality groups of `magical' supergravities \cite{Bossard:2023ajq}. A systematic study is carried out in \cite{Cederwall:2017fjm}.}
A first condition simply reads
\begin{equation}\label{Xconstr}
`F_\,MN^P` \de_P = 0\,,
\end{equation}
where of course $\de_P$ is on section, 
while a second one takes in general a rather convoluted form, see eq. (3.16) of \cite{Inverso:2017lrz}. 
Such expression must be simplified on a case-by-case basis depending on the duality group.
For \En ExFTs it turns out to be equivalent to a much simpler condition on its $\overline\Rv$ component:
\begin{equation}\label{Cconstr simple}
`Y^MN_RS` \,`F_MP^P` \,\partial_N = 0\,.
\end{equation}
Furthermore, one checks that if we pick the 11d solution of the section constraint, \eqref{Cconstr simple} is redundant, namely it is implied by \eqref{Xconstr}.
Beyond the algebraic constraints above, $`F_MN^P`$ must satisfy a Bianchi identity guaranteeing integrability of those entries that are not massive/gauged deformations, as well as closure of the gauge algebra if a gauging is present.
It reads \cite{Inverso:2017lrz}
\begin{align}\label{BIF0 upto7}
& F_{MP}{}^R F_{NR}{}^Q
- F_{NP}{}^R F_{MR}{}^Q
+ F_{MN}{}^R F_{RP}{}^Q\\&\nonumber
+\de_M F_{NP}{}^Q
-\de_N F_{MP}{}^Q
+\de_P F_{MN}{}^Q
-`Y^QR_SP`\, \de_R  F_{MN}{}^S
\ =\ 0\,.
\end{align}
In section~\ref{sec: 3d uplifts} we compute the \E8 version of conditions \eqref{Xconstr}, \eqref{Cconstr simple} and \eqref{BIF0 upto7} and identify its \GL(d) irrep content for 11d and type II supergravities.

Within the possible components of $`F_MN^P`$, massive and gauged deformations play a special role.
They correspond to the \GL(d) singlets in the irrep decomposition of $`F_MN^P`$.
We shall use the different symbol $`\sfF_0\,MN^P`$  to capture massive and gauged deformations, to stress that they cannot be reabsorbed into local field redefinitinons and that they effect the global structure of generalised diffeomorphisms.
The associated generalised Lie derivative then reads
\begin{equation}\label{genlie F0}
\twL{\sfF_0}_{\Lambda} (V)^M = \cL_{\Lambda}V^M -\Lambda^P V^Q `\sfF_0\,PQ^M`\,,
\end{equation}
where $`\sfF_0\,MN^P`$ satisfies \eqref{Xconstr}, \eqref{Cconstr simple}, \eqref{BIF0 upto7} (in place of $`F_MN^P`$) and, again, only contains \GL(d) singlets.

All other components of $`F_MN^P`$ are reabsorbed into a redressing of fields and generalised vectors with an element of $\cS$, the stabiliser of the section solution defined in~\eqref{GLd and S}.
Indeed, elements of $\cS$ capture precisely local shifts of the $p$-forms as well as coordinate-dependent $\Gup\times\Rup$ field redefinitions.
To show how this works we introduce the notation (useful for the \E8 case later on)
\begin{equation}\label{local dressing upto7}
{^S\!\Lambda\!}^M = \Lambda^N S_N{}^M\,,\qquad `S_M^N`\in\cS\,.
\end{equation}
One then can always write that
\begin{equation}
\twL{\sfF_0}_{^S\!\Lambda} {^SV}{}^M
=
{}^S\!\big(\twL{F}_{\Lambda} V^M\big) 
\,. 
\end{equation}
We then compute $`F_MN^P`$ explicitly to find
\begin{equation}
`F_MN^P` = \tors{S}_{MN}{}^P +`S_M^R``S_N^S` \, `\sfF_0\,RS^T` \big(S^{-1}\big)`[s]{}_T^P`\,,
\end{equation}
where the first term has the same structure as a generalised torsion defined in \eqref{tors def upto7}, \eqref{tors of any E} and we can indeed write it as
\begin{equation}
\tors{S}_{MN}{}^P = \bbT\big[\weitz{S}_{MN}{}^P\big]\,,\qquad
\weitz{S}_{MN}{}^P = `S_N^R` \de_M \big(S^{-1}\big)`[s]\!_R^P`\,.
\end{equation}
Compared to \eqref{tors def upto7} and \eqref{tors of any E}, there is no distinction between standard and `spectator' indices.
We have also used $`S_M^N`\partial_N=\partial_M$.

While all non-singlet \GL(d) components of $`F_MN^P`$ can be reabsorbed into field and parameter redefinitions, the massive/gauged ones must be kept as they correspond to physically inequivalent theories.
Therefore, whenever $`\sfF_0\,MN^P`\neq0$ the gSS reduction must take it into account.
The gSS ansatz \eqref{gSS ansatz} does not change.
The condition \eqref{gss cond no F0}, on the other hand, generalises to
\begin{equation}\label{gSS cond with F0}
\twL{\sfF_0}_{\hat E_A} \hat E`_B^M` 
\ =\  -`X_AB^C` \hat E`_C^M`\,,\qquad 
`X_AB^C`= \text{constant.}
\end{equation}
This is the main equation that one has to solve to find gSS reductions.
If we start with a given embedding tensor $`X_AB^C`$ and want to find a gSS uplift,  we must solve \eqref{gSS cond with F0} for $\hat{E}`_A^M`$ \emph{and} $\sfF_0`_MN^P`$.

\subsection{Patching and global definiteness}
\label{sssec:global patch upto7}

A notion that will be important later on is global definiteness of several geometrical objects introduced in the construction of ExFTs and of gSS reductions.
In particular, the frame $\hat E`_A^M`$ as defined in the previous section must extend to a parallelisation of a generalised version of the tangent bundle over the internal manifold~\cite{Lee:2014mla}.

Once we fix a choice of solution to the section constraint \eqref{SCpartial}, we identify generalised vectors as sections of a generalised tangent bundle $\mathsfit{E}$ over the internal manifold. Locally, such bundle decomposes into a direct sum of terms reproducing the branching of $\Rv$ under \GL(d).
For instance, if we denote by $\mf{M}$ the internal manifold in a compactification of 11d supergravity, one has
\begin{equation}\label{local isom gentanbundle}
\mathsfit E \ \overset{\str{loc.}}{=}\  T\mf M  + \mathsf{\Lambda}^2 T^*\mf M + \mathsf{\Lambda}^5 T^*\mf M+\ldots \,,
\end{equation}
where we identify on the right-hand side the generators of diffeomorphisms and three- and six-form gauge transformations, reflecting the decomposition \eqref{Rv deco youngtab}.

The isomorphism between $\mathsfit{E}$  and a direct sum of tangent and (products of) cotangent bundles does not extend globally.
On any coordinate patch, the isomorphism is induced by the local value of the $(D+d)$-dimensional supergravity $p$-form potentials along the internal space, which parametrise a non-constant element of $\cP$.
If we denote by $\widetilde\Lambda^M(x,y)$ the local expression of a section of the direct sum in \eqref{local isom gentanbundle} (or analogous ones for other ExFTs/sections), then we have
\begin{equation}\label{untwisting}
\Lambda^M(x,y) =  {^\sfC\widetilde\Lambda^M}(x,y) = \widetilde\Lambda^N(x,y) `\sfC_N^M`(x,y)\,,\qquad `\sfC_N^M`(x,y)\in\cP\,.
\end{equation}
While at the local level this relation looks very similar to \eqref{local dressing upto7}, notice that matrix appearing here is the same $`\sfC_N^M`$ appearing in the generalised metric \eqref{cV param} and that it encodes entirely the internal $p$-form potentials in the theory.
Also, notice that elements in $\Gup\times\Rup$ are not allowed.

In presence of non-trivial fluxes, the $p$-form potentials encoded in $`\sfC_N^M`$ are not globally defined but rather are patched together by non-trivial gauge transformations. 
This patching is therefore inherited by generalised vectors, so that the global structure of $\mathsfit E$ also encodes the flux content of a compactification.
The fluxes can be resurfaced in the generalised Lie derivative by `untwisting' the generalised vectors, i.e. expressing them in terms of their tilded versions as in \eqref{untwisting}.
Taking also $V^M(x,y) = \widetilde V^N(x,y) `\sfC_N^M`(x,y)$, one has the relation
\begin{equation}\label{genLie flux twisting}
\twL{\sfF_0}_{\Lambda} V^M = \Big(\twL{\sfF_0}_{\widetilde\Lambda}\widetilde V^N - \widetilde\Lambda^P \widetilde V^Q `\sfF_{PQ}^N`\Big) \sfC_N{}^M 
= \big(\twL{\sfF}_{\widetilde\Lambda}\widetilde V^N \big) `\sfC_N^M` \,,
\end{equation}
with
\begin{equation}
`\sfF_MN^P` = \tors{\sfC}_{MN}{}^P +`\sfC_M^R``\sfC_N^S` \, `\sfF_0\,RS^T` \big(\sfC^{-1}\big)`[s]{}_T^P`\,,
\end{equation}
which defines a generalised Lie derivative $\twL{\sfF}$, `twisted' by $`\sfF_MN^P`$.
This is of course analogous to \eqref{genlie local F}, however $`\sfF_MN^P`$ only encodes the globally defined $p$-form field strengths and any massive/gauged deformations that may be present.
Twists by $\Gup\times\Rup$ are not considered for global patching.\footnote{Notice however that $\sfF_0`_\,MN^P`$ may also encode a gauging of a $\Gup\times\Rup$ subgroup.}

Under (finite) $p$-form gauge transformations, $\sfC_M{}^N$ transforms as\footnote{Notice that infinitesimal $p$-form gauge transformations are nothing but generalised diffeomorphisms generated by a vector $\Lambda^M$ with vanishing $T\mf{M}$ component.} 
\begin{align}\label{C to CGamma}
`\sfC_M^N` \to `\sfC_M^P``\Gamma_P^N`\,,\qquad `\Gamma_P^N`\in\cP\,,\qquad
0 = \tors{\Gamma}_{MN}{}^P +`\Gamma_M^R``\Gamma_N^S` \, `\sfF_0\,RS^T` \big(\Gamma^{-1}\big)`[s]{}_T^P`\,.
\end{align}
The condition on the torsion of $`\Gamma_M^N`$ expresses the fact that the $p$-form field strengths (hence, $`\sfF_MN^P`$) are invariant under gauge transformations.
Notice that this statement properly takes into account massive/gauged deformations through the contribution of $`\sfF_0\,MN^P`$.

\smallskip

We conclude that a generalised vector is patched together on overlaps of coordinate patches by  elements $\Gamma_M{}^N\in\cP$ that are `exact' in the sense of \eqref{C to CGamma}.
In other words, transition functions on $\mathsfit{E}$ take values in $\GL(d)\ltimes\cP$, encoding coordinate transformation in the \GL(d) factor and $p$-form gauge transformations in the $\cP$ factor.
In the gSS ansatz \eqref{gSS ansatz}, the frame $\hat{E}_A{}^M$ must extend to a collection of globally defined generalised vectors and hence must patch as described above.

\subsection{Review of general construction up to \texorpdfstring{\E7}{E₇₍₇₎}}
\label{ssec:review2017}

Necessary and sufficient `uplift conditions' for existence of a solution to the gSS condition \eqref{gSS cond with F0} for a given embedding tensor $`X_AB^C`$ were found in \cite{Inverso:2017lrz}, which also gives the explicit construction of the frame $\hat{E}_A{}^M$.
Global definiteness is also proved.
The uplift conditions take a \En invariant form and the whole construction is valid up to \E7 or more generally, for any extended field theory that does not require the introduction of ancillary parameters.
Deformations encoded in $`\sfF_0\,MN^P`$, such as the Romans mass in IIA supergravity, are taken into account.
We now review the main results of \cite{Inverso:2017lrz}, that we plan to generalise to \E8 ExFT in the rest of this paper.

\subsubsection{Covariant uplift conditions}
\label{sssec:uplift conds 2017}

An embedding tensor $`X_AB^C`$ satisfying the quadratic constraint \eqref{QCX} defines a gauge Lie algebra which we denote by $\hggauge$, with abstract generators $\widehat{T}_A$ satisfying the commutator relations%
\footnote{
$`X_AB^C`$ are the structure constants of a Leibniz algebra of dimension $\mathrm{dim}\Rv$.
The Lie algebra $\hggauge$ is defined as the quotient of such Leibniz algebra by its elements $v_A$ that satisfy $`X_(AB)^C`v_C=0$.
}
\begin{equation}\label{hatT Lie algebra}
\big[\widehat{T}_A\,,\,\widehat{T}_B\big] = -`X_AB^C` \widehat{T}_C\,\qquad
`X_(AB)^C` \widehat{T}_C=0\,,
\end{equation}
where the second equation is a consequence of the first one.
In general, \hggauge is not a subalgebra of $\en+\dsR$.
There can be a non-empty centre $\mf{z}\subset\hggauge$ such that the algebra  contained in $\en+\dsR$ is the quotient algebra $\ggauge = \hggauge/\mf{z}$.
The simplest example is the ungauged theory, with $`X_AB^C`=0$, leading to $\hggauge = \mf{z} = \bigoplus^{\scriptscriptstyle\mathrm{dim}\Rv}\mf{u}(1)$.
We shall denote $\hGgauge$ the group generated by $\hggauge$ and faithfully realised on the supergravity fields and gauge potentials.
Quotienting by the central subgroup $Z$ generated by $\mf z$, we obtain a group denoted $\Ggauge \subset \En\times\dsR^+$ which is the one realised on the covariant field strengths.
Importantly, the $\Rv$ representation of the generators $\widehat{T}_A$ is given by the embedding tensor itself:
\begin{equation}
\rho_{\Rv}\big( \widehat T_A \big)`[s]{}_B^C` = `X_AB^C`\,.
\end{equation}

The gSS condition \eqref{gSS cond with F0} implies that \hGgauge acts transitively on the internal manifold $\mf{M}$.
Namely, $\mf M$ is necessarily a coset space
\begin{equation}
\mf{M} = \frac\hGgauge\hHgauge
\end{equation}
for some subgroup $\hHgauge\subset\hGgauge$~\cite{Grana:2005tf,Inverso:2017lrz}.

Not every choice of subgroup \hHgauge allows to solve \eqref{gSS cond with F0}.
Rather, there are very strict requirements for a consistent uplift.
To better formulate them, we introduce adjoint indices $a,b,\ldots$ for \hggauge, $i,j,\ldots$ for \hhgauge, and a basis
\begin{equation}\label{That splitting upto7}
\hggauge = 
 \big\langle\, \{\widehat{T}_{a}\}\,\big\rangle
= \big\langle\, \{\widehat{T}_{\ul m}\}\,\big\rangle + \big\langle\, \{\widehat{T}_{i}\}\,\big\rangle \,,\qquad
\hhgauge = \big\langle\, \{\widehat{T}_{i}\}\,\big\rangle \,,
\end{equation}
where $\widehat{T}_{\ul m}$ are coset generators.
The following construction does not depend on how we choose the coset generators.
We introduce coefficients $\widehat\Theta_A{}^{a}$,
which play for the extended gauge algebra a role similar to $`\Theta_A^\alpha`$ and $\vartheta_A$ in \eqref{XtoTheta upto7}, such that
\begin{equation}
\widehat{T}_A = 
\widehat\Theta_A{}^{a}  \widehat{T}_{a} 
= 
\widehat\Theta_A{}^{\ul m}  \widehat{T}_{\ul m} 
+
\widehat\Theta_A{}^{i} \widehat{T}_{i}\,,
\end{equation}
where in the second equality we have separated coset generators and \hhgauge elements.
Notice that $\widehat\Theta_A{}^{a}$ is determined from the embedding tensor $`X_AB^C`$ by requiring $`X_(AB)^C`\widehat\Theta_C{}^{a}=0$ and choosing a basis of $\hggauge$.
In reverse, $`X_AB^C`$ is entirely determined by specifying $\widehat\Theta_C{}^{a}$ together with the embedding of $\ggauge$ in $\en+\dsR$.\footnote{
More pragmatically, notice that if we denote $`t_a\,A^B` = \rho_{\Rv}\!\big(\widehat{T}_a\big)_B{}^C $ the \hggauge generators in the $\Rv$ representation, we have 
$\rho_{\Rv}\!\big(\widehat{T}_A\big)_B{}^C 
= \widehat\Theta_A{}^a `t_a\,B^C`
=
`X_AB^C`$,  which can be further expanded using \eqref{XtoTheta upto7}. 
}

For a consistent uplift to exist, two conditions must be satisfied.
They are algebraic in the embedding tensor and depend on the choice of coset space through the coefficients $\widehat\Theta_A{}^{\ul m}$.
The first condition states that $\widehat\Theta_A{}^{\ul m}$ solves the section constraint just as the section matrix we introduced originally in \eqref{SCE}:
\begin{equation}\label{Thetahat SC}
`Y^AB_CD` \, \widehat\Theta_A{}^{\ul m}\,  \widehat\Theta_B{}^{\ul n} = 0\,.
\end{equation}
Indeed, $\widehat\Theta_A{}^{\ul m}$ determines what solution of the section constraint should be used to solve the gSS condition and one can write without loss of generality%
\footnote{One can always rotate the choice of section by a constant $\En\times\dsR^+$ element.}
\begin{equation}\label{E=hatTheta}
`\cE_M^m` = \delta_M^{\,A} \, \widehat\Theta_A{}^{\ul m} \, \delta_{\ul m}^m\,.
\end{equation}

The second algebraic condition was given in equation (3.16) of \cite{Inverso:2017lrz}.
It is a rather complicated expression if one wants to study generic extended field theories.
Luckily, for \En ExFTs it simplifies drastically and reduces to
\begin{equation}\label{extracstr simple upto7}
`Y^AB_CD` \, \Big( \vartheta_A - \omega\, \widehat\Theta_G{}^{\ul m} `t_{\ul m\,A}^G` \Big) \, \widehat\Theta_B{}^{\ul n} = 0
\end{equation}
where $\omega$ is the coefficient appearing in \eqref{genlie with P, upto7} and $`t_{\ul m\,A}^B`$ are the coset generators written in the $\Rv$ representation.
Since by definition $\mf z$ is trivially represented in $\Rv$, the generators $`t_{\ul m\,A}^B`$ span a subspace of $\en+\dsR$.
Notice that this condition is analogous to \eqref{Cconstr simple} and indeed it descends from it.
We prove it in section~\ref{ssec:gss uplift cond e8}.

It should be stressed that the possible contributions of elements of $\mathfrak{z}$ to the uplift conditions only matters if one is trying to uplift to a supergravity of dimension as high as possible. 
One can indeed always absorb any $Z$ element into $\hHgauge$, in which case $\hGgauge/\hHgauge\simeq\Ggauge/\Hgauge$ and one can simply work with the standard embedding tensor $\Theta_M{}^\alpha$ and the gauge group as realised on $\Rv$.
There are however cases where, in order to prove existence of an uplift specifically to ten or eleven dimensions, one must keep track of the contribution of central elements.
See for instance the geometric uplift to eleven dimensions of the full family (with four mass parameters) of Cremmer--Scherk--Schwarz gaugings described in~\cite{DallAgata:2019klf}.

\subsubsection{Construction of the generalised frame}
\label{sssec:constr genframe 2017}

The construction of the generalised frame proceeds as follows.
The internal coordinates are denoted $y^m$ with $m=1,\ldots,\mathrm{dim}\big(\hGgauge/\hHgauge\big)$.
The section constraint \eqref{SCE} is solved by \eqref{E=hatTheta}.\footnote{%
Given the equivalence \eqref{E=hatTheta}, by a small abuse of notation we will write the section matrix either with `curved' indices $M,\,m$ or `flat' ones $A,\,\ul m$.}
We pick a coset representative for $\hGgauge/\hHgauge$, denoted $L(y)$ and transforming as
\begin{equation}
L(y) g = h(y') L(y')\,,\qquad g\in\hGgauge\,, \ \ h(y)\in\hHgauge 
\end{equation}
under an \hGgauge transformation $g$ connecting the points $y^m$ and $y'^m$.
When realised in the $\Rv$ representation, the coset representative is denoted
$
L(y)`_A^B`
$
and by all effects parametrises $\Ggauge/\Hgauge$, where the denominator is the quotient of \hHgauge by its intersection with $Z$.
Using standard coset space techniques we construct the natural \hHgauge frame on $T\mf{M}$, denoted $\mathring{e}_{\ul m}{}^m$ with inverse $\mathring{e}_{m}{}^{\ul m}$:
\begin{equation}\label{coset CM deco}
\partial_mL L^{-1} = 
\mathring{e}_{m}{}^{\ul m} \widehat{T}_{\ul m} + Q_m^i \widehat{T}_{i} 
\end{equation}
where $Q_m^i$ is the \hHgauge connection.
Since $\mathring{e}_{\ul m}{}^m$ is invertible, we can regard it as a \GL(d) element and embed it into $\En\times\dsR^+$ as described in and around \eqref{GLd and S}.
This simply means that we construct the matrix $\mathring{e}_A{}^M$ following the \GL(d) branching of $\Rv$, such that in particular%
\footnote{
For instance, the branching \eqref{Rv deco youngtab} leads to the block-diagonal matrix
$$
\mathring{e}_A{}^M = \left(\begin{smallmatrix}
\mathring{e}_{\ul m}{}^m \ \ & & \\
&\ \  \mathring{e}_{[m_1}{}^{\ul m_1}\cdots\mathring{e}_{m_6]}{}^{\ul m_6} \ \ & \\
&&\ \ \ddots\ \ 
\end{smallmatrix}\right)\,.
$$
}
\begin{equation}\label{oe through cE}
\mathring{e}_A{}^M \cE_M{}^m = \cE_A{}^{\ul m}\mathring{e}_{\ul m}{}^m\,.
\end{equation}

In \cite{Inverso:2017lrz} it is proved that the generalised frame always takes the form
\begin{equation}\label{gSS frame upto7}
\hat{E}`_A^M` = \big(L^{-1}\big)_A{}^B \, \mathring{e}_B{}^N \, `S_N^M`\,,\qquad `S_N^M`\in\cS\,,
\end{equation}
where $`S_N^M`$ is not constructed directly but is rather obtained by an integration as we now describe.
The matrix $`S_M^N`$ is denoted $`C_M^N`$ in \cite{Inverso:2017lrz}.
To simplify the notation we define a local frame with $S`_M^N`$ absent:
\begin{equation}\label{local frame E}
E`_A^M` = \big(L^{-1}\big)_A{}^B \, \mathring{e}_B{}^M\,.
\end{equation}
Then, we define the expression
\begin{equation}\label{F=X-T}
`F_MN^P` = E`_M^A `E`_N^B`  \big( `X_AB^C` - \tors{E}_{AB}{}^C \big) E`_C^P`\,.
\end{equation}
The crucial point of the construction is that, if and only if \eqref{Thetahat SC} and \eqref{extracstr simple upto7} are satisfied, then \eqref{F=X-T} satisfies the consistency conditions \eqref{Xconstr}, \eqref{Cconstr simple} and \eqref{BIF0 upto7}.
This means that $`F_MN^P`$ is a valid deformation of the generalised Lie derivative and one indeed has
\begin{equation}\label{gSS nohat e8}
\twL{F}_{E_A}`E_B^M` = -`X_AB^C` `E_C^M`\,.
\end{equation}
Following the same reasoning as above \eqref{Xconstr}, this means that $`F_MN^P`$ can only contain \GL(d) irreps corresponding to background $p$-form field strengths, scalar currents and a local `trombone flux', as well as massive/gauged deformations.
All but the latter are integrable because of \eqref{BIF0 upto7}, hence we have
\begin{equation}
`F_MN^P` = \tors{S}_{MN}{}^P +`S_M^R``S_N^S` \, `\sfF_0\,RS^T` \big(S^{-1}\big)`[s]{}_T^P`\,,
\end{equation}
which defines the $S`_M^N`$ appearing in the gSS frame \eqref{gSS frame upto7}.
While this expression may appear very formal, its decomposition in \GL(d) irreps yields the familiar definitions of $p$-form field strengths and scalar field currents, with $`S_M^N`$ parametrised in analogy with \eqref{cV param}.
For instance, in analogy with \eqref{11d Ctwist expansion upto7}, in the case of an uplift to 11d supergravity we would write (in index-free notation for the overall matrix)
\begin{equation}
S = \tilde{r}\exp\Big( \tilde c_{mnp} t^{mnp} \Big) \exp\Big( \tilde c_{mnpqrs} t^{mnpqrs} \Big) 
\end{equation}
where $\tilde c_{mnp}$ and $\tilde c_{mnpqrs}$ are local integrals of the four-form and seven-form encoded in $`F_MN^P`$ and $\tilde{r}\in\Rup$ satisfies 
\begin{equation}
(1+\omega)(\mathrm{dim}\Rv)\,\tilde{r}^{-1}\partial_M\tilde{r} = -`F_MP^P`\,.
\end{equation}
Notice how condition \eqref{Cconstr simple} is crucial for this last relation to admit a solution.

\smallskip
The proof of global definiteness of the gSS frame \eqref{gSS frame upto7} amounts to studying how the patchings of $`L_A^B`$ and $\mathring{e}`_A^M`$ compensate each other, as well as the existence of an Iwasawa decomposition for $`L_A^B`$ as an element of $\En\times\dsR^+$. Details are found in sections~3.3 and~A of~\cite{Inverso:2017lrz}.

\section{Twisting \texorpdfstring{\E8}{E₈₍₈₎} generalised diffeomorphisms}
\label{sec: 3d uplifts}

The aim of this section is to lay the ground to generalise the above construction of gSS reductions to \E8 ExFT and $D=3$ maximal supergravity.
We begin by reviewing \E8 generalised diffeomorphisms and the generalised Scherk--Schwarz ansatz in presence of ancillaries. 
We then switch to the main technical part of the paper, identifying twistings and deformations of these gauge structures.

\subsection{Gauge structure of \texorpdfstring{\E8}{E₈₍₈₎} exceptional field theory}
\label{ssec: E8 structures}
\label{sssec: e8 gendiffeo}

In \E8 ExFT, $\Rv$ equals the representation ${\bf248}_{+1}$ of $\E8\times\dsR$, where $\bf248$ is the adjoint of~\e8.
The commutation relations and structure constants are denoted\footnote{We keep using indices $M,N,\ldots$, $A,B,\ldots$ and $\ul M,\ul N,\ldots$ to denote the \e8 ${\bf248}$ representation, the choice of notation corresponding to ExFT `curved' indices, indices of gauged supergravity objects and local \Spin(16) indices, respectively.
The $\dsR^+$ charge is specified separately for each object.
}
\begin{equation}
\big[t_M,t_N\big] = `f_MN^P` t_P
\end{equation}
and we raise/lower indices with the invariant
\begin{equation}
\eta_{MN} = \frac{1}{60}`f_MP^Q``f_NQ^P`\,.
\end{equation}

The \E8 ExFT~\cite{Hohm:2014fxa} differs from its lower-rank siblings in one crucial aspect. 
Even after imposing the section constraint, generalised diffeomorphisms do not close unless we add an extra set of `ancillary' gauge parameters, denoted $\Sigma_M$, which are algebraically constrained in their $\Rv$ index to satisfy the same condition as the section constraint.
To make this precise and set up our notation, we begin by reproducing the expression of the generalised Lie derivative, acting on a generalised vector $V^M$ 
\begin{align}
\label{genLie e8}
\cL_{(\Lambda,\Sigma)}V^M = \Lambda^N\partial_N V^M -\big( `f_P^R_S` \partial_R\Lambda^S + \Sigma_P \big) `f^PM_N` V^N
+ \partial_N\Lambda^N V^M\,.
\end{align}
where the last term corresponds to a density weight which is set to $\omega=1$ for generalised vectors.
By comparison, we identify the $`Y^MN_PQ`$ tensor \eqref{Ytens} which appeared in earlier expressions with
\begin{equation}
\label{Ytens e8}
`Y^MN_PQ` = \delta^M_Q \delta^P_N + \delta^M_N \delta^P_Q -`f_RQ^M``f^R_P^N` \,.
\end{equation}
The section constraint \eqref{SCpartial} translates into the vanishing of the ${\bf1}+{\bf248}+{\bf3875}$ representations in the tensor product $\partial_M\otimes\partial_N$.
We can write this explicitly as follows
\begin{align}
\label{SC e8}
\eta^{MN}\partial_M\otimes\partial_N = 0\,,\quad
`f^MNP` \partial_M\otimes\partial_N = 0\,,\quad
(\bbP_{\str\bf3875})`_PQ^MN` \partial_M\otimes\partial_N = 0\,,
\end{align}
where the latter projector reads
\begin{equation}
(\bbP_{\str\bf3875})`_PQ^MN` = \frac17 \delta^{M}_{(P} \delta^{N}_{Q)} 
-\frac{1}{56} `\eta^MN``\eta_PQ`
-\frac1{14} `f_R(P^M` `f^R_Q)^N`\,.
\end{equation}
Several projector identities are found for instance in~\cite{Koepsell:1999uj} and~\cite{Koepsell:2000xg}.
The ancillary parameters such as $\Sigma_M$ in \eqref{genLie e8} must satisfy the same algebraic constraints as $\partial_M$, i.e. we may substitute either or both derivatives in \eqref{SC e8} with ancillary parameters:
\begin{equation}\label{SC ancillary}
\big(\bbP_{\str{{\bf1}+{\bf248}+{\bf3875}}}\big)`_PQ^MN` \ \Sigma_{1}{}_{\,M}\,\Sigma_{2}{}_{\,N} =
\big(\bbP_{\str{{\bf1}+{\bf248}+{\bf3875}}}\big)`_PQ^MN` \ \Sigma_{M}\,\partial_N = 0
\end{equation}
These conditions can be rephrased in terms of a `section matrix' as in \eqref{SCE}, in which case we write
\begin{equation}
\partial_M = `\cE_M^m`\frac{\partial}{\partial y^m}\,,\qquad
\Sigma_M = `\cE_M^m`\Sigma_m\,,
\end{equation}
and
\begin{equation}
\label{SCE e8} 
\big(\bbP_{\str{{\bf1}+{\bf248}+{\bf3875}}}\big)`_PQ^MN`\ `\cE_M^m`\,`\cE_N^n` = 0\,.
\end{equation}
Ancillary parameters are taken with weight equal to 0 and their generalised Lie derivative reduces to the following expression upon using the section constraints:
\begin{equation}
\cL_{\Lambda_1,\Sigma_1} \Sigma_2 =
\Lambda^N\,\partial_N\Sigma_{{\str2}\,M} + 
\partial_N\Lambda^N\,\Sigma_{{\str2}\,M} + 
\partial_M\Lambda^M\,\Sigma_{{\str2}\,N}\,.
\end{equation}

While generalised diffeomorphisms based on couples of parameters $(\Lambda\,,\,\Sigma)$ do close onto themselves, the generalised Lie derivative for \E8 ExFT fails to satisfy the Leinbiz identity \eqref{Leib id upto7}.
The gauge structure of the theory is better represented by a generalised Dorfman product~\cite{Hohm:2017wtr}.
This is defined on couples of parameters, and entails an indecomposable modification the the gauge transformation of ancillary parameters, so that when acting on a couple $(\Lambda_{\it i}\,,\,\Sigma_{\it i})$, the transformation of $\Sigma_{{\str\it i}\,M}$ depends on $\Lambda_{\it i}^M$ as well.
Explicitly, we have
\begin{align}\label{Dorfman on couples}
&\big( \Lambda_{1}\,,\,\Sigma_{1} \big) \dorf
\big( \Lambda_{2}\,,\,\Sigma_{2} \big) =
\Big(\,\cL_{(\Lambda_{1},\Sigma{\str1})} \Lambda_{2}\ ,\ 
\cL_{(\Lambda_{1},\Sigma{\str1})} \Sigma_{2} + \Delta\Sigma_{12}
\Big)\,,\\[1ex]
&\Delta\Sigma_{{\str12}\,M} = \Lambda_{2}^P \, \partial_M \big( 
  `f_P^R_S` \partial_R\Lambda_{1}^S  + \Sigma_{{\str1}\,P} \big)\,.
\end{align}
For convenience, we will often use doublestruck symbols to denote indecomposable gauge parameter couples, and also define a shorthand for the \e8 generator acting on tensors in the generalised Lie derivative \eqref{genLie e8}:\footnote{Depending on context and clarity, we may use either an indexless notation $\bbLambda = \big(\Lambda\,,\,\Sigma\big)$ for gauge parameter couples, or introduce a dummy index in the parenthesis, e.g. $\bbLambda = \big(\Lambda^M\,,\,\Sigma_M\big)$. The latter notation is only used if unambiguous.}
\newcommand{\sqb}[1]{{\smash{[\![}#1\smash{]\!]}\vphantom{[]}}}
\begin{equation}\label{bbLambda def}
\bbLambda = \big( \Lambda^M\,,\,\Sigma_{M} \big)\,,\qquad
 \sqb{\bbLambda}_M = `f_M^P_Q` \partial_P\Lambda^Q  + \Sigma_M\,,
\end{equation}
With this notation, the expression of the Dorfman product becomes
\begin{equation}\label{dorfman on bbLambdas}
\bbLambda_{1}\dorf\bbLambda_{2} = 
\cL_{\bbLambda_{1}} \bbLambda_{2} + \big( 0 \,,\, 
\Delta\Sigma_{12}{}
\big)\,,\qquad
\Delta\Sigma_{12\,M} = \Lambda_{2}^N \partial_M\sqb{\bbLambda_1}_N\,.
\end{equation}
Importantly, the Dorfman product satisfies the Leibniz identity provided the section constraints \eqref{SC e8}, \eqref{SC ancillary} are satisfied:
\begin{equation}\label{dorfleib}
\bbLambda_{1}\dorf\big(\bbLambda_{2}\dorf\bbLambda_{3}\big)
-\bbLambda_{2}\dorf\big(\bbLambda_{1}\dorf\bbLambda_{3}\big)
-\big(\bbLambda_{1}\dorf\bbLambda_{2}\big)\dorf\bbLambda_{3}
\ =\ 0\,.
\end{equation}

\subsection{The generalised Scherk--Schwarz ansatz}
\label{sssec: e8 gSS}

The data for a gSS reduction of \E8 ExFT is still given by \eqref{gSS data upto7}, which we repeat here:
\begin{equation}\label{EUr}
\cU(y)`_M^A` \in\E8\,,\qquad
\rho(y)>0\,,\qquad 
\hat E(y)`[s]{}_A^M`  =   \rho(y)^{-1} \big(\cU(y)^{-1}\big)`[s]{}_A^M` \,.
\end{equation}
Let us also reintroduce the Weitzenb\"ock connection for the frame:
\begin{equation}\label{weitz hatE e8}
\weitz{\hat{E}}_{\,AB}{}^C = \hat E`[s]{}_A^M` \hat E`[s]{}_A^M` \partial_M \hat E`[s]{}_N^C` 
=
\weitz{\hat{E}}_{AD}`f^D_B^C` + \frac12\weitz{\hat{E}}_A  \, \delta_B^C\,,
\end{equation}
where $\hat E`[s]{}_N^C`\hat E`[s]{}_C^M` = \delta_N^M $.

Under the gSS ansatz, generalised vectors (and hence, the vector fields constituting their gauge connection) decompose as
\begin{equation}\label{gss ansatz e8 components}
\Lambda^M(x,y) = \lambda^A(x) \hat E`[s]{}_A^M`(y)\,,\qquad
\Sigma_M(x,y) =  -\lambda^A(x) \, `\cU_M^B`(y) \weitz{\hat{E}}_{BA}(y)\,.
\end{equation}
Importantly, ancillary parameters factorise into the same coefficients $\lambda^A(x)$ as generalised vectors, reflecting the fact that the gauged supergravity tensor hierarchy does not require the presence of covariantly constrained `ancillary' vector fields.

We may rewrite the relation above as follows:
\begin{equation}
\bbLambda(x,y) = \lambda^A(x)\,\hat{\bbmE}_A(y)\,,\qquad
\hat{\bbmE}_A = \Big(
\hat E`[s]{}_A^M`(y)\,,\, -`\cU_M^B`(y) \weitz{\hat{E}}_{BA}(y)
\Big)\,.
\end{equation}
Then, the \E8 version of the gSS condition \eqref{gss cond no F0} is
\begin{equation}\label{gss e8 noF0}
\hat{\bbmE}_A\dorf\hat{\bbmE}_B = -`X_AB^C` \hat{\bbmE}_C\,,\qquad `X_AB^C`= \text{constant.}
\end{equation}
The embedding tensor $`X_AB^C`$ of $D=3$ gauged maximal supergravity decomposes into the ${\bf1}+{\bf248}+{\bf3875}$ representations, where the $\bf248$ corresponds to $\dsR^+$ trombone gaugings, while $\Rtheta={\bf1}+{\bf3875}$ parametrise Lagrangian gaugings.
Explicitly,
\begin{equation}\label{X to Theta e8}
`X_AB^C` = 
\Theta_{AD} `f^D_B^C` 
+ \theta \, `f_AB^C`
- \frac12 `f_GA^D``f^G_B^C` \, `[s]\vartheta_D` 
+ `[s]\vartheta_A` \, \delta_B^C\,,
\end{equation}
where $\Theta_{AB}=\Theta_{BA}$ sits in the $\bf3875$ representation.
In terms of the Weitzenb\"ock connection~\eqref{weitz hatE e8},
the embedding tensor is obtained by a projection analogous to~\eqref{tors def upto7}:
\begin{align}\label{tors def e8}
`X_AB^C` 
= \bbT\big[ \weitz{\hat E}_{AB}{}^C\big]
&= \weitz{\hat{E}}_{AB}{}^C 
+ \weitz{\hat{E}}_{FA}{}^F \delta_B^C
-`f_GE^F``f^G_B^C` \weitz{\hat{E}}_{FA}{}^E
+\frac{1}{60} \weitz{\hat{E}}_{DF}{}^E  `f_AE^F``f^D_B^C` \\[1ex]
&= \weitz{\hat{E}}_{AB}{}^C 
- \weitz{\hat{E}}_{BA}{}^C 
+`Y^CF_EB` \weitz{\hat E}_{FA}{}^E
+\frac{1}{60} \weitz{\hat{E}}_{DF}{}^E  `f_AE^F``f^D_B^C`\,.
\end{align}
We have defined this expression to be the \E8 version of the torsion projection $\bbT\big[\,\big]$ on a Weitzenb\"ock connection.\footnote{With a slight abuse of language, we call $\bbT\big[\,\big]$ a projection although it is really a linear combination of the projectors on $\mathbf1+\mathbf{248}+\mathbf{3875}$.}
We notice that the last term in \eqref{tors def e8} is special to \E8 and was not present in \eqref{tors def upto7}.
It can be traced back to the contribution of the ancillary parameter in \eqref{gss ansatz e8 components}.
The embedding tensor components are then computed to be
\begin{align}\label{Theta to W}
\vartheta_A &=  \weitz{\hat{E}}_A +\weitz{\hat{E}}_{EF} f^{EF}{}_A\,,\\\nonumber
\Theta_{AB} + \theta \,\eta_{AB} &= \weitz{\hat{E}}_{AB} + \weitz{\hat{E}}_{BA} - `f_E(A^C``f^E_B)^D`\,\weitz{\hat{E}}_{CD} \,.
\end{align}

The quadratic constraint \eqref{QCX} is automatically satisfied for any $`X_AB^C`$ obtained from a gSS reduction. It descends straightforwardly from the Leibniz identity~\eqref{dorfleib}.

One may ask how the ansatz for ancillary parameters in \eqref{gss ansatz e8 components} comes to be.
The requirement that only the coefficients $\lambda^A$ should appear descends from the structure of gauged supergravities, which do not include any ancillary parameters.
The precise structure of the ansatz may be fixed by an analysis analogous to the one carried out for \ensuremath{\mathrm{E}_9} ExFT in section~3.2 of~\cite{Bossard:2023wgg}.
In contrast with the \ensuremath{\mathrm{E}_9} case, it is rather straightforward to find that \eqref{gss ansatz e8 components} is the most general ansatz leading to an embedding tensor sitting in the correct irreps, as above.\footnote{
More explicitly, one could try to add an extra term to \eqref{gss ansatz e8 components} so that $\Sigma_M(x,y) =  -\lambda^A(x) \,`\cU_M^B`(y) \big( \weitz{\hat{E}}_{BA}(y) + \tilde{H}_{BA}(y)\big)$.
One then finds that $(\bbP_{\str{\bf1}+{\bf248}+{\bf3875}})`^AB_CD`\tilde{H}_{AB}=\tilde{H}_{AB}$ in order to reproduce the correct irrep content of the embedding tensor, as required by supersymmetry. However, any non-vanishing $\tilde{H}_{AB}$ satisfying the above condition leads to a violation of the section constraint \eqref{SC ancillary} and must be discarded.
}

\subsection{A generalised torsion for \texorpdfstring{\E8}{E₈₍₈₎}}
\label{sssec: e8 gentors}

In section~\ref{sssec:gss upto7} we interpreted the embedding tensor as a constant generalised torsion for the frame $\hat{E}`_A^M`$.
We can give a similar interpretation here.\footnote{
One should be aware, however, that outside of the scope of gSS reductions a different ansatz for the ancillaries could also be considered. This possibility will not play any role in our computations and we can safely ignore it.}
The computation is analogous to the lower rank case, but rendered more subtle by the presence of the ancillary components in the Dorfman product.
An analogous computation was carried out for \ensuremath{\mathrm{E}_9} ExFT in~\cite{Bossard:2023wgg}.
For a generic local frame $`E_A^M`$, we write in analogy with \eqref{EUr}
\begin{equation}\label{generic local frame}
U`_M^A`(y) \in\E8\,,\qquad
r(y)>0\,,\qquad 
E`[s]{}_A^M`(y)  =   r(y)^{-1} U`[s]{}_A^M`(y) \,.
\end{equation}
Let us also reintroduce the Weitzenb\"ock connection for this frame:
\begin{equation}\label{weitz E e8}
\weitz{{E}}_{\,AB}{}^C =  E`[s]{}_A^M` E`[s]{}_A^M` \partial_M E`[s]{}_N^C` 
=
\weitz{E}_{AD} \, `f^D_B^C` + \frac12\weitz{E}_A  \, \delta_B^C\,.
\end{equation}
With these definitions, we then have a generic doubled frame
\begin{equation}\label{generic doubled frame}
\bbmE_A = \big( `E_A^M`\,,\, -U`_M^B` \weitz{E}_{BA} \big)\,,
\end{equation}
We shall also use again the shorthand notation introduced in section~\ref{sssec:constr genframe 2017}:
\begin{equation}
\tors{E}_{AB}{}^C = \bbT\big[ \weitz{E}_{AB}{}^C\big]\,,
\end{equation}
where the torsion projection is the \E8 specific one defined in~\eqref{tors def e8}.
We then find
\begin{equation}\label{dorf to torsion}
{\bbmE}_A\dorf{\bbmE}_B 
= -\tors{E}_{AB}{}^C {\bbmE}_C - \Big(\,0\ ,\ \frac{1}{60}\,r^{-1}\, \partial_M\,\tors{E}_{AC}{}^D\,`f_BD^C`\,\Big)
\end{equation}
where the extra term vanishes for constant torsion, as is the case in gSS reductions.
Notice that its structure is analogous to the extra term appearing in the torsion projection~\eqref{tors def e8}. This observation will play an important role later on.

The torsion $\tors{E}_{AB}{}^C$ satisfies a Bianchi identity analogous to \eqref{torsion BI upto7}, which is the generalisation of the quadratic constraint \eqref{QCX} to non-constant torsion.
It descends from the Leibniz identity~\eqref{dorfleib}:
\begin{equation}
\bbmE_A\dorf(\bbmE_B\dorf\bbmE_C)
-\bbmE_B\dorf(\bbmE_A\dorf\bbmE_C)
-(\bbmE_A\dorf\bbmE_B)\dorf\bbmE_C
= 0\,.
\end{equation}
Taking the $\Lambda$ component of this relation and substituting \eqref{dorf to torsion}, one finds\footnote{Several of the computations in this section and in the appendix were carried out also with the help the xAct package for Wolfram Mathematica~\cite{xActWebsite,Martin-Garcia:2008ysv}}
\begin{align}
\label{torsion BI e8}
0=\ &\tors{E}_{\,AC}{}^F\tors{E}_{\,BF}{}^D
-\tors{E}_{\,BC}{}^F\tors{E}_{\,AF}{}^D
+\tors{E}_{\,AB}{}^F\tors{E}_{\,FC}{}^D
+`[s]E_A^M`\de_M\tors{E}_{\,BC}{}^D
\\&\nonumber\hspace{-2ex}
-`[s]E_B^M`\de_M\tors{E}_{\,AC}{}^D
+`[s]E_C^M`\de_M\tors{E}_{\,AB}{}^D
-`Y^DF_GC`\, `E_F^M`\de_M \tors{E}_{\,AB}{}^G
-\frac1{60} \, `E_G^M`\partial_M \tors{E}_{AF}{}^E `f_BE^F``f^G_C^D` \,,
\end{align}
which differs from \eqref{torsion BI upto7} only in the last term, which is again of the same form as the extra contribution in \eqref{tors def e8}.
Indeed, we observe that the second line of \eqref{torsion BI e8} equals the torsion projection $\bbT$ defined in \eqref{tors def e8}, applied to the indices $B$, $C$ and $D$ of $`E_B^M`\de_M\tors{E}_{\,AC}{}^D$. 
This will prove very useful later on.

\subsection{Fluxes, deformations and (un)twisting }
\label{ssec: e8 twist defo}

The presentation above does not take into account global properties of \E8 generalised vectors and also ignores any deformation required to take into account e.g. the Romans mass in type IIA supergravity.
The global patching of gauge parameter couples in \E8 has not been discussed in the literature so far.
Similarly, while the specific deformation of the generalised Lie derivative necessary to account for the Romans mass could be easily deduced from known results in lower-rank ExFTs, a general analysis of arbitrary deformations as in \eqref{genlie F0} is lacking for \E8.
We need to address both these issues in order to determine the gSS uplift conditions for $D=3$ gauged maximal supergravities.
We begin in this section by studying the possible (un)twistings of generalised vectors and how they are reflected on the Dorfman product.

\subsubsection{Local twisting of gauge parameters}
\label{sssec: local C and F}

As a first step, we shall not attempt to study the global patching of generalised vectors and ancillaries.
Rather, we shall just derive algebraic and differential relations between gauge parameters related thorugh dressing by an element of the subgroup  $\cS\subset\E8\times\dsR^+$ which preserves the choice of section.
The relations found here will be of use later on, both for determining global patching conditions and for explicitly constructing the gSS frame, in analogy with sections~\ref{sssec:F0 defo upto7}, \ref{sssec:global patch upto7} and~\ref{sssec:constr genframe 2017} for lower-rank ExFTs.

The decomposition \eqref{GLxS schematic deco} still holds for \E8, because it simply descends from the structure of maximal supergravities obtained from Kaluza--Klein reduction of the ten- and eleven-dimensional theories.
We thus begin by considering a local dressing of gauge parameters based on an element $S_M{}^N$ in $\cS=(\Gup\times\Rup)\ltimes\cP$, i.e. the stabilizer of the choice of section:
\begin{align}\label{C W and Omega}
`S_M^N` \partial_N = \partial_M\,,\qquad
\weitz{S}_{MN}{}^P = `S_N^R` \de_M \big(S^{-1}\big)`[s]\!_R^P`
= \weitz{S}_{MR}`f^R_N^P` + \frac12\weitz{S}_M  \,\delta_N^P\,.
\end{align}
We have also introduced an associated (locally defined) Weitzenb\"ock connection.
Its decomposition into an \e8 and an $\dsR$ element is also displayed and we denote by $\weitz{S}_{MN}$, $\weitz{S}_M$ these components.
With these definitions in place, we consider the local dressing of gauge parameters
\begin{equation}\label{Cdressing}
\bbLambda=\big(\Lambda^M\,,\,\Sigma_M\big)
\ \to\ 
{^{S\!}\bbLambda} = \Big(\,
\Lambda^N `S_N^M`\,,\,
s^{-1}\,\Sigma_M - s^{-1}\,\Lambda^P \weitz{S}_{MP}
 \,\Big)
\end{equation}
Where both $\bbLambda$ and $S_M{}^N$ depend on the internal coordinates $y^m$.
We have introduced a weight factor $s$ reflecting the trombone component of $S_M{}^N$:
\begin{equation}
s = \det\big(S_M{}^N\big)^{-1/248}\,,
\end{equation} 
in analogy with $\rho$ and $r$ in \eqref{EUr} and \eqref{generic local frame}.
The choice of redefinition for the ancillary parameter is based on the form of the gSS ansatz \eqref{gss ansatz e8 components}.
Eventually, to construct the solution to the gSS condition we will need to switch between the frame $\hat{E}`_A^M`$ and a version $`E_A^M`$ differing by an element in $\cS$, just as in the discussion following \eqref{gSS frame upto7}.
The expression \eqref{Cdressing}, applied to $\hat\bbmE_A$, correctly extends the relation between $\hat{E}`_A^M`$ and $`E_A^M`$ in section~\ref{sssec:constr genframe 2017}.
The rescaling of $\Sigma$ by $s$ can also be deduced more simply from covariance under $\dsR^+$, because generalised vectors have weight~1 while ancillaries have weight~0.

One first computes the generalised Lie derivative and finds
\begin{equation}
\cL_{{^{\,S\!}\bbLambda_1}}\big(\Lambda_2^N `S_N^M`\big) =
\big(\twL{F}_{\bbLambda_1} \Lambda_2^N\big) `S_N^M` =
\big(\cL_{\bbLambda_1} \Lambda_2^N - \Lambda_1^P \Lambda_2^Q `F_PQ^N`\big) `S_N^M` \,,
\end{equation}
with
\begin{equation}
`F_MN^P` = \bbT\big[\weitz{S}_{\,MN}{}^P\big]\,,
\end{equation}
which gives a definition of the deformed Lie derivative $\twL{F}$ that  is entirely analogous to the lower-rank cases and extends by covariance to other tensors, including ancillary parameters.
In fact, anticipating \eqref{Xconstr local F e8} below we it is worth stressing that we shall find 
\begin{equation}
\twL{F}_{\bbLambda_1}\Sigma_{2\,M}=\cL_{\bbLambda_1}\Sigma_{2\,M}\,.
\end{equation}
In analogy with section~\ref{sssec:F0 defo upto7}, we have employed the torsion projection as defined in \eqref{tors def e8}, although on ExFT indices $M,N,\ldots$ rather than the `spectator' indices $A,B,\ldots$ that we use for gauged supergravity objects.

We need to follow the same logic not just for the generalised Lie derivative but for the Dorfman product as well.
A more involved computation, presented below, leads to the result
\begin{equation}\label{dorfman Ctwist}
{^S\bbLambda_1}\dorf{^S\bbLambda_2}
\ =\ 
{\vphantom{\big(}}^S\!\big(\, \bbLambda_1\dorf[F]\bbLambda_2 \,\big) 
\end{equation}
where the deformed Dorfman product is
\begin{align}\label{dorfman F}
\bbLambda_1\dorf[F]\bbLambda_2\ =\ &
\twL{F}_{\bbLambda_1}\bbLambda_2 +\big(\,0\,,\,\Delta\Sigma_{12}\,\big)\,,\\[1ex]\nonumber
\Delta\Sigma_{12\,M}\ =\ &
\Lambda_{2}^N \partial_M\sqb{\bbLambda_1}_N
-\frac1{248} \Sigma_{2\,M} \, \Lambda_1^N`F_NP^P`
-\frac1{60}\Lambda_2^Q\,\partial_M\big(\Lambda_1^P `F_PR^S`\big)\,`f_QS^R`\,.
\end{align}
This expression should be compared with the original definition of the Dorfman product in~\eqref{dorfman on bbLambdas}.
We stress that the action of $S_M{}^N$ on the right-hand side of \eqref{dorfman Ctwist} also involves a shift of the ancillary component of $(\bbLambda_1\dorf[F]\bbLambda_2)$ by its vector component $\twL{F}_{\bbLambda_1}\Lambda_2^M$, just as in \eqref{Cdressing}.

The interpretation of $`F_MN^P`$ is analogous to section~\ref{sssec:F0 defo upto7}.
It corresponds to introducing background values for the field strengths of the $p$-forms of the $(D+d)$-dimensional theory.
It also allows to introduce further deformations associated to dressing all fields by a coordinate dependent element of the global symmetry group $\Gup\times\Rup$.
Such dressings can indeed be rewritten in terms of extra couplings in analogy with background values for $p$-form field strengths.
We shall sometimes improperly call `trombone flux' the $\Rup$ component of such deformations and `scalar currents' the $\Gup$ components.
In a \GL(d) decomposition of $`F_MN^P`$, these appear as algebra-valued one-forms.
We can indeed compute that $`F_MN^P`$ satisfies the same algebraic equations as in \eqref{Xconstr}, \eqref{Cconstr simple}.
Since $`S_M{}^N`\partial_N=\partial_M$ by definition, we have $\weitz{S}_{\,MN}{}^P\partial_P=0$ and $\weitz{S}_{\,PN}{}^P=0$ as well.
Combining this with the section constraint, it is immediate to find indeed 
\begin{align}\label{Xconstr local F e8}
&`F_MN^P` \de_P = 0\,,
\\[1ex]
\label{Cconstr local F simple e8}
&`Y^MN_RS` \,`F_MP^P` \,\partial_N = 0\,,
\end{align}
\begin{align}
\label{BI local F e8}
&F_{MP}{}^R F_{NR}{}^Q
-F_{NP}{}^R F_{MR}{}^Q
+F_{MN}{}^R F_{RP}{}^Q
+\de_M F_{NP}{}^Q
\\&\nonumber
-\de_N F_{MP}{}^Q
+\de_P F_{MN}{}^Q
-`Y^QR_SP`\, \de_R F_{MN}{}^S
-\frac1{60} \, \partial_T F_{MS}{}^R `f_NR^S``f^T_P^Q` 
\ =\ 0\,.
\end{align}
Just as observed right after \eqref{torsion BI e8}, we recognise that the second line of the latter relation corresponds a torsion projection \eqref{tors def e8} acting on the indices $N,P,Q$.
Also notice that this relation differs from \eqref{BIF0 upto7} only by the last term.

It must be stressed that so far we have only shown that relations \eqref{Xconstr local F e8}, \eqref{Cconstr local F simple e8} and \eqref{BI local F e8} are necessary for some object $`F_MN^P`\in{\bf1}+{\bf248}+{\bf3875}$ to define a consistent (local) deformation of the generalised Lie derivative and Dorfman product.
Sufficiency is proved in section~\ref{sssec: F0 defo e8} below.

\subsubsection{Derivation of the deformed Dorfman product}

The $\Lambda^M$ component of \eqref{dorfman Ctwist} is straightforward to compute, because there is a single derivative involved which either acts on the gauge parameters, reproducing the generalised Lie derivative, or on $`S_N^M`$, reproducing its torsion as in the first component of \eqref{dorf to torsion}
\begin{align}\label{FirstComponent}
\big({}^S\bbLambda_1 \circ {}^S\bbLambda_2\big)^{M} &= 
 \Big[\cL_{\bbLambda_1}\Lambda_2^{P} 
- F_{RS}{}^P\,\Lambda_1^R\,\Lambda_2^S\Big] S_P{}^M\,,
\end{align}
where we have $`F_MN^P` = \tors{S}_{\,MN}{}^P$.

For the ancillary $\Sigma_M$ component, combining \eqref{Cdressing} and \eqref{dorfman Ctwist} we can write
\begin{align}\label{proof dorfman dress ancillary 01}
s\,\big({^S\bbLambda_1}\dorf{^S\bbLambda_2}\big)_M
\ =\ &
\big(\, \bbLambda_1\dorf[F]\bbLambda_2 \,\big)_M
- \big(\, \bbLambda_1\dorf[F]\bbLambda_2 \,\big)^P \weitz{S}_{MP}
\\\nn\ =\ &
\big(\, \bbLambda_1\dorf\bbLambda_2 \,\big)_M
+ {\mathbbm{\Delta}_{12}}_M
- \big(\, \bbLambda_1\dorf[F]\bbLambda_2 \,\big)^P \weitz{S}_{MP}
\end{align}
where ${\mathbbm{\Delta}_{12}}_M$ denotes the $`F_MN^P`$ dependent terms in the ancillary component of $\bbLambda_1\dorf[F]\bbLambda_2$, which we now compute.
Derivatives of $`S_M^N`$ can always be written in terms of $\weitz{S}_{MN}{}^P$ using \eqref{C W and Omega}.
Then, ${\mathbbm{\Delta}_{12}}_M$ reduces to terms containing at most one more derivative.
Looking first at those terms where this derivative acts on $\weitz{S}_{MN}{}^P$, we compute
\begin{align}
\big({\mathbbm{\Delta}_{12}}_M\big)\Big|_{\partial W} 
=\ &
\Lambda_1^P\Lambda_2^Q \Big(
\frac12`f_PQ^N`\partial_M\weitz{S}_N
-\partial_M\weitz{S}_{QP}
-\partial_P\weitz{S}_{MQ}
+`f_TP^S``f^T_Q^R`\partial_M\weitz{S}_{RS}
\Big)
\\=\ &\nn
\Lambda_1^P\Lambda_2^Q \Big(
`f_Q^RS`\weitz{S}_{MR}\weitz{S}_{PS}
-\frac12`f_{PQ}^N` \partial_M\weitz{S}_N
-2\partial_M\weitz{S}_{[PQ]}
+`f_TP^S``f^T_Q^R`\partial_M\weitz{S}_{RS}
\Big)
\\=\ &\nn
-\frac{1}{60}\Lambda_1^P\Lambda_2^Q  \partial_M`F_PR^S``f_QS^R`
+\Lambda_1^P\Lambda_2^Q `f_Q^RS`\weitz{S}_{MR}\weitz{S}_{PS}
\end{align}
where going from the first to the second line we have used the identity
\begin{equation}
2\partial^{\vphantom{A}}_{[M}\weitz{S}_{N]P} =
`f^RS_P` \weitz{S}_{MR}\weitz{S}_{NS}
\end{equation}
on the last term of the first line. 
To get to the third line we have used the Jacobi identity on the last term of the second line, together with the relations
\begin{equation}\label{WSd=0}
\weitz{S}_{MN}{}^P\partial_P = 0 = \weitz{S}_{PN}{}^P\,,
\end{equation}
which follow from $`S_M^N`\partial_N=\partial_M$.
At this point, we find that all the remaining terms quadratic in $\weitz{S}_{MN}{}^P$ cancel out upon using the Jacobi identity, the section constraint and \eqref{WSd=0}.
In particular, one uses the relations
\begin{align}
`f^MN_P` \weitz{S}_{QM}\weitz{S}_N &= \frac12\weitz{S}_P\weitz{S}_Q\,,\\
`f^MN_P` \weitz{S}_{QM}\weitz{S}_{NR} &= \frac12\weitz{S}_{PR}\weitz{S}_Q\,.
\end{align}

Using analogous manipulations it is then rather straightforward to compute that the remaining terms in ${\mathbbm{\Delta}_{12}}_M$ -- involving either a derivative acting on a gauge parameter or an ancillary, combine to give
\begin{equation}
{\mathbbm{\Delta}_{12}}_M =
-\frac1{248} \Lambda_1^N \Sigma_{2\,M} `F_NP^P`
-\frac{1}{60}\Lambda_2^Q\,\partial_M\Big( \Lambda_1^P `F_PR^S` \Big) `f_QS^R`\,,
\end{equation}
thus reproducing \eqref{dorfman Ctwist},~\eqref{dorfman F}.

\subsubsection{Arbitrary deformations of the Dorfman product}
\label{sssec: F0 defo e8}

We now ask the reverse question than in the previous section and show that conditions \eqref{Xconstr local F e8}--\eqref{BI local F e8} are also sufficient for an object $`F_MN^P`\in{\bf1}+{\bf248}+{\bf3875}$ to define a consistent deformation of the Dorfman product.
We introduce a yet undetermined deformation $`F_MN^P`$ just as in \eqref{dorfman F}, but this time we do not assume that it comes from the dressing of gauge parameters by an element in $\cS$ as we did above.
We thus impose the Leibniz identity on the deformed Dorfman product
\begin{equation}\label{dorfleib F}
\bbLambda_{1}\dorf[F]\big(\bbLambda_{2}\dorf[F]\bbLambda_{3}\big)
-\bbLambda_{2}\dorf[F]\big(\bbLambda_{1}\dorf[F]\bbLambda_{3}\big)
-\big(\bbLambda_{1}\dorf[F]\bbLambda_{2}\big)\dorf[F]\bbLambda_{3}
\ =\ 0\,.
\end{equation}
Assuming the section constraint \eqref{SC e8} is satisfied, only terms proportional to $`F_MN^P`$ and its derivatives survive. 
The generalised vector component of \eqref{dorfleib F} then reads
\def\L#1#2{{\Lambda_{#1}^{#2}}}
\def\S#1#2{{\Sigma_{#1\,#2}}}
\begin{align}
0=\,\,&
- `F_PS^M``F_QR^P` \L1Q\L2R\L3S
- `F_QS^P``F_RP^M`\L1Q\L2R\L3S
+ `F_QP^M``F_RS^P`\L1Q\L2R\L3S
\\[1ex]\nonumber&
+ `f^MPQ``F_RSP` \S1Q \L2R\L3S  
- `f^P_Q^R``F_SP^M`\S1R\L2S\L3Q
- `f^P_Q^R``F_PS^M`\S1R\L2Q\L3S
\\\nonumber&
-\frac1{248} `f^M_P^Q``F_SR^R` \L1S\S2Q\L3P
- `f^M_P^Q``F_RS^P`\L1R\L2S\S2Q
+ `f^P_Q^R``F_SP^M` \L1S\S2R\L3Q
\\\nonumber&
- \partial_P`F_QR^M`\L1P\L2Q\L3R
+ \partial_Q`F_PR^M`\L1P\L2Q\L3R
+\frac1{60}`f^MPQ``f_TR^S` \partial_Q`F_US^T` \L1U\L2R\L3P
\\\nonumber&
+\frac1{60}`f^MPQ``f_TR^S` `F_US^T` \partial_Q\L1U\L2T\L3P
+ \partial_R`F_PQ^R` \L1P\L2Q\L3M
+ `f_P^M_R^Q`\partial_R`F_TU^Q`\L1T\L2U\L3P
\\[.5ex]\nonumber&
+ `f_P^M_Q^R``F_TU^P`\partial_R\L1Q\L2T\L3U
+ `F_QP^R`\partial_R\L1Q\L2P\L3M
+ `F_QP^M`\partial_R\L1Q\L2R\L3P
\\[1ex]\nonumber&
+ `F_PQ^M`\partial_R\L1R\L2P\L3Q
+ -`f_P^M_Q^R``F_UT^Q` \partial_R\L1U\L2T\L3P
+ -`f_P^M_Q^R``F_UT^P`\L1U\partial_R\L2Q\L3T
\\[1ex]\nonumber&
+ `F_PQ^R`\L1P\partial_R\L2Q\L3M
+ `f_PS^M``f^S_Q^R``F_UT^P`\L1U\partial_R\L2T\L3P
+ `F_PQ^R`\L1P\L2Q\partial_R\L3M
\\[1ex]\nonumber&
+ `f_Q^P_R^S``F_UP^M` \partial_S\L1R\L2Q\L3U
+ `f_Q^P_R^S``F_PU^M` \partial_S\L1R\L2U\L3Q
- `f_Q^P_R^S``F_UP^M` \L1U\partial_S\L2R\L3Q\,.
\end{align}
where we introduced the shortcut notation $`f_M^N_P^Q` = `f_SM^N``f^S_P^Q`$.

Since the gauge parameters are arbitrary, we can choose them appropriately in order to separate several pieces of this expression. 
By setting $\L{1,2,3}M$ to constants and the ancillaries to~0, we reproduce the Bianchi identity~\eqref{BI local F e8}.
The constraint~\eqref{Xconstr local F e8} is then easily recovered as well.
Notice that, being algebraic, it also holds when we substitute the derivative $\partial_P$ for any other object on section, such as an ancillary parameter $\Sigma_P$.
We can then reinstate $\S1M$ to find the extra condition
\begin{align}
\label{F extra cond}
  &\frac1{248} `f_R^QM``F_PS^S`\Sigma_Q
+ `f_P^QT``F_TR^M` \Sigma_Q =
\\&\qquad\nonumber
= 
\Sigma_Q\big( `f^Q_P^T` `F_TR^M` + `f^Q_R^T``F_PT^M` - `f^Q_T^M``F_PR^T`   \big)
= 0\,,
\end{align}
which states that $`F_MN^P`$ is invariant under the \e8 subalgebra generated by ancillary parameters.
The remaining terms give a constraint that generalises to \E8 the `C-constraint' found in \cite{Ciceri:2016dmd,Inverso:2017lrz} for extended field theories without ancillaries.
It reads
\begin{align}
\label{full Cconstr e8}
\Big(
  `F_(PQ)^M`\delta_S^N
- `Y^MN_TS``F_(PQ)^T`
-\frac12 `Y^TN_PQ``F_TS^M`
+\frac1{120}`f_S^NM``f_QT^U``F_PU^T` 
\Big)\,\partial_N = 0\,.
\end{align}
One should now simplify this expression, for instance by separating each independent irrep.
Tracing with $\delta_M^P$ reproduces the simple constraint~\eqref{Cconstr local F simple e8}.
In fact, it turns out by direct computation that both the algebraic constraints~\eqref{F extra cond} and~\eqref{full Cconstr e8} are satisfied if \eqref{Xconstr local F e8} and~\eqref{Cconstr local F simple e8} are.\footnote{We have found it easier to obtain this result by using \texttt{Wolfram Mathematica} and explicitly constructing the set of linear equations described here, for all inequivalent solutions of the section constraint.}

A rather long computation (see the appendix for more details) shows that the ancillary component of~\eqref{dorfleib F} vanishes if the constraints above are satisfied.
Namely, the ancillary component of~\eqref{dorfleib F} only contains the constraints above and their derivatives.
We conclude that the conditions found in the previous section, namely \eqref{Xconstr local F e8}, \eqref{Cconstr local F simple e8} and~\eqref{BI local F e8}, are necessary and sufficient for a deformed Dorfman product~\eqref{dorfman F} to be consistent and satisfy the Leibniz identity.

Furthermore, since in this section we never assumed that $`F_MN^P`$ comes from the torsion associated to some matrix $`S_M^N`$, the conditions~\eqref{Xconstr local F e8}, \eqref{Cconstr local F simple e8} and~\eqref{BI local F e8} are all that is needed to also capture deformations of the Dorfman product associated to massive and gauged versions (if any exist) of the $(D+d)$-dimensional theory that is being described.
Such \emph{global} deformations are denoted $\sfF_0$ as in section~\ref{sssec:F0 defo upto7} and only entail \GL(d) singlets.
For later convenience, let us then summarize
\begin{align}\label{dorfman F0}
\bbLambda_1\dorf[\sfF_0]\bbLambda_2\ =\ &
\Big(
\cL_{\bbLambda_1}\Lambda_2^M -\Lambda_1^P\Lambda_2^Q \sfF_0`_PQ^M`\ ,\ 
\cL_{\bbLambda_1}\Sigma_2^M + \Delta\Sigma_{12\,M}
\Big)\,,
\\[1.5ex]\nonumber
\Delta\Sigma_{12\,M}\ =\ &
\Lambda_{2}^N \partial_M\sqb\bbLambda_N
-\frac1{248} \Sigma_{2\,M} \, \Lambda_1^N \sfF_0`_NP^P`
-\frac1{60}\Lambda_2^Q\,\partial_M\big(\Lambda_1^P \sfF_0`_PR^S`\big)\,`f_QS^R`\,.
\end{align}
which is the same expression as~\eqref{dorfman F} but with $\sfF_0`_MN^P`$ in place of $`F_MN^P`$.

A computation analogous to the one in the previous section and in the appendix shows that a generic $`F_MN^P`$ can in fact be separated into an integrable piece, obtained as the torsion of an element in $\cS$, and a massive/gauged deformation:
\begin{equation}
`F_MN^P` = \tors{S}\!\!`_MN^P` + `S_M^R``S_N^S` `\sfF_0\,RS^T` (S^{-1})`_T^P`\,,
\end{equation}
and also
\begin{equation}\label{dorf F0 to dorf F}
{^S\bbLambda_1}\dorf[\sfF_0]{^S\bbLambda_2} = {}^S\big(\bbLambda_1\dorf[F]\bbLambda_2\big)\,.
\end{equation}

\subsubsection{Table of Dorfman twists for 11d and type II supergravities}
\label{sssec: explicit F0 11d type II}

Let us look explicitly at the solutions to \eqref{Xconstr local F e8} and \eqref{Cconstr local F simple e8} to see how they are interpreted.
A convenient parametrisation of \e8 is based on the decomposition\footnote{We do not need to display the \e8 commutation relations in terms of this decomposition, but they can be found for instance in equations (2.6) and (A.6) of \cite{Bossard:2023jid}.}
\begin{equation}\label{e8 to sl9}
\e8 \ =\ \sl(9) + \mathbf{84} + \mathbf{84}'\,.
\end{equation}
We shall use indices $a,b,c,\ldots$ for the $\bf9$ of \sl(9).\footnote{Notice that we used $a,b,\ldots$ to denote the adjoint of \hggauge in section~\ref{sssec:uplift conds 2017}. This should cause no confusion.}
We shall then have \sl(9) generators $`t^a_b`$ with $`t^a_a`=0$ and $t_{abc}=t_{[abc]}$, $t^{abc}=t^{[abc]}$ in the $\bf84$ and ${\bf84}'$, respectively.
The decomposition applies to any object in the $\bf248$, hence we have for instance
\begin{equation}
\partial_M\ =\ \big(\,`\partial^a_b`\,,\ \partial_{abc}\,,\ \partial^{abc}  \,\big)\,.
\end{equation}

The two maximal solutions to the section constraint are as follows.
For 11d supergravity, it suffices to isolate the \gl(8) subalgebra of \sl(9) such that 
\begin{equation}
\sl(9)\ =\ \gl(8) + \mathbf{8}_{+1}+\mathbf{8}'_{-1}
\end{equation}
and pick derivatives in the ${\bf8}_{+1}$ exclusively:
\begin{equation}\label{11dsec}
\text{11d section:}\qquad \partial`^9_m` \neq 0\,,\quad \partial`^m_n`=0=\partial_{abc}=\partial^{abc}\,,\qquad m,n=1,\ldots,8\,.\qquad\quad\ \ \,{}
\end{equation}
We stress that the central element of this \gl(8) algebra does \emph{not} correspond to the one in the structure group \GL(d) appearing in the decomposition \eqref{GLxS schematic deco}. The latter involves a linear combination with the $\dsR^+$ charge.
Of course, the type IIA section is obtained by dropping one internal direction, e.g. taking $m,n=1,\ldots,7$.
The type IIB section is obtained by selecting
\begin{equation}\label{IIBsec}
\text{IIB section:}\qquad \partial_{m89}\neq0\,,\quad
\partial_{mn8}=0=\partial_{mnp}=\partial`^a_b`=\partial^{abc}\,,\qquad
m,n=1,\ldots,7\,,
\end{equation}
which breaks $\sl(9)\to\gl(7)+\sl(2)$.

We decompose $F`_MN^P`$ just as we do the embedding tensor in \eqref{X to Theta e8}:
\begin{equation}\label{F irrep deco}
F_{MN}{}^P = 
(\Phi_{MR} + \phi\,\eta_{MR})\,f^{R}{}_N{}^P
+\varphi_M\,\delta_N^P 
- \frac{1}{2}\,f^R{}_N{}^P\,f_{RM}{}^S\,\varphi_S 
\end{equation}
with $\Phi_{MR}=\Phi_{(MR)}\in\mathbf{3875}$.
The decomposition of the latter irrep is
\begin{equation}\label{3875sl9deco young}
\begin{array}{rcccccccccccc}
\mathbf{3875}&\ \to\ &
\underset{\yng(2,1,1,1,1,1,1,1)}{\vphantom{\sum}\mathbf{80}}&+&
\underset{\yng(2,1)}{\vphantom{\sum}\mathbf{240}}&+&
\underset{\yng(2,2,2,2,2,2,2,1)}{\vphantom{\sum}\mathbf{240}'}&+&
\underset{\yng(2,2,2,2,1,1,1,1) }{\vphantom{\sum}\mathbf{1050}}&+&
\underset{\yng(2,1,1,1,1)}{\vphantom{\sum}\mathbf{1050}'}&+&
\underset{\yng(2,2,1,1,1,1,1)}{\vphantom{\sum}\mathbf{1215}}
\\
&& \vphantom{} `\xi_a^b`
&& \vphantom{} Z_{ab,c}
&& \vphantom{} W^{ab,c}
&& \vphantom{} `A^a_bcde`
&& \vphantom{} `B_a^bcde`
&& \vphantom{} `\Xi^ab_cd`
\end{array}
\end{equation}
where we have also displayed each Young tableaux and the associated tensor we shall use in the following.
Every object is traceless and furthermore $Z_{[ab,c]}=Z^{[ab,c]}=0$ with $Z_{(ab),c}=0=Z^{(ab),c}$, $`A_a^bcde`=`A_a^[bcde]`$ (same for $B$) and $`\Xi^ab_cd`=`\Xi^[ab]_[cd]`$.

We then write
\begin{align}\label{3875sl9deco vars}
`[s]\Phi^a_b\,^c_d` &= `\xi^a_d``\delta^c_b` + `\xi^c_b``\delta^a_d` + `\Xi^ac_bd`\,,&
`[s]\Phi^abc_def`  &= {9}\,\Xi_{\vphantom[}^{[ab}{}^{\vphantom[}_{[de}\delta^{c]}_{f]}
                     + 6 \xi_{\vphantom[}^{[a}{}^{\vphantom{[}}_{[b}\delta^{bc]}_{ef]}\,,\\
`[s]\Phi^a_b,_def` &= `A^a_bdef` + \delta^a_{[d}\, Z^{\vphantom ]}_{ef],b}\,,
&
`[s]\Phi^a_b\,^def` &= `B_b^adef` + \delta_b^{[d}\, W_{\vphantom [}^{ef],a}\,,
\\
`[s]\Phi_abc\,def` &= \frac1{8} `A_[a^{g_1g_2g_3g_4}`\,\epsilon_{bc]defg_1g_2g_3g_4}\,,&
`[s]\Phi^abc\,def` &= \frac1{8} `B^[a_{g_1g_2g_3g_4}`\,\epsilon^{bc]defg_1g_2g_3g_4}\,,
\end{align}
and of course $\varphi_M = (\varphi`^a_b`\,,\ \varphi`_abc`\,,\ \varphi`^def`)$. 
Choosing the 11d section \eqref{11dsec}, we find that only the following components survive, corresponding to the expected fluxes ($m,\,n,\ldots=1,\ldots,8$):
\begin{align}\label{11d F0 fluxes}
`A^9_{mnpq}`\quad&\text{4-form flux,}\\\nonumber
W^{m9,9}\quad&\text{7-form flux,}\\\nonumber
\,\xi`^9_m`=-\frac12\varphi`^9_m` \quad&\text{`trombone flux'}\,.
\end{align}
For type IIA supergravity, we simply use the 11d section \eqref{11dsec} but drop the $y^8$ coordinate, so that $m,n,\ldots=1,\ldots,7$.
In this case, we find the following non-vanishing entries
\begin{align}\label{IIA F0 fluxes}
`W^89,8`\quad&F_0\,,\\\nonumber
\Xi`^89_mn`\quad&F_2\,,\\\nonumber
`A^9_mnp8`\quad&H_3\,,\\\nonumber
`A^9_mnpq`\quad&F_4\,,\\\nonumber
`W^m9,9`\quad&F_6\,,\\\nonumber
`W^89,9`\quad&H_7\,,\\\nonumber
\Xi`^p9_qm`=2`\delta^p_[q`\big(\frac{1}{2}\varphi`^9_m]`+\xi`^9_m]`\big)\quad&\text{`trombone flux' + twist by dilaton shift smmetry,}\\\nonumber
\xi`^9_8`=-\frac12\varphi`^9_8`\quad&\text{embedding tensor: gauging of the IIA trombone.}
\end{align}
Notice in particular that beyond the $p$-form and other `fluxes` which can be locally reabsorbed into a field redefinition (for some matrix $`S_M^N`$ in ExFT language), we find both deformations of IIA supergravity: the Romans mass $F_0$ and the gauging of the type IIA trombone symmetry~\cite{Howe:1997qt}.
We conclude that, in particular,~\eqref{dorfman F} can consistently capture the \E8 generalised geometry for massive IIA supergravity.

For type IIB supergravity, we use the section~\eqref{IIBsec}, introduce the notation ${\sf i}, {\sf j} = 8,9$ to capture the indices of the \SL(2) basic representation and find the following non-vanishing components:
\begin{align}\label{IIB F0 fluxes}
`Z_m(\sf i,j)` \quad&\text{\SL(2) twist}\,,\\\nonumber
`B_{\sf i}^mnpq` \quad&(F_3\,,\,H_3)\text{ doublet}\,,\\\nonumber
`\Xi^mn_{\sf ij}` \quad&F_5\,,\\\nonumber
Z_{\sf ij,k}\quad&(F_7\,,\,H_7)\text{ doublet}\,,\\\nonumber
`Z_m[{\sf i,j}]` = -\frac34\,\varphi_{m\sf ij} \quad&\text{`trombone flux'\,.}
\end{align}

\subsection{Comments on global patching}
\label{sssec:global patch e8}

We are now ready to discuss the global patching of gauge parameters in \E8 exceptional geometry.
In analogy with section~\ref{sssec:global patch upto7}, we take generalised vectors to be sections of a generalised tangent bundle that is locally a direct sum of tangent and (products of) cotangent bundles reproducing the \GL(d) decomposition of $\Rv$.
It will be useful to keep track of these decompositions explicitly, including the trombone and \GL(1) charges.
For the 11d solution of the section constraint, the \e8 adjoint representation branches as follows
\begin{align}
\label{248 11d branch}
\mathbf{248}_{0}\ \mathrlap{\quad\overset{\str\GL(8)}{\longrightarrow}}\hphantom{\overset{\str\GL(7)\times\SL(2)}{\longrightarrow}}\ &
\boxed{
{\bf8}'{}_{\!\!-9}
+ {\bf28}_{-6} 
+{\bf56}'{}_{-3}
}
+ ({\bf63}+{\bf1})_{0} 
+ 
\underbrace{
{\bf56}_{3} 
+ {\bf28}'{}_{6} 
+{\bf8}_{9}
}_{\smash{\mf p}} \,,
\intertext{and generalised vectors in $\Rv=\mathbf{248}_{+1}$ branch in the same way but with degrees shifted by $+8$, while $\partial_M$ sits in ${\bf248}_{-1}$ and is shifted by $-8$. We remind the reader that ancillaries sit in ${\bf248}_0$. For the IIB solution of the section constraint, we have instead}
\label{248 IIB branch}
\mathbf{248}_{0}\ {\overset{\str{\GL(7)\times\SL(2)}}{\longrightarrow}}\ &
\boxed{(\mathbf{7}',\mathbf{1})_{-8}
+ (\mathbf{7},{\bf2})_{-6}
+ (\mathbf{35},{\bf1})_{-4}
+ (\mathbf{{21}}',{\bf2})_{-2}
}
\\\nonumber&
\,+ (\mathbf{48}+1,{\bf1})_0 
+ (\mathbf{1},{\bf3})_0
+
\underbrace{
(\mathbf{{21}},{\bf2})_{2}
+ (\mathbf{{35}}',{\bf1})_{4}
+ (\mathbf{7}',{\bf2})_{6}
+ (\mathbf{7},\mathbf{1})_{8}
}_{\smash{\mf p}} \,.
\end{align}
In this case, generalised vectors have \GL(1) charges shifted by $+7$ and $\partial_M$ is shifted by $-7$.
In both cases we have framed the entries which, for generalised vectors, correspond to the tangent space and to the gauge parameters of standard $p$-forms.
All other components act trivially either by themselves, or when combined with a suitable choice of ancillary parameter~\cite{Hohm:2014fxa}.
The branching for IIA supergravity is deduced from the 11d one by KK reduction:
\begin{align}
  \label{248 IIA branch}
  \mathbf{248}_{0}\ \mathrlap{\quad\overset{\str\GL(7)}{\longrightarrow}}\hphantom{\overset{\str\GL(7)\times\SL(2)}{\longrightarrow}}\ &
  \boxed{
  {\bf7}'{}_{-8}
  +{\bf1}_{-7}
  +{\bf7}_{-6} 
  +{\bf21}_{-5}
  +{\bf35}_{-3}
  +{\bf21}'_{-2}
  +{\bf7}'_{-1}
  }
  \\&\nonumber
  +( {\bf48}+{\bf1} )_0
  \underbrace{
  +{\bf1}_0
  +{\bf7}_{+1}  
  +{\bf21}_{+2}
  +{\bf35}_{+3}
  +{\bf21}'_{+5}
  +{\bf7}'_{+6} 
  +{\bf1}_{+7}
  +{\bf7}_{\!+8}
  }_{\smash{\mf p}} \,.
\end{align}
Again, the \GL(1) gradings for generalised vectors and their conjugate representation are shifted by $+7$ and $-7$, respectively.

In order to relate twisted and untwisted gauge parameters, we need to identify the group $\cP$ introduced in~\eqref{GLxS schematic deco}.
Its algebra of generators is denoted $\mf p$ and is also identified above.
For \E8 ExFT the unipotent group $\cP$ necessarily contains an abelian subgroup associated to the scalars $\varphi_m$ dual to the Kaluza--Klein vector fields.
They correspond to the highest-graded component in the branchings above.
In a standard Kaluza--Klein dimensional reduction, these scalars would be interpreted as coming from a dual graviton.
In ExFT they are in fact gauged away by invariance under the local, finite transformations generated by ancillary parameters $\Sigma_M$~\cite{Hohm:2013pua}. 
In studying the global patching conditions of \E8 generalised vectors, we must ask whether and how these extra scalars contribute to the \E8 analogue of the (un)twisting introduced in \eqref{untwisting}, \eqref{genLie flux twisting} for lower rank ExFTs.

In analogy with section~\ref{sssec:global patch upto7}, we introduce untwisted generalised vectors $\widetilde\Lambda^M$ and ancillaries $\widetilde\Sigma_M$.
The former sit globally in the direct sum of tangent and (products of) cotangent spaces corresponding to the \GL(d) branchings above.
Being subject to the section constraint, the ancillaries only sit in the highest-degree element of the \GL(d) branching.
We then posit that on each coordinate patch one has
\begin{equation}
\bbLambda = {^\sfC\widetilde\bbLambda}
= \Big(\, 
\widetilde\Lambda^N `\sfC_N^M`  \,,\,  \mathsf{c}\,\widetilde\Sigma_M - \mathsf{c}\,\widetilde\Lambda^N\, \weitz{\sfC}_{\,MN}
 \,\Big)  \,,\qquad 
`\sfC_M^N` \in \cP   \,,\quad
\mathsf{c} = \det\big(\sfC_M{}^N\big)^{1/248}\,.
\end{equation}
The generalised Dorfman product is then twisted accordingly:
\begin{equation}\label{dorfman global twisting}
{^\sfC\bbLambda_1} \,\dorf[\sfF_0]\, {^\sfC\bbLambda_2}
= 
{^\sfC}\Big( \widetilde\bbLambda_1 \,\dorf[\sfF]\, \widetilde\bbLambda_2 \Big)\,,\qquad
\sfF`_MN^P` = \tors{\sfC}_{\,MN}{}^P+\sfF_0`_MN^P`\,.
\end{equation}

On overlaps between two coordinate patches, $\sfC$ transforms as in \eqref{C to CGamma} which we reproduce for convenience:
\begin{equation}\label{C to CGamma e8}
`\sfC_M^N` \to `\sfC_M^P``\Gamma_P^N`\,,\qquad `\Gamma_P^N`\in\cP\,,\qquad
0 = \tors{\Gamma}_{MN}{}^P +`\Gamma_M^R``\Gamma_N^S` \, `\sfF_0\,RS^T` \big(\Gamma^{-1}\big)`[s]{}_T^P`\,,
\end{equation}
where the torsion projection is now defined in \eqref{tors def e8}.
The matrices $\Gamma`_M^N`$ defined on each overlap correspond to the action on generalised vectors and ancillaries of the finite version of a generalised diffeomorphism, with vanishing tangent space component.
They include gauge transformations for the interanl $p$-forms of the $(D+d)$-dimensional theory, as well as the exponential of those highest degree \e8 generators in \eqref{248 11d branch} and \eqref{248 IIB branch} that correspond to the dual graviton.
The latter are parametrised as
\begin{equation}
\Gamma_{\!\text{d.g.}} = \exp\big(\sigma_M t^M\big)
\end{equation}
with $\sigma_M$ on section just as all ancillaries.
We must check that $\Gamma_{\text{d.g.}}$ does not contribute to the torsion condition~\eqref{C to CGamma e8}.
Under $\sfC`_M^N`\to\sfC`_M^P`\Gamma`_P^N`$ the Weitzenb\"ock connection becomes
\begin{equation}
\weitz{\sfC}_{MN}{}^P \to \weitz{\sfC}_{MN}{}^P + \sfC`_M^R`\sfC`_N^S`\weitz{\Gamma}_{\,RS}{}^T\sfC`_T^P`
\end{equation}
so one needs to check that $\Gamma_{\text{d.g.}}$ has vanishing torsion and that it leaves $\sfF_0`_\,MN^P`$ invariant.
The latter descends from the algebraic constraint  \eqref{F extra cond} that $\sfF_0`_\,MN^P`$ satisfies, while for the former one has
\begin{equation}
\weitz{\Gamma_{\text{d.g.}}}_{MN} = -\partial_M \sigma_N\,,\qquad 
\weitz{\Gamma_{\text{d.g.}}}_{M} = 0\,,
\end{equation}
and by the section condition the torsion projection of the first term onto ${\bf1}+{\bf3875}$ vanishes.

We conclude that the gauged shift symmetries associated to ancillary transformations, as well as the associated potentials within $\sfC$ do not contribute to the twisting~\eqref{dorfman global twisting} of the generalised Dorfman product.
Some further analysis taking into account multiple overlap conditions is presented in appendix~\ref{app:cocycle}.
Since ancillaries are gauge symmetries of ExFT, we can safely choose a gauge on each coordinate patch (of a good cover) such that 
the patching of generalised vectors and ancillaries is entirely determined by finite $p$-form gauge transformations, encoded into $\Gamma\in\cP$ generated by the positive elements in the branchings \eqref{248 11d branch}, \eqref{248 IIB branch}, except the one of top degree.

\section{Uplift conditions for $D=3$ gauged maximal supergravities}
\label{ssec:gss uplift cond e8}
\subsection{Coset construction of gSS ansaetze}

We are now ready to derive the existence conditions for a gSS ansatz giving rise to a specific embedding tensor $`X_AB^C`$ of $D=3$ maximal supergravity, extending~\cite{Inverso:2017lrz} to \E8.
In order to also take into account the Romans mass in IIA supergravity (and other deformations for uplifts to $D<10$ supergravities), we will look at the gSS condition
\begin{equation}\label{gss with F0 e8}
\hat\bbmE_A\dorf[\sfF_0]\hat\bbmE_B = - `X_AB^C`\hat\bbmE_C\,.
\end{equation}

The presence of ancillaries in \E8 ExFT does not enter the proof that the internal space is a coset space.
For that, one only uses~\cite{Grana:2008yw,Inverso:2017lrz} that the tangent space components of the frame are globally defined and realise the Lie algebra $\hggauge$.
This is simply because for any gauge parameters,
\begin{equation}
\big(\cL_{(\Lambda,\Sigma)\,}V^M\big)\cE_M{}^m = \Lambda^n\partial_n V^m - V^n\partial_n \Lambda^m \,,
\end{equation}
where $\Lambda^m=\Lambda^M\cE_M{}^m$ selects the tangent space component of a generalised vector, and the right-hand side is of course the standard Lie derivative between two standard vectors.
The presence of an ancillary parameter on the left-hand side has no effect because of the section constraint.
We can then follow~\cite{Inverso:2017lrz} to conclude that the internal space is $\mf{M}=\hGgauge/\hHgauge$ and that the necessary condition \eqref{Thetahat SC} also holds for \E8.
The same reasoning is sufficient to conclude that the generalised frame always takes the form \eqref{gSS frame upto7}, with $S`_M^N`\in\cS$ to be determined by integrating an associated flux deformation $`F_MN^P`$.
More precisely, we consider the local frame defined in \eqref{local frame E} with associated doubled frame $\bbmE_A$ as defined in \eqref{generic doubled frame} and solve for $`F_MN^P`$ in  
\begin{equation}\label{EdfE=-XE e8}
\bbmE_A\dorf[F]\bbmE_B = -`X_AB^C`\bbmE_C\,.
\end{equation}
Then, $`F_MN^P`$ is determined algebraically as in \eqref{F=X-T}
\begin{equation}\label{F=X-T e8}
  `F_MN^P` = E`_M^A `E`_N^B`  \big( `X_AB^C` - \tors{E}_{AB}{}^C \big) E`_C^P`\,.
\end{equation}
and this equality holds also for the ancillary component of the gSS condition.

The non-trivial part of the proof of the existence and explicit construction of the generalised frame is to determine algebraic conditions on the embedding tensor such that $`F_MN^P`$ as defined in \eqref{F=X-T e8} satisfies the consistency conditions and Bianchi identity \eqref{Xconstr local F e8}--\eqref{BI local F e8}.
The proof of \eqref{Xconstr local F e8} is identical to \cite{Inverso:2017lrz} and we do not repeat it.
To prove \eqref{Cconstr local F simple e8}, we begin by noticing that $\mathring{e}`_A^M`$ is a \GL(d) element and can thus be brought through the section matrix as in~\eqref{GLd and S}.
The object that needs to satisfy the section constraint is then 
\begin{equation}\label{Cconstr simple e8 proof 01}
\mathring{e}`_A^M``F_MP^P`
= 
`L_A^B`\big( \tors{E}_{\,BC}{}^C-`X_BC^C` \big)
= 
`L_A^B`\tors{E}_{\,BC}{}^C-`X_AC^C`\,,
\end{equation}
where we have used gauge invariance of the embedding tensor in the second step.
Using the first of \eqref{tors def e8}, we have
\begin{equation}
\tors{E}_{\,BC}{}^C = \weitz{E}_{\,BC}{}^C+248\,\weitz{E}_{\,CB}{}^C\,.
\end{equation}
Notice that the extra term in the torsion projection for \E8 does not contribute at all, so this proof is valid for lower-rank ExFTs as well upon substituting the appropriate dimension of $\Rv$ and weight $\omega$ of a generalised vector.
Dressing the first term $\weitz{E}_{\,BC}{}^C$ with $`L_A^B`$, it is automatically on section and hence does not contribute to \eqref{Cconstr local F simple e8}.
As for the second term, we compute
\begin{align}
`L_A^B`\weitz{E}_{\,CB}{}^C
&=
\widehat\Theta`_C^{\ul m}` \mathring{e}`_{\ul m}^m` \mathring{e}`_A^N`\partial_m\mathring{e}`_N^C`
+
\widehat\Theta`_C^{\ul m}`\mathring{e}`_{\ul m}^m`
\partial_m`L_A^B`L^{-1}`_B^C`\\\nonumber
&=
\widehat\Theta`_A^{\ul m}`  \mathring{e}`_{\ul m}^m`\mathring{e}`_{\ul n}^n`\partial_m\mathring{e}`_n^{\ul m}`
+
\widehat\Theta`_C^{\ul m}` t_{\ul m}`_A^C`
+
\widehat\Theta`_C^{\ul m}` \mathring{e}`_{\ul m}^m` Q_{\ul m}`_A^C`
\end{align}
where we remind the reader that $t_{\ul m}`_A^B`$ are the coset generators written in the $\Rv$ representation and $Q_{\ul m}$ is the \hHgauge connection, see \eqref{coset CM deco}.
In the second line we have used again the fact that $\mathring e$ can be brought through the section matrix as in \eqref{GLd and S}.
Furthermore, the \hhgauge projection of $\dd LL^{-1}$ is also an element of the Lie algebra of $\GL(d)\ltimes\cS$, as follows from gauge invariance of the embedding tensor  and the identification of the section matrix with $\widehat\Theta`_A^{\ul m}`$~\cite{Inverso:2017lrz}.
This means that it can also be brought through the section matrix.
The first and the last terms are therefore manifestly on section while the middle one is not.
Collecting this surviving term and plugging back into \eqref{Cconstr simple e8 proof 01} and the latter into \eqref{Cconstr local F simple e8}, we reproduce the uplift condition \eqref{extracstr simple upto7} with $\omega=1$.

We will now prove that the Bianchi identity \eqref{BI local F e8} is satisfied by \eqref{F=X-T e8} if the embedding tensor satisfies \eqref{Thetahat SC} and \eqref{extracstr simple upto7}.
The strategy is analogous to the lower-rank cases in \cite{Inverso:2017lrz}, but we must now take into account the extra terms in the torsion definition \eqref{tors def e8} as well as in the Bianchi identity itself.
As a first step, we manipulate the Lie derivative of any generic $`F_MN^P`$ satisfying the constraints \eqref{Xconstr local F e8} and \eqref{Cconstr local F simple e8}:
\begin{align}
\cL_{\bbLambda}`F_MN^P`
=\ &
\Lambda^R\partial_R`F_MN^P`
-`F_MN^R`\partial_R\Lambda^P
+\partial_M\Lambda^R`F_RN^P`
+\partial_N\Lambda^R`F_MR^P`\\&\nonumber
+`Y^PR_ST`\partial_R\Lambda^S`F_MN^T`
-`Y^TR_SN`\partial_R\Lambda^S`F_MT^P`
-`Y^TR_SM`\partial_R\Lambda^S`F_TN^P`\,.
\end{align}
Since we are imposing \eqref{Xconstr local F e8} and \eqref{Cconstr local F simple e8}, we have just proved above that \eqref{Xconstr local F e8} an \eqref{Cconstr local F simple e8} are satisfied. 
This in turn imples~\eqref{F extra cond}, which guarantees that the ancillary parameter does not contribute to the Lie derivative above.
Now we apply to the last term the relation~\eqref{full Cconstr e8} (which also holds for the same reasons) and bring $`F_MN^P`$ through the $`Y^MN_PQ`$ tensor in the second to last term to find
\begin{equation}\label{Lie F as torsion}
\cL_\bbLambda `F_MN^P` =
\Lambda^R\partial_R`F_MN^P`
+
\bbT\Big[
  \partial_M\Lambda^R`F_RN^P`
\Big]\,,
\end{equation}
where $\bbT$ denotes the torsion projection~\eqref{tors def e8}, acting on the three free indices $M,N$ and $P$.
Using this result and contracting the Bianchi identity \eqref{BI local F e8} with $`\Lambda^M`$ one finds that it is equivalent to the relation
\begin{equation}\label{BI of F as genLie}
\cL^{\scriptscriptstyle[F]}_{\bbLambda} `F_MN^P` =
\bbT\Big[ \partial_M\big(`\bbLambda^R``F_RN^P`\big) \Big]\,,
\end{equation}
for any $\bbLambda$.
An important point is that the `extra terms' appearing in~\eqref{tors def e8}, \eqref{full Cconstr e8} and \eqref{BI local F e8} compared to their counterparts for lower-rank ExFTs all conjure to produce \eqref{BI of F as genLie}.
The conclusion that \eqref{F=X-T e8} satisfies the Bianchi identity then follows the same steps as in \cite{Inverso:2017lrz}.
We substitute $\bbmE_A$ in place of $\bbLambda$ in \eqref{BI of F as genLie} and expand \eqref{F=X-T e8}.
Using the $\Lambda$ component of \eqref{EdfE=-XE e8} to expand the Lie derivative of $`E_A^M`$, and taking into account the quadratic constraint \eqref{QCX} one finds after some algebra
\begin{align}
0=  &\tors{E}_{\,AC}{}^F\tors{E}_{\,BF}{}^D
-\tors{E}_{\,BC}{}^F\tors{E}_{\,AF}{}^D
+\tors{E}_{\,AB}{}^F\tors{E}_{\,FC}{}^D
+`[s]E_A^M`\de_M\tors{E}_{\,BC}{}^D
-\bbT\big[ `E_B^M`\partial_M \tors{E}_{AC}{}^D \big]
\,,
\end{align}
where the torsion projection acts on the indices $B,\,C$ and $D$.
This is indeed the Bianchi identity satisfied by the torsion of the frame $`E_A^M`$, as noted under~\eqref{torsion BI e8}.

\bigskip

To summarise, we have proved that a gauging of $D=3$ maximal supergravity admits a gSS uplift to a higher-dimensional supergravity theory \emph{if and only if} one can find a subalgebra $\hhgauge\subset\hggauge$ such that conditions \eqref{Thetahat SC} and \eqref{extracstr simple upto7} are satisfied (with $\omega=1$ and $\mathrm{dim}\Rv=248$), which we display again here for convenience:
\begin{align}\label{Thetahat SC e8}
    `Y^AB_CD` \, \widehat\Theta_A{}^{\ul m}\,  \widehat\Theta_B{}^{\ul n} &= 0\,,
\\[1ex]
\label{extracstr simple upto7 e8}
    `Y^AB_CD` \, \big( \vartheta_A - \widehat\Theta_G{}^{\ul m} `t_{\ul m\,A}^G` \big) \, \widehat\Theta_B{}^{\ul n} &= 0\,.
\end{align}
Then, the internal space is a coset space $\hGgauge/\hHgauge$ and the generalised frame is explicitly constructed from \eqref{gSS frame upto7}:
\begin{equation}\label{gSS frame e8}
  \hat{E}`_A^M` = \big(L^{-1}\big)_A{}^B \, \mathring{e}_B{}^N \, `S_N^M`\,,\qquad `S_N^M`\in\cS\,,
  \end{equation}
The $S`_M^N`$ matrix appearing in the construction, as well as any massive or gauged deformation of the uplift theory encoded in $\sfF_0`_MN^P`$ are obtained by integration of $`F_MN^P`$, defined in \eqref{F=X-T}, so that \eqref{dorf F0 to dorf F} is satisfied.

Having proved in section~\ref{sssec:global patch e8} that gauge parameters in \E8 ExFT are patched along the internal space by elements of $\cP$ just as in lower-rank ExFTs, we can use the results in~\cite{Inverso:2017lrz} to conclude that the generalised frame $\hat\bbmE_A$ constructed above on a patch-by-patch basis is automatically globally well-defined.
In fact, we are free to apply a finite ancillary gauge transformation to the frame on each patch (of a good cover) in order to guarantee that the transition functions are determined exclusively in terms of $p$-form fluxes on the internal geometry.

\subsection{Uplift conditions as linear constraints}\label{sec:linear uplift conds}

Classifying all gauged maximal supergravities with an uplift is a daunting task that has not yet been acheived.
In fact, there is no known classification of inequivalent gaugings for dimensions 7 or lower, regardless of the existence of an uplift.
The difficulty lies not just in finding all solutions of the quadratic constraints but also in classifying them into independent duality orbits.

The duality invariant uplift conditions \eqref{Thetahat SC}, \eqref{extracstr simple upto7} are most useful if we are searching for uplifts of a specific gauged supergravity model, namely for a given embedding tensor that satisfies the quadratic constraint.
If one wants to try and tackle the task of classifying all gauged models with an uplift, however, classifying all gauged supergravities first to only later test for the existence of an uplift would be even in principle too inefficient.
We can instead use the requirement of existence of an uplift to greatly reduce the amount of independent entries within the $\RTheta+\overline\Rv$ representations of the embedding tensor.
One could therefore restrict to such components and only then attempt to solve for the quadratic constraints, knowing that every independent solution is guaranteed to admit an uplift.
The way this restriction on the components of the embedding tensor is achieved is by rephrasing the upilft conditions \eqref{Thetahat SC}, \eqref{extracstr simple upto7} in terms of a set of section-dependent linear constraints.
%
%
In practice, we shall fix a choice of solution of the section constraint \eqref{SCE}, breaking $\En\times\dsR^+$ according to \eqref{GLxS schematic deco}.
We shall then identify which $\GL(d)\times\Gup$ irreps of the embedding tensor can be turned on for an uplift to exist.
We shall prove that the resulting linear uplift constraints are sufficient, and necessary `up to duality orbit'.
The latter qualifier means that one needs to study the \En orbit of an embedding tensor to determine whether one representative of the orbit falls entirely within the set of allowed $\GL(d)\times\Gup$ irreps.
The procedure must be carried out separately for each independent solution of the section constraint.%
\footnote{Alternatively, we may think of the set of linear uplift constraints combined with the section constraint itself, the section matrix being itself a variable to be solved for. This gives a set of duality invariant conditions, which however are no longer linear.}

\subsubsection{General proof}

The following discussion is valid for any \En ExFT up to and including \E8.\footnote{This approach to uplift conditions was presented in the online seminar~\cite{InversoEGSS}.}
We begin by pointing out \cite{Inverso:2017lrz} that the form of the generalised frame \eqref{gSS frame upto7}, combined with the gSS condition (\ref{gSS cond with F0}~or~\ref{gss with F0 e8}) imply
\begin{equation}\label{Xsym uplift cstr}
`X_(AB)^C`\cE_C{}^m = 0\,,
\end{equation}
where here and in the following we will often write the section matrix with flat $\Rv$ indices, which is indeed motivated by the proof of \eqref{Xsym uplift cstr}.
Projecting the gSS condition \eqref{gSS cond with F0} with $\cE_M{}^m$ and symmetrising in $A,\,B$, the left-hand side vanishes because it reduces to the standard Lie derivative between two vectors.
On the right-hand side, we use \eqref{gSS frame upto7} and notice that $L$ can be passed through $`X_AB^C`$ because the latter is $\Ggauge$ invariant, while the other constituents of the frame can pass through the section matrix, thus giving \eqref{Xsym uplift cstr} as a necessary uplift condition up to \En orbit.
This is only to hold up to \En orbit because we could obviously rotate $`X_AB^C`$ by a constant \En element.
This would amount to a field redefinition in the gauged supergravity theory and cannot change whether or not the theory admits an uplift.
Indeed, the uplift can obtained by dressing the `flat' index of the frame by the inverse \En element, without changing the choice of ExFT section. 
This same reasoning on duality orbits applies to the discussion below.

We now introduce the projectors
\begin{equation}
\Pi`_M^N`\,,\ \ \overline\Pi`_M^N`=\delta_M^N-\Pi`_M^N`\,
\qquad
\Pi`_M^N``\cE_N^m`=`\cE_M^m`\,,\ \ 
\overline\Pi`_M^N``\cE_N^m`=0\,,
\end{equation}
and notice (with reference to \eqref{hatT Lie algebra}) that $\Pi`_A^B`\widehat T_B$ selects a set of coset generators because of the identification \eqref{E=hatTheta} and hence $\overline\Pi`_A^B`\widehat T_B$ generate \hHgauge.
Closure of the latter then implies
\begin{equation}\label{Hclosure uplift cond}
\overline\Pi`_A^E`\overline\Pi`_B^F` `X_EF^G` \Pi`_G^C` = 0\,,
\end{equation}
which is the another necessary condition up to \En orbit.
Notice that we can reduce closure of the \hhgauge Lie algebra to a linear constraint only because we assume that $`X_AB^C`$ satisfies the quadratic constraint~\eqref{QCX}.
Also notice that we are again mixing `flat' and curved indices by exploiting the invariances of the objects at hand.
Finally, we can rewrite \eqref{extracstr simple upto7} straightforwardly as
\begin{equation}\label{trombone uplift cond}
`Y^AB_CD` \Big( \vartheta_A+\omega\,`X_AF^G``\Pi_G^F` \Big) \Pi`_B^E` = 0\,.
\end{equation}

To see that the above constraints are also sufficient, it is enough to observe that \eqref{Xsym uplift cstr} implies that $\Pi`_A^B`\widehat{T}_B$ selects a subset of generators of \hGgauge and then, \eqref{Hclosure uplift cond} implies that $\overline\Pi`_A^B`\widehat{T}_B$ span a subalgebra $\hhgauge\subset\hggauge$.
Projecting the \hggauge index of $\widehat\Theta`_A^a`$ onto the vector space generated by $\Pi`_A^B`\widehat{T}_B$ defines an object $\widehat\Theta`_A^{\ul m}`$ where $\ul m$ runs over the dimension of such vector space.
We see that $\widehat\Theta`_A^{\ul m}`$ satisfies the covariant uplift condition \eqref{Thetahat SC}.
Finally, we now have by definition $\Pi`_A^D``X_DB^C` = \widehat\Theta`_A^{\ul m}``t_{\ul m}\,B^C`$ which maps \eqref{trombone uplift cond} back to \eqref{extracstr simple upto7}.

\bigskip

The conditions \eqref{Xsym uplift cstr}, \eqref{Hclosure uplift cond} and \eqref{trombone uplift cond} have appeared, for lower-rank ExFTs and in a rather different language, in \cite{Bugden:2021wxg,Bugden:2021nwl,Hulik:2022oyc,Hulik:2023aks}.
An embedding tensor $`X_AB^C`$ satisfying the quadratic constraint defines an `elgebra' in the language of \cite{Bugden:2021wxg}.
The choice of a solution to the section constraint, such that $\overline\Pi_A{}^B$ selects a subalgebra of \hggauge, corresponds to a choice of a co-Lagrangian subalgebra $V$.
Condition \eqref{Xsym uplift cstr} corresponds to the requirement $\mathrm{Im}\mathcal{D}\subset V$ there.
Finally, the condition \eqref{trombone uplift cond} does not appear in \cite{Bugden:2021wxg}, because it is redundant for uplifts to 11d supergravity.
An analogous trace condition appears instead in \cite{Bugden:2021nwl} for type IIB supergravity.

An alternative approach to finding the \GL(d) covariant uplift constraints above is to consider the torsion projection of an undetermined Weitzenb\"ock connection $`W_MN^P`$ and taking into account that its first index must be on section \cite{Hassler:2022egz,Bossard:2023jid}.
One can scan for the resulting \GL(d) irreps and state that an embedding tensor admitting a gSS uplift must necessarily sit in such representations \emph{up to duality orbit}.\footnote{One must in fact use again the expression \eqref{gSS frame upto7} in order to transfer any conclusions which apply to the torsion projection of $`W_MN^P`$ \emph{with curved indices} to its version with `flat' indices, $`W_AB^C`$~\cite{Bossard:2023jid}.}
Concluding that such linear requirements are also sufficient entails a repetition of the explicit construction of the frame carried out in \cite{Inverso:2017lrz} and here.
Notice also that this approach has only been applied to massless theories such as 11d supergravity and type IIB, but not for massive IIA supergravity, because the Romans mass cannot be obtained from a Weitzenb\"ock connection unless one violates the section constraint~\cite{Ciceri:2016dmd}.
We have explicitly computed for all \En, $n\le8$ that the set of linear constraints obtained by projection of a Weitzenb\"ock connection are equivalent to the uplift conditions above.\footnote{We find a mismatch with the counting of allowed components performed for \E8 in the appendix of~\cite{Hassler:2022egz}. 
We have computed independently the projection of teh Weitzenb\"ock connection for this case, and checked that the irrep content found from such projection matches the one obtained from \eqref{Xsym uplift cstr}, \eqref{Hclosure uplift cond} and \eqref{trombone uplift cond}, which is a non-trivial cross-check of our counting.}

\subsubsection{Components with uplift for $D=3$ maximal supergravity}

If we are interested in classifying all $D$ dimensional gauged maximal supergravities admitting a gSS uplift we can, without loss of generality, impose the uplift conditions \eqref{Xsym uplift cstr}, \eqref{Hclosure uplift cond} and \eqref{trombone uplift cond} first and then attempt a classification of orbits of solutions of the quadratic constraint~\eqref{QCX} under the residual symmetry group $\GL(d)\ltimes\cS$.
This is generally a daunting task and has not been carried out yet.
We can however at least list the \GL(d) irreps within $`X_AB^C`$ which solve the uplift conditions.
We shall use the same irrep decomposition and variable names as in \eqref{3875sl9deco young} and \eqref{3875sl9deco vars}, with $\Theta_{AB}$ in place of $\Phi_{AB}$.

\subsubsection*{Gaugings from 11d supergravity}

Choosing the 11d supergravity section, the allowed independent entries for a Lagrangian gauging are given by the 890 entries
\begin{align}\label{components uplift 11d}
`\xi^9_m`\,,\quad
`A^9_mnpq`\,,\quad
`B_m^npq9`\,,\quad
`W^m9,9`\,,\quad
`Z_mn,p`\,,\quad
`\Xi^m9_pq`\,.
\end{align}
For trombone gaugings, one further allows for the following 156 variables
\begin{align}
`\vartheta^m_n`\,,\quad
`\vartheta^9_m`\,,\quad
`\vartheta_mnp`\,,\quad
`\vartheta^{mn9}`\,,
\end{align}
as well the following linear identifications
\begin{align}\nonumber
\theta&=\frac12`[s]\vartheta^m_m` \,, &
`[s]\xi^m_n` &= \frac{9}{14}`[s]\vartheta^m_n` -\frac27\delta^m_{\,n}\, `\vartheta^p_p` \,, & 
`[s]\Xi^mn_pq` &= \frac47`[s]\delta^{[m}_{[p}``[s]\vartheta^n]_q]`-\frac17\delta^{mn}_{\,p\,q}\,`\vartheta^r_r` \,, \\
W^{mn,9} &= -2 \vartheta^{mn9} \,, &
W^{m9,n} &= - \vartheta^{mn9} \,, &
`[s]A^m_npqr`&=\frac16 \delta^m_{\,n}\, \vartheta_{pqr} \,. &
\end{align}
Notice that Lagragian gaugings with uplift to 11d supergravity have $\theta=0$, i.e. they necessarily live entirely in the $\bf3875$. This is a duality invariant statement.

\subsubsection*{Gaugings from IIB supergravity}
In this case we have 807 allowed independent entries for Lagrangian gaugings
\begin{align}\label{components uplift IIB}
&`A^m_pq\,\sf ij`\,,\quad
`B_m^npqr`\,,\quad
`B_{\sf i}^mnpq`\,,\quad
`Z_{\mathsf{i}\,m,n}`=`Z_{\mathsf{i}\,n,m}`\,,\quad
`Z_{m{\mathsf i},\mathsf{j}}`\,,\quad
`Z_{\mathsf{ij},\sf l}`\,,\quad
`\Xi^mn_\sf ij`\,,\quad
`\Xi^mn_p\,\sf i`\,,
\end{align}
with $`B_r^npqr`=0=`\Xi^mp_p\,\sf{i}`$ and $`Z_{mn,{\sf i}}`=2`Z_{{\sf i}[m,n]}`$
We remind the reader that $m,n,\ldots$ run from 1 to 7 and $\mathsf{i},\mathsf{j},\ldots=8,9$ denote the \SL(2) fundamental.
Notice again that Lagragian gaugings with uplift to IIB supergravity necessarily live entrirely in the $\bf3875$.

Allowing for gaugings of the trombone, we furthermore have 147 extra allowed components
\begin{align}
`\vartheta^m_n`\,,\quad 
`\vartheta^m_\sf i`\,,\quad 
`\vartheta^mnp`\,,\quad 
`\vartheta_mn\,\sf i`\,,\quad 
`\vartheta_m\,\sf ij`\,,\quad 
\end{align}
together with the linear relations
\begin{align}
&\nonumber
\theta=-\frac{3}{16}`\vartheta^m_m`\,,\quad
`\vartheta^\sf i_\sf j` = -\frac12 \delta^{\sf i}_{\sf j} `\vartheta^m_m`\,,\quad
`\xi^m_n`=\frac{3}{14}`\vartheta^m_n`\,,\quad 
`\xi^m_\sf i`=\frac{3}{14}`\vartheta^m_\sf i`\,,
\\& \nonumber
`B_m^mpqr` = - `\vartheta^pqr`\,,\quad
`B_{\sf i}^{\mathsf j\,}pqr` = -\frac12 \delta^{\sf i}_{\sf j} \, `\vartheta^pqr`\,,\quad 
`Z_{mn,\sf i}`=-2`Z_{\mathsf{i}\,[m,n]}` =\vartheta_{mn\,\sf i}\,,
\\&\nonumber
`[s]\Xi^mn_pq` = -\frac87`[s]\delta^{[m}_{[p}``[s]\vartheta^n]_q]`+\frac18\delta^{mn}_{\,p\,q}\,`\vartheta^r_r` \,,\\&\nonumber
`[s]\Xi^{mp}_{p\,\mathsf{i}}` = \frac57 `\vartheta^m_{\sf i}` \,,\\&
`[s]\Xi^{m\mathsf{i}}_{n\mathsf{j}}` = \frac57\delta^{\sf i}_{\sf j} \Big( `\vartheta^m_n` - \frac{1}{16}\delta^m_{\,n}\,`\vartheta^p_p` \Big) \,,
\end{align}

\subsubsection*{Gaugings from IIA supergravity}

For type IIA we have 807 entries of Lagrangian gaugings
\begin{align}\label{components uplift IIA}
&\nonumber
`\xi^9_m`\,,\quad
`\xi^9_8`\,,\quad
`A^9_mnpq`\,,\quad
`A^9_mnp8`\,,\quad
`B_m^npq9`\,,\quad
`B_m^np89`\,,\quad
`Z_mn,k`\,,\quad
`Z_m8,n`=`Z_n8,m`\,,
\\&
`W^m9,9`\,,\quad
`W^89,8`\,,\quad
`W^89,9`\,,\quad
`\Xi^m9_pq`\,,\quad
`\Xi^m9_n8`\,,\quad
`\Xi^89_mn`\,.
\end{align}
The singlet is again ruled out for Lagrangian gaugings.
Allowing for trombone gaugings we furthermore have 148 extra entries
\begin{align}
`\vartheta^m_n`\,,\quad
`\vartheta^8_m`\,,\quad
`\vartheta^9_m`\,,\quad
`\vartheta^9_8`\,,\quad
`\vartheta_mnp`\,,\quad
`\vartheta_mn8`\,,\quad
`\vartheta^mn9`\,,\quad
`\vartheta^m89`\,,
\end{align}
with the following extra identifications
\begin{align}
&\nonumber
\theta=\frac14`\vartheta^m_m`\,,
\\&\nonumber
`\xi^m_n` = \frac{9}{14}`\vartheta^m_n` -\frac{1}{7} \delta^m_n `\vartheta^p_p`\,,\quad
`\xi^8_m` = \frac{9}{14}`\vartheta^8_m`\,,\quad
`\xi^8_8` = -\frac17`\vartheta^p_p`\,,\quad
`\xi^9_9` = \frac12`\vartheta^p_p`\,,
\\&\nonumber
`A^m_npqr` = \frac16 \delta^m_n `\vartheta_pqr`\,,\qquad
`A^m_npq8` = \frac16 \delta^m_n `\vartheta_pq8`\,,\qquad
`A^m_8pq8` = -\frac16 `\vartheta_pq8`\,,
\\&\nonumber
`W^mn,9` = -2`\vartheta^mn9`\,,\quad
`W^m9,n` = -`\vartheta^mn9`\,,\quad
`W^m8,9` = -2`\vartheta^m89`\,,\quad
`W^m9,8` = -`\vartheta^m89`\,,
\\&\nonumber
`[s]\Xi^mn_pq` = \frac47`[s]\delta^{[m}_{[p}``[s]\vartheta^n]_q]`
-\frac{1}{14}\delta^{mn}_{\,p\,q}\,`\vartheta^r_r` \,, \quad
`[s]\Xi^m8_pq` = \frac27`[s]\delta^{m}_{[p}``[s]\vartheta^8_q]`\,,\quad
`[s]\Xi^m8_n8` = \frac17`[s]\vartheta^m_n` 
- \frac{1}{28} \delta^m_n `\vartheta^p_p`\,,
\\&
`B_p^pmn9` = \frac16`\vartheta^mn9`\,,\quad
`Z_{mn,8}`=-2`Z_{8[m,n]}` =-\vartheta_{mn8}\,.
\end{align}

\subsubsection*{Gaugings from $D\leq9$ supergravities}

This analysis can be carried out also for uplifts to $D=9,8,...$ supergravities, by choosing the appropriate solution to the section constraint.
We do not display here the whole set of solutions but notice that upliftable Lagrangian gaugings always require a vanishing singlet $\theta=0$.

\section{Applications}\label{sec:examples}

\subsection{Compactness conditions}\label{ssec: compactness}

The conditions presented in the sections above do not guarantee compactness of the internal manifold $\hGgauge/\hHgauge\,$.
We shall now point out that one can impose compactness by implementing some extra linear conditions on the embedding tensor, together with the uplift conditions \eqref{Xsym uplift cstr}, \eqref{Hclosure uplift cond} and \eqref{trombone uplift cond}.
These constraints account for all manifolds that are the product of compact (topologically) homogeneous spaces and circles.
We will also account for situations where there are non-trivial fluxes and/or monodromies by some global symmetry along some circles (such as the S-fold solutions of type IIB supergravity \cite{Inverso:2016eet}).
We do not consider, however, the case where a non-compact group manifold or coset space (such as an hyperboloid) is quotiented by a discrete subgroup of isometries in order to make it compact.
Usually such quotients are incompatible with the gSS truncation ansatz, because the discrete quotient group does not commute with the generalised vectors defining the truncation.
An exception are group manifold reductions, where the truncation is based on vectors generating the (say) right isometries and which are therefore invariant under the left ones. A quotient by a discrete subgroup within the left isometries is thus possible.
It is not possible to take into account discrete quotients by the linear constraints we shall now display.
However, it is rather straightforward to amend the constraints to allow for arbitrary group manifolds as we shall comment at the end of this section.

Of course, once we impose the linear constraints to be displayed shortly we are still faced by the much harder task of solving the quadratic constraints \eqref{QCX} and classifying the duality orbits of such solutions.
We do not attempt to do so exhaustively here, but shall provide a few simple examples.

\bigskip

To begin, we restrict to Lagrangian gaugings, because (as argued for instance in \cite{Bossard:2023wgg}) gSS ansaetze giving rise to trombone gaugings violate the conditions for integration by parts of the higher-dimensional (pseudo) action~\cite{Hohm:2014qga}, which means that they necessarily involve spaces where some field blows up at the boundary. By homogeneity, this boundary must be at infinity and hence the space must be non-compact.

First of all, we notice that we can focus on the gauge group \Ggauge as it is embedded in \E8, because if any central element plays a role in the coset construction, it can be associated to an $S^1$.
We then observe that if ${\Ggauge}\big/{\Hgauge}$ is compact, there must be a choice of coset generators that belong to the maximal compact subalgebra of the generators of \Ggauge.
Since \ggauge must embedded inside \e8 for a gSS reduction to exist, then, up to conjugation by an \E8 element, its maximal compact subalgebra must be contained within \spin(16), associated to anti-Hermitian generators.
We can therefore guarantee compactness of \Ggauge\big/\Hgauge by imposing
\begin{equation}\label{compact 1}
\Pi_A{}^F `X_FB^C`+\Pi_A{}^F `X_FC^B`=0\,,
\end{equation}
which makes sure that all coset generators belong to \spin(16).
One may worry that the projector in this equation is not \E8 invariant.
However, we can always perform an Iwasawa-like decomposition of any \E8 element into a \Spin(16) one times an element of $\GL(d)\ltimes\cS$.
The latter preserves $`\Pi_A^B`$ and the former can be dropped for this argument.\footnote{For the same reason, to study the duality orbits of the linear uplift conditions \eqref{Xsym uplift cstr}, \eqref{Hclosure uplift cond} and \eqref{trombone uplift cond}, it is enough to look at \Spin(16) rotations.}

By itself, condition \eqref{compact 1} is too strict, as it excludes for instance reductions on tori with (constant) fluxes, or reductions on circles with duality twists.
These situations are associated to coset generators belonging to the algebra generating $\cS$, rather than \spin(16).\footnote{In the construction of the generalised frame, coset generators within $\cS$ can be omitted from the parametrisation of the coset representative and reabsorbed into $S$, because they are normalised by \GL(d) and hence can be brought through the reference frame $\mathring{e}$. This is important because it means that we do not need to worry about the duality orbit of coset generators within $\cS$, as we did above for compact coset generators.}
To take them into account, it is convenient to define the pseudoinverse $\cE`_m^M`$ such that $\Pi`_M^N` = \cE`_M^m`\cE`_m^N`$ and then impose
\begin{align}\label{compact conditions}
&\cE`_m^A` \big(`X_AB^C`+`X_AC^B`)=0\,, && m=1,\ldots,p 
&&1\leq p\leq d\,.
\nonumber\\
&\cE`_m^A` `X_AB^C` \cE`_C^n` =0\,, && m=p+1,\ldots,d\,.
\end{align}
Notice that one must consider each value of $p$ separately.
Also notice that these equations are invariant under an $\SO(p)\times\SO(d-p)$ subgroup of the \GL(d) group preserved by the choice of section.

\bigskip
Fully exploring the landscape of solutions of such constraints is beyond the scope of this article, but we can look at a few examples.
For eleven-dimensional supergravity, taking $p=d=8$ in \eqref{compact conditions} we find the following conditions among the allowed components \eqref{components uplift 11d}:
\begin{align}\label{11d comp p=0}
`B_m^npq9` = `A^9_mnpq`\,,\qquad
Z_{pm,n} = \delta_{mn} W^{p9,9}\,,\qquad
\Xi`^m9_np` = - \Xi`^n9_mp` \,,
\end{align}
where the last condition is easily interpreted as the compactness condition for the structure constants of a group manifold.
All other embedding tensor components vanish.
A couple simple solutions to the quadratic constraint are give by choosing the following non-vanishing components subject to \eqref{11d comp p=0}:
\begin{equation}
\begin{array}{rclcl}
\boldmath\text{\bf 11d on }S^7\times S^1 &\to & \SO(8)\ltimes\dsR^{28} &:& W^{m9,9}\neq0\,,\\[2ex]
\boldmath\text{\bf 11d on }S^4\times S^3 \times S^1 &\to & \SO(5)\times\SO(3)\times N^{43} &:& `A^9_1234` \neq0\,,
\\[1ex]&&&& 
\Xi`^59_67`=\Xi`^69_75`=\Xi`^79_56`\neq0\,,
\end{array}
\end{equation}
where $N^q$ denotes a unipotent group of dimension $q$.
Notice that the possibility of realising the full \SO(4) isometry group of the three-sphere in $S^4\times S^3$ is ruled out by the no-go result of~\cite{Galli:2022idq}.

Setting $p=6$ we find a gauging of $\SO(4)\times\SO(4)\times N^{44}$ arising from the coset reduction on $S^3\times S^3$.
This seems to differ in the nilpotent part from the gaugings obtained from the same geometry in~\cite{Eloy:2023zzh}.
We do not display here the general identifications analogous to \eqref{11d comp p=0} but directly the set of non-vanishing components for this example:
\begin{equation}
\begin{array}{rclcl}
\boldmath\text{\bf 11d on }S^3\times S^3 \times T^2 &\to &  \SO(4)\times\SO(4)\times N^{44} &:& 
`B_1^2389`{=}`B_2^3189`{=}`B_3^1289`{=}A`^9_1238`=g_1\,,
\\[1ex]&&&& 
`B_4^5689`{=}`B_5^6489`{=}`B_6^4589`{=}A`^9_4568`=g_2
\end{array}
\end{equation}
where $g_{1,2}$ are the gauge couplings of the two \SO(4) factors.
Notice that all these examples can be regarded as massless type IIA reductions.
We can also straightforwardly reproduce the $S^6\times S^1$ reduction of type IIA supergravity with or without a non-vanishing Romans mass $F_0$. 
These models are obtained from circle KK reduction of the $D=4$ \ISO(7) gaugings \cite{DallAgata:2011aa,Guarino:2015vca}:
\begin{equation}
\begin{array}{rclcl}
\hspace*{-5em}\boldmath\text{\bf IIA on } S^6\times T^2 &\to &  \SO(7)\times N^{34} &:& 
Z_{i7,j}=\delta_{ij} W^{79,9}\,,\ \
W^{89,8} \propto F_0\,,
\end{array}
\end{equation}
with $i,j=1,\ldots,6$.

\smallskip
For type IIB supergravity, we have for instance
\begin{equation}
\begin{array}{rclcl}
\boldmath\text{\bf IIB on }S^7 &\to &  \SO(8)\times \dsR^{28} &:& 
Z_{m8,n}=\delta_{mn} \,,\ Z_{89,9}=-1\,,\ \  m,n=1,\ldots,7\,,
\end{array}
\end{equation}
where we have used the global \SL(2,\dsR) to rotate $Z_{m\,\mathsf{i},n}$ to $\mathsf{i}=8$.
We can also reproduce the KK reduction of the S-folds on $S^5\times S^1$ constructed in \cite{Inverso:2016eet}:
\begin{equation}
\begin{array}{rclcl}
\boldmath\text{\bf IIB on }S^5\times S^1 &\to &  \SO(6)\times X\times N^{40} &:& 
\qquad
`B_1^2345`=-`B_2^3451`=\ldots=F_5\,,
\\[1ex]&&&& \qquad
Z_{6\,\mathsf{i},\mathsf{j}}=Z_{6\,\mathsf{j},\mathsf{i}}\neq0
\end{array}
\end{equation}
where $Z_{6\,\mathsf{i},\mathsf{j}}$ determines whether $X=\SO(2),$ \SO(1,1) or $\dsR$. Several other examples are obtained by contraction of the ones above, giving other \CSO(p,q,r) gaugings and their siblings, in analogy with \cite{Hohm:2014qga}.

We could also find the T-dual of the~$S^3\times S^3$ reduction above, as well as reduction on~$S^3\times S^2$ and~$S^4\times S^2$:
\begin{equation}
  \hspace{-.5em}\begin{array}{rclcl}
    \boldmath\!\!\!\!\text{\bf IIB on }S^3\times S^3 \!\!\!&\to & \!\!\!  \SO(4)\times\SO(4)\times N^{44}\!\!\!  &:& 
  \!\!
  `B_8^1237`{=}-`\Xi^45_69`{=}-`\Xi^64_59`{=}-`\Xi^56_49`=g_1
  \\[1ex]&&&& \!\!
  `B_8^4567`=`\Xi^12_39`=`\Xi^23_19`=`\Xi^31_29`=g_2
\\[3ex]
    \boldmath\!\!\!\!\text{\bf IIB on }S^3\times S^2 \!\!\!&\to & \!\!\!  \SO(4)\times\SO(3)\times N^{47}\!\!\!  &:& 
  \!\!
  `B_8^4567`=`\Xi^12_39`=`\Xi^23_19`=`\Xi^31_29`=g_1
  \\[1ex]&&&& \!\!
  `B_8^1234`=-`\Xi^57_69`=`\Xi^67_59`=g_2
  \\[3ex]
      \boldmath\!\!\!\!\text{\bf IIB on }S^4\times S^2 \!\!\!&\to & \!\!\!  \SO(5)\times\SO(3)\times N^{43}\!\!\!  &:& 
    \!\!
    `B_i^jkl7`= g_1 \, \epsilon_{ijkl}\,,\quad i,j,\ldots=1,2,3,4
    \\[1ex]&&&& \!\!    
    `\Xi^56_89`=-g_1
    \\[1ex]&&&& \!\!
    `B_8^1234`=-`\Xi^57_69`=`\Xi^67_59`=g_2
  \end{array}
\end{equation}
We have displayed these in a specific \SL(2) frame for simplicity.

\bigskip

Finally, we point out that in order to allow for arbitrary group manifold reductions, which may be rendered compact by some discrete group quotient, it is sufficient to modify the second of \eqref{compact conditions} to
\begin{equation}
\cE`_m^A`  {\overline\Pi}`_B^F` `X_AF^C` \cE`_C^n` =0\,, \qquad m=p+1,\ldots,d\,.
\end{equation}
which selects generators of $\GL(d)\ltimes\cS$ rather than $\cS$ only.

\subsection{Further observations and no-go results}

Several no-go results have been spelled out in the literature that allow to rule out a higher dimensional origin for some $D=3$ gauged maximal supergravities.
A first observation in this sense was made in \cite{Nicolai:2003bp}, which point out that consistent truncations usually yield Lagrangians where the vector fields have a standard Yang--Mills kinetic terms. However, $D=3$ gauged supergravities with semisimple gaugings only admit Larangians of Chern--Simons type. This seems to rule out a gSS uplift for any $D=3$ gauged supergravity with semisimple gauge group.
Notice however that the Lagrangian \E8 ExFT is itself of Chern--Simons type (although it involves ancillary vector fields)~\cite{Hohm:2014fxa}.
Applying a gSS ansatz, a gauged supergravity Lagrangian of Chern--Simons type is obtained.
It seems therefore less obvious how one can conclude whether or not gSS uplifts for semisimple gaugings should be ruled out.

More recently, \cite{Galli:2022idq} proved that the compact part of a gauge group cannot be `larger than' \SO(9), and ruled out gSS reductions on products of spheres of total dimension 7 or 8, if one requires that the full isometry group of the spheres is realised in the gauge group.

Another recent result is the invariant uplift condition identified in~\cite{Eloy:2023zzh}.
We have computed this condition using our conventions to find the expression
\begin{equation}\label{Malek constraint}
\Theta_{AB}\Theta^{AB} -21 \vartheta_A\vartheta^A + 280 \, \theta^2 = 0\,.
\end{equation}
We have checked that this condition is satisfied for any embedding tensor subject to the linear uplift conditions of section~\ref{sec:linear uplift conds}, for any choice of section, even if we only require existence of an uplift to a (possibly gauged) $D=4$ maximal supergravity.
It is rather straightforward to see that this must be the case, at least for Lagrangian gaugings.
Indeed, since \eqref{Malek constraint} is an \E8 singlet, it is also a singlet under \E7. Restricting the embedding tensor of $D=3$ supergravity to the $\bf912$ of \E7 corresponding to gaugings of $D=4$ supergravity, \eqref{Malek constraint} should reduce to an \E7 singlet quadratic condition on the $D=4$ embedding tensor.
However, there is no singlet in ${\bf912}\times{\bf912}$ and hence such contraction vanishes identically.

Having at our disposal the solutions \eqref{components uplift 11d}, \eqref{components uplift IIB} and \eqref{components uplift IIA} of the linear uplift conditions \eqref{Xsym uplift cstr}, \eqref{Hclosure uplift cond} and \eqref{trombone uplift cond}, we have also computed that \eqref{Malek constraint} generalises to 
\begin{equation}
  \Tr\big(\Theta_{\wedge}^{\,n}\big)\,,\quad 
  \forall n>0\,,\quad 
  \Theta_{\wedge}{}_{\,A}{}^B = \Theta_{AC}\eta^{CB}\,,
\end{equation}
for Lagrangian gaugings ($\vartheta_A=0$) and for any choice of section constraint (i.e., the conditions hold also for uplifts to $D=9,\ldots,4$).
Notice that we have already substituted the covariant condition $\theta=0$.
We have obtained these conditions exactly for $n=3,4$, which we have also checked to be independent from the quadratic condition at the level of representation theory.
We have tested all other $n$ numerically.\footnote{More precisely, we have carried out these computation by explcitly solving the linear uplift constraints in terms of arrays in Mathematica. We then verified for random values of the allowed entries in the emebdding tensor that \mbox{$\Tr(\exp\Theta_{\wedge})=248$} and tested the first few hundred values of $n$. Presumably only a finite set of these conditions are independent.  Also notice that the following alternative cubic contraction is not independent: $\Theta_{AD}\Theta_{BE}\Theta_{CF}`f^ABC``f^DEF` = \frac{20}{3}\mathrm{Tr}\big(\Theta_{\wedge}^{\,3}\big)$.}

\bigskip
Based on the results of the previous sections, we can derive a few interesting no-go results.
A first simple observation we can make, based on the uplift condition \eqref{Thetahat SC}, is that compact gaugings cannot admit a gSS uplift.
To see this, notice that such gaugings must be contained in the $\bf3875$ (trombone gaugings are non-compact and we have proved above that the singlet must vanish for an uplift to exist).
We can ignore central elements in the gauge algebra, because as pointed out in \cite{Inverso:2017lrz} one can always enlarge \hHgauge to include all of them.
We therefore focus on the gauge group \Ggauge as embedded into \E8.
Projecting one index of $\Theta_{AB}$ onto a set of coset generators, \eqref{Thetahat SC} requires that the other index must select a solution of the section constraint, which at the same time must correspond to a subset of generators of the gauge group.
But such solutions correspond to nilpotent $\dsR^d$ subalgebras of \e8.
Clearly, they are not contained in the gauge algebra if the latter is compact.
Notice that while one may hope that the same conclusion applies more generally to semisimple gaugings, groups such as \SL(d) do contain $\dsR^d$ subalgebras solving the section constraint.
A more careful analysis of the interplay between the section constraint~\eqref{Thetahat SC} and the representation and quadratic constraints~\eqref{QCX} on the embedding tensor is required to determine whether or not some semisimple gaugings may admit an uplift.

Another simple application of our results is to check for the existence of an uplift for a large set of $\cN=(8,0)$ AdS$_3$ vacua found in \cite{Deger:2019tem}. 
There, such vacua are found in half-maximal supergravity, but a rather large subset is shown to admit embedding into the maximal theory.
Several of them can be immediately ruled out because they require a non-vanishing singlet component. 
The remaining gauge groups that are not immediately excluded are $\SO(8)^2$, $\SO(7,1)^2$, $\SO(6,2)^2$, $\SO(5,3)^2$ and some contractions $\SO(7)^2$, $\SO(6)^2\times\SO(2)^2$,  and $\SO(5)^2\times\SO(3)^2$.
Direct construction of these gaugings shows that the compact ones are all different truncations of the $\SO(8)\times\SO(8)$ gaugings of maximal supergravity, hence they do not admit an uplift.
The $\SO(p,q)^2$ gaugings embed into maximal supergravity without any modification to the gauge group.
We then rule out the existence of a gSS uplift by computing $\Theta_{AB}\Theta^{AB}\neq0$.
We conclude that none of the $\cN=(8,0)$ AdS$_3$ gaugings constructed in~\cite{Deger:2019tem} that admit embedding into maximal supergravity can be obtained from a gSS reduction.

\section{Conclusions}\label{sec:outro}

In this work we have proved necessary and sufficient conditions for a $D=3$ gauged maximal supergravity to admit a gSS uplift to a higher-dimensional theory.
We have reworked these conditions in terms of linear constraints on the embedding tensor, subject to a choice of solution to the section constraint adn tabulated their solutions.
Any embedding tensor that sits within the components solving these linear constraints, even if just \emph{up to duality rotations}, and that satisfies the quadratic constraint, defines a gSS reduction.
We have also discussed how to impose compactness of the internal space and derived several no-go results.

There are many directions in which the results of this paper can be applied and extended.
A first, natural application is to attempt a classification of gauged maximal supergravities with geometric uplift to ten and eleven dimensions.
This has not yet been attempted even in higher dimensions.
To make things even more interesting, it is important to notice that the linear uplift conditions \eqref{Xsym uplift cstr}, \eqref{Hclosure uplift cond} and \eqref{trombone uplift cond} are invariant under the section-preserving group $\GL(d)\ltimes\cS$, and that this group contains a full Borel subalgebra of \En. This means that we can globally parametrise the scalar manifold $\En/\KEn$ in terms of $\GL(d)\ltimes\cS$.
When searching for vacuum solutions, the critical value of the scalar fields determines a constant coset representative that can thus be taken to belong to $\GL(d)\ltimes\cS$ and reabsorbed into the embedding tensor.
There is thus no loss of generality in searching for vacua only at the origin of the scalar manifold if one takes as unknowns the general solutions of the uplift constraints tabulated in section~\ref{sec:linear uplift conds}, or analogous ones for other \En.
This opens the way to using the techniques of~\cite{DallAgata:2011aa,Dibitetto:2011gm} to carry out an algebraic classification not only of gaugings, but also of vacua with a \emph{guaranteed} uplift to ten and eleven dimensions.
For instance, it would certainly be very interesting to classify gauged maximal supergravities with supersymmetric AdS$_3$ vacua and some amount of supersymmetry, and/or to carry out a similar search for vacua in $D\ge4$, where the lower number of embedding tensor entries might make it possible to carry out full classifications.

Along similar lines, it would be highly desirable to classify all gaugings admitting an uplift on a compact internal space, by using the linear compactness conditions described in section~\ref{ssec: compactness}.
These compactness conditions apply to $D\ge4$ as well, so again it would be desirable to carry out such classifications in different dimensions.
So far, all known \emph{compact} internal spaces supporting gSS reduction have been quotients of group manifolds and/or products of spheres. It would be highly interesting to determine whether or not there exist other classes of compact geometries supporting a gSS frame.

The analysis of twistings and deformations of \E8 generalised diffeomorphisms carried out here may also pave the way to a study of extended versions of a notion of algebroid for \E8, along the lines of \cite{Bugden:2021wxg,Bugden:2021nwl,Hulik:2022oyc,Hulik:2023aks}, and to phrase in that language the conditions for the existence of gSS reductions.
Another outstanding question is how the explicit construction of gSS reductions and associated frames described in \cite{Inverso:2017lrz} and here can help construct consistent truncations based on non-identity generalised $G$-structures and preserving fewer supersymmetries, along the lines of~\cite{Cassani:2019vcl}.

A natural further step in the study of consistent Kaluza--Klein truncations and gauged supergravities is to carry out the same analysis done here for  $D=2$ gauged maximal supergravities~\cite{Samtleben:2007an,Ortiz:2012ib} and generalised Scherk--Schwarz reductions of E$_9$ exceptional field theory~\cite{Bossard:2017aae,Bossard:2018utw,Bossard:2021jix,Bossard:2022wvi,Bossard:2023jid,Bossard:2023wgg}.
While the embedding tensor of $D=2$ maximal supergravity is infinite-dimensional, it was proved in \cite{Bossard:2023jid} that for Lagrangian gaugings, only a finite amount of components can admit a geometric uplift. 
Nonetheless, with the exception of the \SO(9) model of~\cite{Ortiz:2012ib}, $D=2$ gauged supergravities are largely unexplored.
Deriving necessary and sufficient algebraic conditions for a gSS uplift would provide a great motivation to start exploring these theories further.

\section*{Acknowledgements}
GI would like to thank Guillaume Bossard, Franz Ciceri and Axel Kleinschmidt for discussions.

\appendix

\section{Computations}

\subsection{Second component of the generalised torsion}

In this paragraph we check that the second component of the generalised torsion \eqref{dorf to torsion} reads
\begin{equation}
\big({\bbmE}_A\dorf{\bbmE}_B\big)_N = -r^{-1} `E_N^F`\,\tors{E}_{AB}{}^C\,\weitz{E}_{FC} - \frac{1}{60}\,r^{-1}\,\de_N\,\tors{E}_{AC}{}^D\,f_{BD}{}^C.
\end{equation}
As a preliminary step, we decompose the first component of the generalised torsion and we derive an integrability condition for the Weitzenb\"ock connection. Define, on the same footing as the embedding tensor in \eqref{X to Theta e8},
\begin{equation}
\tors{E}_{AB}{}^C = \vartheta[E]_A\,\delta^C_B - \frac{1}{2}\,f_{DB}{}^C\,f^D{}_A{}^E\,\vartheta[E]_E + f^D{}_B{}^C\,\vartheta[E]_{AD}.
\end{equation}
Therefore
\begin{align}
\vartheta[E]_A &= \frac{1}{248}\,\tors{E}_{AB}{}^B,\label{ThetaT1}\\
\vartheta[E]_{AB} &= \frac{1}{60}\,\tors{E}_{AC}{}^D\,f_{BD}{}^C - \frac{1}{2}\,f_{AB}{}^C\,\vartheta[E]_C,\label{ThetaT2}
\end{align}
or, in terms of the components of the Weitzeb\"ock connection,
\begin{align}
\vartheta[E]_A &= \weitz{E}_A + f_A{}^{ED}\,\weitz{E}_{ED},\label{ThetaW1}\\
\vartheta[E]_{AB} &= \weitz{E}_{AB} +\weitz{E}_{BA} - f_{(A}{}^{EF}\,f_{B)}{}^D{}_F\,\weitz{E}_{ED},\label{ThetaW2}
\end{align}
as in \eqref{Theta to W}. 
Notice however $\vartheta[E]_{AB}$ belongs to the $\bf1+3875$ and we do not separate the singlet.
Given the definition of the Weitzenb\"ock connection in terms of the derivatives of the frame, one can show that the following is an identity:
\begin{equation}
E_{[A}{}^N\,\de_N\,\weitz{E}_{B]F}{}^C + \weitz{E}_{[A|F}{}^E\,\weitz{E}_{B]E}{}^C - \weitz{E}_{[AB]}{}^E\,\weitz{E}_{EF}{}^C = 0.
\end{equation}
If we define $\weitz{E}_{AB}{}^C =: E_A{}^M\,\tilde{W}^{[E]}_{MB}{}^C = E_A{}^M\,(\tilde{W}^{[E]}_M)_B{}^C$, it assumes a simpler form:
\begin{equation}\label{FlatnessW}
\Big(2\,\de^{\vphantom{\sum}}_{[M}\,\tilde{W}^{[E]}_{N]} + [\tilde{W}^{[E]}_M,\tilde{W}^{[E]}_N\big]\Big)\!{}^{\vphantom{\sum}}_B{}_{\vphantom{\sum}}^A = 0,
\end{equation}
which shows that the Weitzenb\"ock connection is flat. Then, the components $E_A{}^M\,\tilde{W}^{[E]}_{MB} = \weitz{E}_{AB}$ and $E_A{}^M\,\tilde{W}^{[E]}_M = \weitz{E}_A$ satisfy
\begin{align}
& \de^{\vphantom{\sum}}_{[M}\,\tilde{W}^{[E]}_{N]} = 0,\\
& \de^{\vphantom{\sum}}_{[M}\,\tilde{W}^{[E]}_{N]E} = \frac{1}{2}\,f^{AB}{}_C\,\tilde{W}^{[E]}_{MA}\,\tilde{W}^{[E]}_{NB}.
\end{align}
To get the second relation, one has to separate the symmetric and the antisymmetric part of \eqref{FlatnessW} in $A, B$ (the symmetric part is trivial) and to use the Jacobi identity. Finally, multiplying the second relation by $E_A{}^M$, one uses this relation, which will be useful later:
\begin{equation}\label{IntegrabilityRelation}
E_A{}^M\,\de_M\,\tilde{W}^{[E]}_{NB} = \de_N\,\big(E_A{}^M\,\tilde{W}^{[E]}_{MB}\big) - \de_N\,E_A{}^M\,\tilde{W}^{[E]}_{MB} + E_A{}^M\,f^{CD}{}_B\,\tilde{W}^{[E]}_{MC}\,\tilde{W}^{[E]}_{ND}.
\end{equation}

Consider now the second component of the generalised torsion. By definition, it is equal to
\begin{align}
\big({\bbmE}_A\dorf{\bbmE}_B\big)_N =\ & 
-r^{-1}\,\tilde{W}^{[E]}_{NB}\,\de_M\,E_A{}^M -
E_A{}^M\,\tilde{W}^{[E]}_{NB}\,\de_M\,r^{-1} -
r^{-1}\,E_A{}^M\,\de_M\,\tilde{W}^{[E]}_{NB} \,+\nn\\
& - r^{-1}\,\tilde{W}^{[E]}_{MB}\,\de_N\,E_A{}^M -
E_B{}^M\,\tilde{W}^{[E]}_{MA}\,\de_N\,r^{-1} -
r^{-1}\,E_B{}^M\,\de_N\,\tilde{W}^{[E]}_{MA} \,+\nn\\
& + r^{-1}\,E_C{}^P\,E_M{}^D\,f_B{}^C{}_D\,\de_N\,\de_P\,E_A{}^M.
\end{align}
Use the following relation to replace the derivative of $r$:\footnote{Consider $\weitz{E}_{AB}{}^C$ and replace $E_A{}^M$ with $r^{-1}\,(U^{-1})_A{}^M$, as in \eqref{EUr}:
\begin{equation*}
\weitz{E}_{AB}{}^C = -U^C{}_N\,(E_A{}^M\,\de_M)\,(U^{-1})_B{}^N -\delta_B^C\,r\,E_A{}^M\,\de_M\,r^{-1},
\end{equation*}
which implies
\begin{equation*}
\frac{1}{2}\,\weitz{E}_A = - r\,E_A{}^M\,\de_M\,r^{-1},\quad 
\weitz{E}_{AD}\,f^D{}_B{}^C = -U^C{}_N\,E_A{}^M\,\de_M\,(U^{-1})_B{}^N.
\end{equation*}}
\begin{equation}
\de_M\,r^{-1} = -\frac{1}{2}\,r^{-1}\,\tilde{W}^{[E]}_M.
\end{equation}
Then, replace the derivatives of $E_A{}^M$ in terms of the $\weitz{E}_{AB}{}^C$ and decompose the last in its components $\weitz{E}_{AB}$ and $\weitz{E}_A$. One gets
\begin{align}
r\,({\bbmE}_A\dorf{\bbmE}_B)_N =\ & 
\frac{1}{4}\,E_C{}^M\,f_{AB}{}^C\,\tilde{W}^{[E]}_M\,\tilde{W}^{[E]}_N +
\frac{1}{2}\,E_B{}^M\,\tilde{W}^{[E]}_N\,\tilde{W}^{[E]}_{MA} +
\frac{1}{2}\,E_A{}^M\,\tilde{W}^{[E]}_N\,\tilde{W}^{[E]}_{MB} \,+\nn\\
& - \frac{1}{2}\,E_C{}^M\,f_A{}^{EF}\,f_B{}^C{}_F\,\tilde{W}^{[E]}_N\,\tilde{W}^{[E]}_{ME} + 
E_C{}^M\,f_A{}^{CE}\,\tilde{W}^{[E]}_{ME}\,\tilde{W}^{[E]}_{NB} \,+\nn\\
& - \frac{1}{2}\,E_C{}^M\,f_A{}^{EF}\,f_B{}^C{}_F\,\tilde{W}^{[E]}_M\,\tilde{W}^{[E]}_{NE} +
E_C{}^M\,f_A{}^C{}^E\,\tilde{W}^{[E]}_{MB}\,\tilde{W}^{[E]}_{NE} \,+\nn\\
& - E_C{}^M\,f_A{}^{DF}\,f_B{}^C{}^G\,f^E{}_{FG}\,\tilde{W}^{[E]}_{MD}\,\tilde{W}^{[E]}_{NE} -
E_A{}^M\,\de_M\,\tilde{W}^{[E]}_{NB} \,+\nn\\
& - E_B{}^M\,\de_N\,\tilde{W}^{[E]}_{MA} -
\frac{1}{2}\,E_C{}^M\,f_{AB}{}^C\,\de_N\,\tilde{W}^{[E]}_M + E_C{}^M\,f_A{}^E{}^F\,f_B{}^C{}_F\,\de_N\,\tilde{W}^{[E]}_{NE},
\end{align}
Now, use the integrability condition \eqref{IntegrabilityRelation} to replace the last term in the last-but-one line, and use the Leibniz identity in the terms in the last line, for example writing $E_C{}^M\,\de_N\,\tilde{W}^{[E]}_M$ as $\de_N\,(E_C{}^M\,\tilde{W}^{[E]}_M) - \de_N\,E_C{}^M\,\tilde{W}^{[E]}_M$. Then, replace again the derivatives of $E_A{}^M$ with $\weitz{E}_{AB}{}^C$. Doing so, one gets 
\begin{align}
r\,({\bbmE}_A\dorf{\bbmE}_B)_N =\ & 
E_A{}^M\,\tilde{W}^{[E]}_M\,\tilde{W}^{[E]}_{NB}+
E_C{}^M\,f_A{}^{CE}\,\tilde{W}^{[E]}_{ME}\,\tilde{W}^{[E]}_{NB} \,+\nn\\
& + E_A{}^M\,f_B{}^{DE}\,\tilde{W}^{[E]}_{ME}\,\tilde{W}^{[E]}_{ND} -
\frac{1}{2}\,E_C{}^M\,f_A{}^{EF}\,f_B{}^C{}_F\,\tilde{W}^{[E]}_M\,\tilde{W}^{[E]}_{NE} \,+\nn\\
& - \frac{1}{2}\,E_C{}^M\,f_{ABF}\,f^{CEF}\,\tilde{W}^{[E]}_M\,\tilde{W}^{[E]}_{NE} -
E_C{}^M\,f_B{}^{CE}\,\tilde{W}^{[E]}_{MA}\,\tilde{W}^{[E]}_{NE} \,+\nn\\
& - E_C{}^M\,f_A{}^{DF}\,f_{BF}{}^G\,f^{CE}{}_G\,\tilde{W}^{[E]}_{MD}\,\tilde{W}^{[E]}_{NE} -
E_C{}^M\,f_A{}^{DF}\,f_B{}^{CG}\,f^E{}_{FG}\,\tilde{W}^{[E]}_{MD}\,\tilde{W}^{[E]}_{NE} \,+\nn\\
& - \frac{1}{2}\,f_{AB}{}^C\,\de_N\,\weitz{E}_C -
2\,\de_ N\,\weitz{E}_{(AB)} + 
f_A{}^{DE}\,f_B{}^C{}_E\,\de_N\,\weitz{E}_{CD}.
\end{align}
Notice that the last line is the total derivative of
\begin{align}
-\frac{1}{60}\,\tors{E}_{AC}{}^D\,f_{BD}{}^C
&= -\vartheta[E]_{AB} - \frac{1}{2}\,f_{AB}{}^C\,\vartheta[E]_C \\\nn
&= -\frac{1}{2}\,f_{AB}{}^C\,\weitz{E}_C -
2\,\weitz{E}_{(AB)} + f_A{}^{DE}\,f_B{}^C{}_E\,\weitz{E}_{CD}
\end{align}
by means of the relations \eqref{ThetaT1}--\eqref{ThetaW2}. So it remains to show that the first four lines and $\tors{E}_{AB}{}^{C}\,\tilde{W}^{[E]}_{NC}$ sum to zero. Using again the relations \eqref{ThetaT1}--\eqref{ThetaW2}, one can write $\tors{E}_{AB}{}^{C}$ in terms of $\tilde{W}^{[E]}_M$ and $\tilde{W}^{[E]}_{MA}$, so that one arrives to
\begin{align}
&r\,({\bbmE}_A\dorf{\bbmE}_B)_N + \tors{E}_{AB}{}^{C}\,\tilde{W}^{[E]}_{NC}\ =\  
E_C{}^M\,\,\tilde{W}^{[E]}_{ME}\,\tilde{W}^{[E]}_{ND}\,\times\\[1ex]\nonumber&
\times\,\Big[f_A{}^{EF}\,f_B{}^{DG}\,f_{CFG} -
f_A{}^{EF}\,f_{BC}{}^G\,f^D{}_{FG} - 
\frac{1}{2}\,f_{AB}{}^F\,f_C{}^{DG}\,f^E{}_{FG} + 
f_C{}^{DF}\,f_{(A}{}^{EG}\,f_{B)FG}\Big],
\end{align}
which vanishes by using the Jacobi identity twice.

\subsection{Deformations/twistings of Dorfman product: Second component}

In this section we show that the second component of the deformed Dorfman product \eqref{dorfman F} satisfies the Leibniz identity \eqref{dorfleib F}, if the conditions \eqref{Xconstr local F e8}--\eqref{BI local F e8} are satisfied. 
It is useful to decompose the Leibniz identity  \eqref{dorfleib F} into its symmetric and antisymmetric parts
\begin{align}
\frac{1}{2}\,(\bbLambda_1 \overset{F}\circ \bbLambda_2 + \bbLambda_2 \overset{F}\circ \bbLambda_1)\,\overset{F}\circ \bbLambda_3 = 0,\\
\bbLambda_1 \overset{F}\circ (\bbLambda_2\overset{F}\circ \bbLambda_3) - \bbLambda_2 \overset{F}\circ (\bbLambda_1\overset{F}\circ \bbLambda_3) - \frac{1}{2}\,(\bbLambda_1 \overset{F}\circ \bbLambda_2 - \bbLambda_2 \overset{F}\circ \bbLambda_1)\,\overset{F}\circ \bbLambda_3 = 0.
\end{align}

It is useful to parametrise the flux in the following way:
\begin{equation}\label{Fparametrisation}
F_{MN}{}^P = \varphi_M\,\delta_N^P - \frac{1}{2}\,f^R{}_N{}^P\,f_{RM}{}^S\,\varphi_S + f^{R}{}_N{}^P\,\varphi_{MR}\,,
\end{equation}
where for convenience we have defined, with reference to \eqref{F irrep deco},
\begin{equation}
\varphi_{MN} = \Phi_{MN} + \phi\,\eta_{MN}\,.
\end{equation}
The inverse relations are
\begin{equation} \label{DualExtraConstraint}
\varphi_M = \frac{1}{248}\,F_{MN}{}^N, \quad
\varphi_{MN} = -\frac{1}{60}\,F_{MR}{}^S\,f_N{}^R{}_S - \frac{1}{2}\,\frac{1}{248}\,f_{MN}{}^R\,F_{RS}{}^S.
\end{equation}
The parametrisation of the flux \eqref{Fparametrisation} is the same of the torsion in terms of $\vartheta[E]_A$ and $\vartheta[E]_{AB}$ in \eqref{ThetaT2}. In particular, we have that $\varphi_{MN} = \varphi_{NM}$ belongs to the ${\bf3875}+{\bf1}$.

The possible terms in the antisymmetric part of the Leibniz identity, which should vanish, are the following:
\begin{itemize}
\item[--] $\de\Sigma_3$, which is zero using \eqref{Xconstr local F e8};
\item[--] $\Sigma_3\,\Lambda_2\,\de\Lambda_1$ and $\Sigma_3\,\Lambda_1\,\de\Lambda_2$, which are zero on the trace of \eqref{Cconstr local F simple e8};
\item[--] $\Sigma_3\,\Lambda_1\,\Lambda_2$, which is zero on the trace of \eqref{BI local F e8};
\item[--] $\Sigma_3\,\Lambda_2\,\Sigma_1$ and $\Sigma_3\,\Lambda_1\,\Sigma_2$, which are zero on the trace of \eqref{F extra cond};
\item[--] $\Lambda_3\,\Lambda_2\,\de\de\Lambda_1$ and $\Lambda_3\,\Lambda_1\,\de\de\Lambda_2$, which are zero using \eqref{Fparametrisation}, \eqref{Cconstr local F simple e8}, the Jacobi identity and the $f$ invariance;
\item[--] $\Lambda_3\,\de\Lambda_2\,\de\Lambda_1$, which is zero using \eqref{Cconstr local F simple e8} and the Jacobi identity;
\item[--] $\Lambda_3\,\Lambda_2\,\de\Sigma_1$ and $\Lambda_3\,\Lambda_1\,\de\,\Sigma_2$, which are zero using \eqref{F extra cond};
\item[--] $\Lambda_3\,\Lambda_2\,\de\Lambda_1$ and $\Lambda_3\,\Lambda_1\,\de\Lambda_2$, which are zero using \eqref{Fparametrisation}, \eqref{Cconstr local F simple e8}, \eqref{BI local F e8} and the Jacobi identity;
\item[--] $\Lambda_3\,\Lambda_2\,\Lambda_1$, which is zero using \eqref{Fparametrisation} and  \eqref{BI local F e8};
\item[--] $\Lambda_3\,\Sigma_1\,\de\Lambda_2$ and $\Lambda_3\,\Sigma_2\,\de\Lambda_1$, which are zero using \eqref{Cconstr local F simple e8}, \eqref{F extra cond} and imposing $\varphi_{MN}$ to be symmetric;
\item[--] $\Lambda_3\,\Sigma_1\,\Lambda_2$ and $\Lambda_3\,\Sigma_2\,\Lambda_1$, which are the derivative of the previous coefficient, so they are zero.
\end{itemize}

Therefore, we conclude that the second component of the flux-deformed Dorfman derivative satisfies the antisymmetric part of Leibniz identity if we impose the same constraints, which are needed the first component to satisfy the Leibniz identity, and if we suppose that the flux sits in the representations of the embedding tensor.

One can proceed similarly for the symmetric part, showing that it is also satisfied. The coefficients of the terms $\de\Sigma_3$, ($\Sigma_3\,\Lambda_2\,\de\Lambda_1$), ($\Sigma_3\,\Lambda_1\,\de\Lambda_2$) and ($\Sigma_3\,\Lambda_2\,\Lambda_1$) vanish thanks to the symmetric part of the constraint \eqref{Xconstr local F e8}, the trace of the constraint \eqref{full Cconstr e8} and the trace of the (symmetric part of the) Bianchi identity \eqref{BI local F e8}; the other possible structures have the same coefficients as the corresponding ones of the antisymmetric part.

As an example, let us show in detail that the coefficient of the terms $(\Lambda_3\,\Sigma_1\,\de\Lambda_2)$ are equal to zero, using the already known constraints. The coefficient we are interested in is the following:
\begin{align}
\frac{1}{2}\,{\Sigma_1}_Q\,\Lambda_3^N\,\de_N\,\Lambda_2^R
\Bigg[\frac{1}{248}\,\delta_M^Q\,F_{RP}{}^P 
+ \frac{1}{60}\,f^Q{}_R{}^P\,\big(F_{PS}{}^T\,f_M{}^S{}_T\big) + \frac{1}{30}\,f^Q{}_M{}^P\,\big(F_{RS}{}^T\,f_P{}^S{}_T\big)\Bigg]\,.
\end{align}
Consider the constraint \eqref{F extra cond}, which we can rewrite as
\begin{equation}
\Big(\frac{1}{248}\,f^{QT}{}_S\,F_{RP}{}^P + f^{QP}{}_R\,F_{PS}{}^T\Big)\,\mathcal{E}_Q{}^q = 0.
\end{equation}
Multiply it by $f_M{}^S{}_T$ and use the Cartan-Killing metric
\begin{equation}\label{DualNewConstraint}
\frac{1}{248}\,\mathcal{E}_Q{}^q\,\delta_M^Q\,F_{RP}{}^P = \frac{1}{60}\,\mathcal{E}_Q{}^q\,f^Q{}_R{}^P\,\big(F_{PS}{}^T\,f_M{}^S{}_T\big)\,.
\end{equation}
Replacing the last term in the coefficient with the previous expression (remembering that $\Sigma_1$ is on section, so that ${\Sigma_1}_Q \propto \mathcal{E}_Q{}^q$),
\begin{equation}
-\frac{1}{2}\,\mathcal{E}_Q{}^q\,{\Lambda_3}^N\,\de_N\,{\Lambda_2}^R\,\frac{1}{30}\,\Big[f^Q{}_R{}^P\,\big(F_{PS}{}^T\,f_M{}^S{}_T\big) + f^Q{}_M{}^P\,\big(F_{RS}{}^T\,f_A{}^S{}_T\big)\Big].
\end{equation}
Now, replace the components of the flux
\begin{equation}
F_{MP}{}^P = 248\,\varphi_M, \quad
F_{MS}{}^T\,f_N{}^S{}_T = -60\,\Big(\varphi_{MN} -\frac{1}{2}\,f_{MN}{}^P\,\varphi_P\Big),
\end{equation}
so that the coefficient becomes
\begin{equation}
-\frac{1}{2}\,\mathcal{E}_Q{}^q\,{\Lambda_3}^N\,\de_N\,{\Lambda_2}^R\,\frac{1}{30}\,\Big(\frac{60}{248}\,f^Q{}_{[R}{}^P\,f_{M]}{}^T{}_P\,\varphi_T + 60\,f^Q{}_{R}{}^P\,\varphi_{PM} + 60\,f^Q{}_{M}{}^P\,\varphi_{RP}
\Big)\,.
\end{equation}
Now, using the Jacobi identity, the first term can be rewritten as 
\begin{equation}
2\,\mathcal{E}_Q{}^q\,f^Q{}_{[R}{}^P\,f_{M]}{}^T{}_P\,\varphi_T = -\mathcal{E}_Q{}^q\,f_{RM}{}^P\,f^{QS}{}_P\,\varphi_S = 0,
\end{equation}
which vanishes as a consequence of \eqref{Cconstr local F simple e8}.

In order to prove the last two terms to vanish, consider the expression \eqref{DualNewConstraint}, replacing the flux components:
\begin{equation}
\varphi_R\,\mathcal{E}_M{}^q = \mathcal{E}_Q{}^q\,f^Q_R{}^P\,\Big(\varphi_{PM} - \frac{1}{2}\,f_{PM}{}^S\,\varphi_S\Big)
\end{equation}
and take the symmetric part in $R,M$:
\begin{equation}
2\,\varphi_{(R}\,\mathcal{E}_{M)}^q = \mathcal{E}_Q{}^q\,f^Q{}_{(R}{}^P\,f_{M)}{}^S{}_P\,\varphi_S -\mathcal{E}_Q{}^q\,f^Q{}_{(R}{}^P\,\varphi_{M)P}.
\end{equation}
But the constraint \eqref{Cconstr local F simple e8} implies the left-hand side to be equal to the first term in the right-hand side, so that we arrive to the following expression
\begin{equation}
\mathcal{E}_Q{}^q\,f^Q{}_{(R}{}^P\,\varphi_{M)P} = 0,
\end{equation}
which is precisely what is needed for the last two terms in the coefficient to vanish, also recalling that $\varphi_{MN}$ is symmetric.

\section{Cocycle conditions and trivial parameters}
\label{app:cocycle}

Let us first consider the simple cases of a vector and of a two-form gauge potential.
We take a good cover of the internal space.
For one-form potentials, on a coordinate patch $U_{\sf a}$ we have $F^{(2)} = \dd A_{\sf a}^{(1)}$ (using $^{(p)}$ to indicate form degree).
On double overlaps $U_{\sf ab} = U_{\sf a}\cap U_{\sf b}$ we find
\begin{equation}
A^{(1)}_{\sf a} = A^{(1)}_{\sf b} + \dd \lambda^{(0)}_{\sf ba}\,,
\end{equation}
and on triple ones $U_{\sf abc}$,
\begin{equation}
\lambda^{(0)}_{\sf ab}+\lambda^{(0)}_{\sf bc}+\lambda^{(0)}_{\sf ca} = \text{constant}\,,
\end{equation}
where the constant identifies the cohomology class of $F^{(2)}$.%
\footnote{For $A^{(1)}$ a \U(1) connection and with standard normalisations the constant is $2\pi\dsZ$. However we do not commit to any normalisations in the following.}

For two-forms, 
on a patch $U_{\sf a}$ we have $H^{(3)} = \dd B_{\sf a}^{(2)}$ and on double overlaps
\begin{equation}\label{appendix B2 patching 01}
B^{(2)}_{\sf a} = B^{(2)}_{\sf b} + \dd \lambda^{(1)}_{\sf ba}\,,
\end{equation}
on triple ones,
\begin{equation}\label{appendix B2 patching 02}
\lambda^{(1)}_{\sf ab}+\lambda^{(1)}_{\sf bc}+\lambda^{(1)}_{\sf ca} = \dd \xi_{\sf abc}^{(0)}\,,
\end{equation}
and finally on quadruple ones
\begin{equation}\label{appendix B2 patching 03}
\xi_{\sf abc}^{(0)} - \xi_{\sf bcd}^{(0)} + \xi_{\sf cda}^{(0)} - \xi_{\sf dab}^{(0)} = \text{constant}\,.
\end{equation}
Higher form versions of these relations work similarly.
Multiple $p$-forms may be intertwined (e.g. the 11d supergravity six-form transforms under the three-form gauge symmetry).

Moving to generalised geometry and extended field theories, let us first focus on the case without ancillaries.
Generalised vectors $V^M$ are patched on double overlaps as
\begin{equation}
V_{\sf a}^M = V_{\sf b}^N \Gamma_{\sf ab}{}_N{}^M
\end{equation}
with $\Gamma_{\sf ab}\in\cP$ satisfying the torsion condition~\eqref{C to CGamma}.
We are leaving as understood that all quantities are written in the same coordinate system (any one valid on the overlap).
Since patching must be done by symmetries of the theory, we conclude that can interpret the action of $\Gamma_{\sf ab}$ as the exponential of a generalised Lie derivative acting on $ V_{\sf b}^M$, with some appropriate choice of gauge parameter
\begin{equation}\label{appendix genvec patching}
V_{\sf a}^M = V_{\sf b}^N \Gamma_{\sf ab}{}_N{}^M = \exp\!\big(\cL_{\Lambda_{\sf ab}}\big) V^M_{\sf b}\,,\qquad \Lambda_{\sf ab}^M\partial_M = 0\,.
\end{equation}
The later requirement is implied by the equivalence with $\Gamma_{\sf ab}\in\cS$ but it is useful to spell it out explicitly.
Notice that, because of it, $\cL_{\Lambda_{\sf ab}}$ acts only algebraically.
Also notice that $\Lambda_{\sf ab}^M$ is only defined by the above relation up to trivial parameters, i.e. $\Lambda_{\sf ab}\simeq\Lambda_{\sf ab}+\widetilde\Lambda_{\sf ab}$ if $\cL_{\widetilde\Lambda_{\sf ab}} = 0$.
This is analogous to $\lambda^{\smash{(1)}}_{\sf ab}$ being defined up to exact pieces in the patching of a two-form potential.

On triple overlaps,~\eqref{appendix genvec patching} implies that 
$
\Lambda^M_{\sf ab}+\Lambda^M_{\sf bc}+\Lambda^M_{\sf ca}
$
must equal a trivial parameter:
\begin{equation}\label{appendix genvec cocycle}
\Lambda^M_{\sf ab}+\Lambda^M_{\sf bc}+\Lambda^M_{\sf ca} = \big(\widehat\partial \xi_{\sf abc}\big)^M\,,
\end{equation}
where we denoted $\widehat\partial$ the linear operator that maps the space of all independent trivial parameters into the $\Rv$ representation.
It is, by definition, the projector that defines the external spacetime two-form contribution to the covariant vector field strengths in ExFT.
It may not necessarily act only differentially, as we shall see below.

As a simple example, we take the patching of a two-form potential and write it in terms of double field theory objects.
The \O(d,d) invariant is 
\begin{equation}
\eta_{MN} = \begin{pmatrix}
&\delta^m{}_n\\\delta_m{}^n&
\end{pmatrix}\,,\qquad m=1,\ldots,d\,,
\end{equation}
with $\Rv$ equal to the \O(d,d) vector representation.
To relate the formalism to the example above, we solve the section constraint by setting $\partial_M=(\partial_m\,,0)$.
Generalised vectors patch as in \eqref{appendix genvec patching}, reflecting a twisting by an internal $B^{(2)}$ potential.
We identify 
\begin{equation}
\Lambda_{\sf ab}^M = \begin{pmatrix}
0\\\lambda_{{\sf ab}\, m}
\end{pmatrix}\,,
\end{equation}
which is the same object appearing in \eqref{appendix B2 patching 01}.
Then, we have on double overlaps
\begin{equation}
\Lambda^M_{\sf ab}+\Lambda^M_{\sf bc}+\Lambda^M_{\sf ca} 
= 
\eta^{MN}\partial_N \xi_{\sf abc}
\end{equation}
which reflects the well-known form of trivial parameters in DFT and reproduces \eqref{appendix B2 patching 02}.
In order to display the final cocycle condition \eqref{appendix B2 patching 03} one needs to define the operator next to $\widehat\partial$ in an exact sequence, as determined by the tensor hierarchy.

A second instructive example is \E7 ExFT.
In this case from \eqref{appendix genvec patching} and \eqref{appendix genvec cocycle} one has
\begin{equation}
\Lambda^M_{\sf ab}+\Lambda^M_{\sf bc}+\Lambda^M_{\sf ca} 
= 
t_\alpha^{MN} \partial_N\xi_{\sf abc}^\alpha + \Omega^{MN} \xi_{{\sf abc}\,M}\,,
\end{equation}
where $`t_\alpha_M^N`$ are the \e7 generators, $\Omega_{MN}$ is its symplectic invariant use to raise/lower indices and $\xi_{{\sf abc}\,M}$ is constrained to be on section on its $\Rv$ index.
The presence of this extra, non-derivative term is needed to render inert the cocycle condition of the components within $\Lambda^M_{\sf ab}$ that would be associated to a dual graviton---namely, the components with highest \GL(1) degree in the decomposition of $\Rv=\bf56_{1}$.
Indeed, \eqref{appendix genvec patching} does not determine such components which may therefore be set to arbitrary values on each double overlap.
Correspondingly, on quadruple overlaps one finds identically
\begin{equation}
\xi_{{\sf abc}\,M}-\xi_{{\sf bcd}\,M}+\xi_{{\sf cda}\,M}-\xi_{{\sf dab}\,M}=0\,.
\end{equation}
The higher order conditions for $\xi_{\sf abc}^\alpha$ are non-trivial.
Following the ExFT tensor hierarchy, they will inevitably involve further constrained components that appear without derivatives.
In general one may need several further levels---higher than the external top-forms in the tensor hierarchy---to encode the cohomology of all fluxes associated to a certain extended generalised geometry.
We do not attempt to give a full description here.

\bigskip

The logic for \E8 ExFT is analogous, provided we work in terms of doubled gauge parameters $\bbLambda=(\Lambda^M\,,\,\Sigma_M)$ and the Dorfman product.
On double overlaps we have
\begin{equation}
\bbmV_{\sf a} = {^{\Gamma_{\!\sf ba}\!}\bbmV_{\sf b} }
= \exp\!\Big( \bbLambda_{\sf ba}\dorf{} \Big) \bbmV_{\sf b}\,,
\end{equation}
and again $\Lambda^M_{\sf ab}\partial_M = 0$.
The cocycle conditions on triple overlaps are then deduced to be
\begin{align}\label{appendix:triple overlap e8}
&\bbLambda_{\sf ab}+\bbLambda_{\sf bc}+\bbLambda_{\sf ca} =
\big( \Lambda_{\sf abc}^M\ ,\ \Sigma^{\vphantom{M}}_{{\sf abc}\,M} \big) =
\\[1ex]\nonumber&\ 
=\Big(\, 
(\bbP_{\str\bf3875})`_PQ^MN` \partial_N `\xi_{\sf abc}^PQ`
+\eta^{MN} `\Xi_{{\sf abc}\,N}`
+`f^MN_P` \Xi'`_{{\sf abc}\,N}^P`
\ \,,\ \,
\partial_M\Xi'`_{{\sf abc}\,P}^P`
+\partial_P\Xi'`_{{\sf abc}\,M}^P`
\, \Big)
\end{align}
where the right-hand side is a linear combination of trivial parameters.
The bare parameters $`\Xi_{{\sf abc}\,M}`$ and $\Xi'`_{{\sf abc}\,M}^N`$ are constrained to be on section in their lower index.
It is rather straightforward to identify their role.
With reference to the $\Rv$ decomposition of the two maximal sections in~\eqref{248 11d branch} and~\eqref{248 IIB branch}, $`\Xi_{{\sf abc}\,M}`$ encodes the arbitrariness in the patching of the highest-grade component within $\Lambda^M_{\sf ab}$.
This is indeed a trivial parameter.
The other components of non-negative degree are also arbitrary and this is encoded in $\Xi'`_{{\sf abc}\,M}^N`$.
In particular, notice that the components of zero degree within $\Lambda^M_{\sf abc}$ are rendered trivial thanks to the contribution of $\Xi'_{\sf abc}$ to the ancillary $\Sigma^{\vphantom{M}}_{{\sf abc}\,M}$.
In fact, this is the only contribution to the ancillary.
This observation implies that if we set to vanish the zero-degree components of $\Lambda^M_{\sf ab}$, which we can do without affecting the patching of gauge parameters and fields, then the ancillary transition functions $\Sigma_{{\sf ab}\,M}$ have trivial triple-overlap:
\begin{equation}
\Sigma_{{\sf ab}\,M}+\Sigma_{{\sf bc}\,M}+\Sigma_{{\sf ca}\,M} = 0\,.
\end{equation}
Therefore, we have $\Sigma_{{\sf ab}\,M} = \Sigma_{{\sf a}\,M} - \Sigma_{{\sf b}\,M}$ for some $\Sigma_{{\sf a}\,M}$ well-defined on each coordinate patch.
Since ancillary parameters encode gauge transformations in ExFT, we are then free to gauge away these $\Sigma_{{\sf a}\,M}$.
We are left with transition functions on simple overlaps being determined by the components of $\Lambda_{{\sf ab}}^M$ corresponding to $p$-form gauge transformations, encoding background fluxes on the internal space just as in lower-rank exceptional generalised geometries.

\small

\providecommand{\href}[2]{#2}\begingroup\raggedright\endgroup
  
\end{document}